\documentclass[usenatbib,useAMS]{mn2e}
\usepackage[pdftex]{graphicx}
\usepackage{epstopdf}
\usepackage{ctable}
\usepackage{url}
\usepackage{times}
\usepackage{xspace}

\newcommand{\ps}{\rm s^{-1}}

\newcommand{\msun}{~\mathrm{M_{\odot}}}
\newcommand{\msunperyr}{~\mathrm{M_{\odot} {\rm ~yr}^{-1}}}

\newcommand{\cs}{c_{\rm s}}
\newcommand{\rmax}{r_{\rm max}}

\newcommand{\gadget}{\textsc{gadget}\xspace}
\newcommand{\gadgettwo}{\textsc{gadget-2}\xspace}
\newcommand{\gadgetthree}{\textsc{gadget-3}\xspace}
\newcommand{\arepo}{\textsc{arepo}\xspace}
\newcommand{\eorbit}{\textsf{e}-orbit\xspace}
\newcommand{\forbit}{\textsf{f}-orbit\xspace}
\newcommand{\urlformovies}{http://www.cfa.harvard.edu/itc/research/arepomerger/}
\newcommand{\mw}{\textsf{MW}\xspace}
\newcommand{\smc}{\textsf{SMC}\xspace}
\newcommand{\sbc}{\textsf{Sbc}\xspace}
\newcommand{\hiz}{\textsf{HiZ}\xspace}
\newcommand{\mwe}{\textsf{MW-e}\xspace}
\newcommand{\smce}{\textsf{SMC-e}\xspace}
\newcommand{\sbce}{\textsf{Sbc-e}\xspace}
\newcommand{\hize}{\textsf{HiZ-e}\xspace}
\newcommand{\mwf}{\textsf{MW-f}\xspace}
\newcommand{\smcf}{\textsf{SMC-f}\xspace}

\newcommand{\hizf}{\textsf{HiZ-f}\xspace}

\newcommand{\acknowledgments}{\begin{small}\section*{Acknowledgments}\end{small}}

\title[Galaxy mergers on a moving mesh]{Galaxy mergers on a moving mesh: a comparison with smoothed-particle hydrodynamics}

\author[C.~C. Hayward et al.]{
	\parbox[t]{\textwidth}{
		Christopher C. Hayward$^{1}$\thanks{E-mail: christopher.hayward@h-its.org}, Paul Torrey$^2$, Volker Springel$^{1,3}$, 
		Lars Hernquist$^2$, and Mark Vogelsberger$^2$}
	\vspace*{6pt} \\
	$^1$Heidelberger Institut f\"ur Theoretische Studien, Schloss-Wolfsbrunnenweg 35, 69118 Heidelberg, Germany \\
	$^2$Harvard-Smithsonian Center for Astrophysics, 60 Garden Street, Cambridge, MA 02138, USA \\
	$^3$Zentrum f\"ur Astronomie der Universit\"at Heidelberg, Astronomisches Recheninstitut, M\"onchhofstrasse 12-14, 69120 Heidelberg, Germany
}

\usepackage{hyperref}

\begin{document}

\date{Accepted for publication in MNRAS}

\pagerange{\pageref*{firstpage}--\pageref*{lastpage}} \pubyear{2014}

\maketitle

\label{firstpage}

\begin{abstract}

Galaxy mergers have been investigated for decades using smoothed
particle hydrodynamics (SPH), but recent work highlighting
inaccuracies inherent in the traditional SPH technique calls into
question the reliability of previous studies.  We explore this
issue by comparing a suite of \gadgetthree~SPH simulations of
idealised (i.e., non-cosmological) isolated discs and galaxy mergers
with otherwise identical calculations performed using the 
moving-mesh code \arepo.  When black hole (BH) accretion and
active galactic nucleus (AGN)
feedback are not included, the star formation histories (SFHs)
obtained from the two codes agree well.  When BHs are included, the
code- and resolution-dependent variations in the SFHs are more
significant, but the agreement is still good, and the stellar mass
formed over the course of a simulation is robust to variations in the
numerical method.  During a merger, the gas morphology and phase
structure are initially similar prior to the starburst phase.
However, once a hot gaseous halo has formed from shock heating and AGN
feedback (when included), the agreement is less good.  In particular,
during the post-starburst phase, the SPH simulations feature more
prominent hot gaseous haloes and spurious clumps, whereas with \arepo, gas
clumps and filaments are less apparent and the hot halo gas can cool
more efficiently.  We discuss the origin of these differences and
explain why the SPH technique yields trustworthy results for some
applications (such as the idealised isolated disc and galaxy merger simulations
presented here) but not others (e.g., gas flows onto galaxies in cosmological
hydrodynamical simulations).

\end{abstract}


\section{Introduction}

Galaxy mergers are a natural and crucial ingredient of the
$\Lambda$CDM hierarchical galaxy formation paradigm.  Although the
fraction of galaxies undergoing a merger at any given time is
relatively small, nearly all galaxies will experience a merger at some
point in their histories \citep[e.g.,][]{Stewart2009,Lotz2011,LopezSanjuan2013}.
Particularly significant are `major'
mergers,\footnote{We adopt the convention
that a `major' merger is one in which the ratio of the baryonic
masses is closer to unity than 1:3 (see, e.g., \citealt{Cox:2008} for 
motivation), but this choice is somewhat
arbitrary.} which can be transformative.  In these cases, mergers
violently alter the orbits of the stars in the galaxies and can
transform rotationally supported discs into dispersion-supported
spheroids \citep[e.g.,][]{Toomre:1972,Toomre:1974,Barnes:1988, Barnes:1992, Hernquist:1992,
Hernquist:1993, Cox:2006}.  Furthermore, tidal torques exerted by
the galaxies upon one another drive gas inwards \citep[e.g.,][]{Barnes:1991, Barnes:1996},
thereby resulting in powerful
starbursts \citep[e.g.,][]{Mihos:1994a, Mihos:1996}, triggering active
galactic nuclei (AGN) \citep[e.g.,][]{DiMatteo:2005}, altering
metallicity gradients \citep{Kobayashi2004,DiMatteo2009,Rupke2010,Torrey:2012Z_grad},
and leaving behind signatures of the starbursts and AGN activity in the form of compact
stellar cores \citep[e.g.,][]{Mihos:1994, Hopkins:2008extra_light}
and supermassive black holes \citep[BHs; e.g.,][]{Hopkins:2007a}.
Also, mergers may drive the size evolution of quiescent galaxies
\citep[e.g.,][]{Buitrago2008,Lani2013}. It has thus been
proposed that various seemingly different observational classes of
objects in the Universe -- including blue star-forming disc galaxies,
irregular galaxies of a variety of morphologies, heavily dust-obscured
(ultra-)luminous infrared galaxies ((U)LIRGs), both obscured and
unobscured AGN, `post-starburst' (aka `K+A') galaxies, and `red and dead' elliptical
galaxies -- may be related in a merger-driven evolutionary sequence
\citep[e.g.,][]{Sanders:1988a, Sanders:1988b, Sanders:1996,
Springel:2005c,Hopkins:2006unified_model,Snyder:2011}.

Galaxy mergers have been studied using numerical simulations for more than
forty years (\citealt{Toomre:1972,Toomre:1974, Kozlov:1974a,
Kozlov:1974b}), and for more than seventy years if one considers the
pioneering laboratory method of \citet{Holmberg:1941}.  Although the
early simulations included only gravity, they provided much insight
into the effects of mergers on galaxy morphologies, the formation of
tidal tails and shells, and the kinematics of merger remnants.  More
sophisticated simulations \citep[e.g.,][]{Hernquist:1989, Barnes:1991,
Barnes:1996, Mihos:1994, Mihos:1996} included also gas dynamical
processes, which can be important for many galaxies because a
significant fraction of the baryonic mass is in gas, and, unlike the
stars, the gas is dissipational.  Hydrodynamic simulations of galaxy
mergers have helped us to understand the driving mechanism of
starbursts \citep[e.g.,][]{Hernquist:1989,Barnes:1991,Mihos:1996,
DiMatteo2007,DiMatteo2008,Cox:2008}, supermassive BH fueling and feedback
\citep[e.g.,][]{DiMatteo:2005,Springel:2005feedback,Kazantzidis2005,Robertson:2006,
Johansson2009, Mayer:2010,Hopkins:2011torus}, feedback from supernovae
\citep[e.g.,][]{Cox:2006feedback}, the
kinematics of merger remnants (e.g., \citealt*{Bendo2000,Naab2003,Naab2006,Naab2006b};
\citealt{Bournaud:2005,Cox:2006,
DiMatteo2009kinematics,Bois2010,Bois2011}), the
survivability of discs during mergers
\citep[e.g.,][]{Barnes2002,Springel:2005disks,Robertson:2006disk_formation,Robertson:2008,
Hopkins:2009disk_survival,Puech2012}, the sizes
of merger remnants (e.g., \citealt{Bournaud2007}; \citealt*{Naab2009};
\citealt{Wuyts:2010,Hilz2013,Perret2013}),
and the formation of local \citep[e.g.,][]{Younger:2009} and high-redshift ULIRGs
\citep[e.g.,][]{Narayanan:2009,Narayanan:2010dog,
Narayanan:2010smg,Hayward:2011smg_selection,
Hayward:2012smg_bimodality,Hayward:2013number_counts,Karl2013,
Snyder:2013}, among many other topics.

Whereas cosmological simulations are routinely performed using both
grid-based Eulerian and particle-based pseudo-Lagrangian methods,
idealised isolated (i.e., non-cosmological) galaxy merger simulations
have almost always been performed using pseudo-Lagrangian smoothed
particle hydrodynamics (SPH; \citealt{Lucy:1977,Gingold:1977}; see
\citealt{Rosswog2009}, \citealt{Springel:2010}, and
\citealt{Price2012review} for recent reviews).  SPH is well-suited to
simulating galaxy mergers because it naturally treats the large bulk
velocities present before the discs coalesce and it concentrates
resolution elements in regions where the mass is located (the galaxy
centre(s); see section 9.4 of \citealt{Springel:2010arepo} for further
discussion). Furthermore, the other primary method used for galaxy
formation studies, adaptive mesh refinement (AMR), does not treat
self-gravity as accurately as particle-based approaches
\citep{OShea:2005,Heitmann:2008}.  For these reasons, relatively few
galaxy merger simulations have been performed using AMR
(to our knowledge, the only published examples of
such simulations are \citealt{Kim:2009, Teyssier:2010,Bournaud:2011,
Chapon:2013,Perret2013,Powell2013}).

Recent studies that used simple,
idealised test problems and cosmological simulations have
highlighted potentially significant problems that are inherent in the standard
formulation of the SPH technique.\footnote{Here and throughout this
work unless otherwise noted, we refer to the standard formulation
of SPH, in which the mass is discretised and thus the density enters
the equations of motion (see, e.g., \citealt{Hopkins:2013SPH} for a
detailed discussion). We do so because the vast majority of
previous SPH simulations of isolated discs and galaxy
mergers have employed this
formulation of SPH, and it is the version that is implemented in frequently used codes such
as \gadget \citep{Springel:2001gadget,Springel:2005gadget}
and {\sc Gasoline} \citep{Wadsley:2004}. Below, we briefly discuss some proposed
modifications to the standard SPH technique that may address some
of its disadvantages.}
\citet{Agertz:2007} demonstrated that the standard
implementation of SPH artificially suppresses the Kelvin--Helmholtz (KH)
and Raleigh--Taylor (RT) instabilities because the method introduces
spurious pressure forces near steep density gradients that allow a gap
to be created between particles, thereby reducing their interactions.
Furthermore, SPH can artificially damp subsonic turbulence
\citep{Bauer:2012} and restricts gas stripping from substructures
falling on to haloes \citep{Sijacki:2012}.  However, fixed grid-based
methods also have drawbacks; for example, these codes are not
Galilean invariant and can produce over-mixing
\citep{Springel:2010arepo}.

Partially motivated by the limitations inherent in the traditional SPH technique,
\citet{Springel:2010arepo} developed a novel moving-mesh
hydrodynamics code known as \arepo.  Although similar approaches have
been proposed earlier, such techniques have not yet seen wide-spread use 
in astrophysical applications.  \arepo~is quasi-Lagrangian because the
unstructured mesh is advected with the flow.  Whereas the motion of the
grid reduces mass exchange between cells compared with a static mesh
\citep[e.g.,][]{Genel:2013}, \arepo~is not strictly Lagrangian because
the mass in a given cell can change with time.  Nevertheless,
\arepo~offers a number of advantages over other methods that make it
attractive for astrophysical applications.  For example, it is
Galilean invariant and naturally concentrates resolution elements in
dense regions, similar to particle-based techniques.  Because
\arepo~uses a finite-volume method to solve the Euler equations, it is
better than traditional SPH at capturing shocks and contact
discontinuities and does not artificially suppress fluid instabilities
\citep[see also][]{Bauer:2012,Sijacki:2012}.  Its ability to better
capture weak shocks than SPH is potentially significant for
cosmological problems because these features are ubiquitous in the
cosmic web \citep[e.g.,][]{Keshet:2003}.  Because of its hybrid
nature, \arepo~performs better than (or at least as well as) both
traditional SPH and grid-based approaches for idealised test
problems. Thus, it is useful to compare the results of
\arepo~simulations to those of simulations performed using other
techniques to investigate what effects, if any, the shortcomings of the
traditional methods have on the results of cosmological and idealised
galaxy merger simulations.

Some comparisons between cosmological simulations performed with \arepo and
\gadgetthree~-- in which all other ingredients, including the gravity
solver and sub-resolution models, are identical -- have already been
made.  In some situations, the baryonic properties of the
simulations performed with the two codes differ strikingly.  For
example, \citet{Vogelsberger:2012} demonstrated that gas cooling is
more efficient in the \arepo~simulations, which results in more star
formation at late times. They attribute the difference to spurious
heating in the outer regions of virialized haloes in the
\gadgetthree~simulations and the inability of conventional SPH to
correctly develop a turbulent cascade to smaller scales.  The
\arepo~simulations produce galaxies with extended, relatively smooth
gas discs, whereas in the \gadgetthree~simulations, the discs are more
compact and clumpy \citep{Keres:2012,Torrey:2012disks}.
\citet{Nelson:2013} demonstrated that the relative fraction of gas supplied to
galaxies in halos of moderate to large size through `cold-mode accretion' (aka `cold flows') 
is dramatically
less in the \arepo~simulations because with traditional SPH, much of the cold gas that reaches the
central galaxies does so because it is locked in `blobs' of purely
numerical origin and because the rate at which gas cools from
galaxies' hot haloes is higher in \arepo because it allows for a
proper cascade of turbulent energy to small scales.

Because of the nature of the aforementioned comparisons, the
differences can only originate from differences in the hydrodynamical
solver. It is sometimes argued \citep[e.g.,][]{Scannapieco:2012} that
the differences caused by inaccuracies in the numerical technique
employed are subdominant to the effects caused by varying the
prescriptions for star formation and stellar and AGN feedback and hence may
be ignored.  However, the results of \citet{Nelson:2013} caution
against this because they find differences in the accretion rate of
hot gas on to galaxies between \arepo~and \gadgetthree~that in some
cases approach two orders of magnitude, which is far greater than the
discrepancies between simulations and observations that feedback
effects are invoked to resolve (see \citealt{Vogelsberger:2013}
and \citealt{Torrey:2013} for comparisons of \arepo
cosmological hydrodynamical simulations with observations).

With the exception of a single modest-resolution merger simulation
presented in \citet{Springel:2010arepo}, the moving-mesh approach has
not yet been used to simulate galaxy mergers. Here, we present
the first detailed study of idealised galaxy merger simulations
performed using moving-mesh hydrodynamics.  We present a small suite
of simulations of equal-mass mergers simulated with both
\gadgetthree~and \arepo.  To isolate differences caused by the
different hydrodynamical solvers, we have kept all other components of
the simulations -- namely, the gravity solver and the sub-resolution
models for star formation, the interstellar medium (ISM), BH
accretion, and AGN feedback -- as similar as possible. Other,
more-comprehensive comparisons, such as the \emph{Aquila}
\citep{Scannapieco:2012} and AGORA \citep{Kim:2014} projects,
allow multiple ingredients of the simulations to vary simultaneously, thereby
yielding a general characterization of the systematic uncertainties due to
different numerical methods and sub-resolution models. Our goal is much
more specific: we aim to isolate effects of the hydrodynamical solver
from those caused by differences in the gravity solver or sub-resolution
models, which is not readily possible in these other comparisons.

In our work, we have chosen to use a `standard' implementation of SPH,
as incorporated in the \gadgetthree~code.  We have not explored recent
variants of SPH that are designed to improve its reliability in some
circumstances \citep[e.g.,][]{Monaghan1997, Ritchie2001, Price:2008,
  Wadsley:2008, Read:2010, Abel:2011, Garcia-Senz:2012, Read:2012,
  Saitoh2013, Hopkins:2013SPH,Hobbs:2013} for several reasons.  The most
important is that we wish to assess the reliability of previous
simulations of gas dynamics in galaxy mergers that were performed
using traditional formulations of SPH. We also note that many of
the SPH modifications that have recently been proposed have not yet
been tested under a wide range of conditions. It is presently thus
still unclear which of the numerous modification should eventually be
adopted in a new `best SPH' variant. Also, a number of problems with
SPH are still unresolved even in the most recent proposed
revisions of the method.  We discuss these issues further below.

The remainder of this paper is organised as follows.  In
Section~\ref{S:methods}, we describe the two different hydrodynamical
methods used and the sub-resolution models for star formation, BH
accretion, and supernova and AGN feedback.  In Section~\ref{S:results_no_BHs}
(\ref{S:results_BHs}), we compare \gadgetthree~and \arepo~simulations
that do not (do) include BH accretion and AGN feedback.
Section \ref{S:tests} presents tests of different methods for treating BH
accretion and AGN feedback that we used to inform our choice of a fiducial
treatment.
In Section~\ref{S:discussion}, we review some of the major issues with SPH,
discuss why SPH works reasonably well for some applications but not
others, and outline which previous work is likely to be robust to the
hydrodynamical solver used and which may need to be reconsidered.
Section~\ref{S:conclusions} presents our conclusions.

\vspace{\baselineskip}

\section{Methods} \label{S:methods}

\ctable[
	caption = {Galaxy models \label{tab:galaxy_models}},
	center,
	star,
	doinside=\small,
	notespar
]{lccccccccccc}{
	\tnote[a]{Disc galaxy identifier.}
	\tnote[b]{Halo concentration.}
	\tnote[c]{Virial velocity.}
	\tnote[d]{Virial mass.}
	\tnote[e]{Initial disc stellar mass.}
	\tnote[f]{Initial bulge stellar mass.}
	\tnote[g]{Initial disc gas mass.}
	\tnote[h]{Initial disc gas fraction.}
	\tnote[i]{Stellar disc scalelength.}
	\tnote[j]{Gaseous disc scalelength.}
	\tnote[k]{Stellar disc scaleheight.}
	\tnote[l]{\citet{Hernquist:1990} profile scalelength for the bulge.}
}{
																																									\FL
			& 			& $V_{200}$\tmark[c]		& $M_{200}$\tmark[d]			& $M_{\star, {\rm disc}}$\tmark[e]		& $M_{\star, {\rm bulge}}$\tmark[f]		& $M_{\rm gas, disc}$\tmark[g]	& $f_{\rm gas}$\tmark[h] 	& $r_{\rm disc}$\tmark[i]	& $r_{\rm gas}$	\tmark[j]	
	& h\tmark[k]		& a\tmark[l]		\NN
Name\tmark[a] 		& c\tmark[b]		& (km s$^{-1}$)		& ($10^{11} \msun$) 	& ($10^9 \msun$)			& ($10^9 \msun$)			& ($10^9 \msun$)		&				& (kpc)			& (kpc)			& (pc)	& (kpc)	\ML
\mw 			& 12		& 190			& 23					& 67				 		& 21						& 13					& 0.16			& 3.0				& 6.0				& 300	& 1.0		\NN
\smc 		& 15		& 46				& 0.32 				& 0.19 					& 0.014					& 1.1			 		& 0.85 			& 0.7 			& 2.1 			& 140	& 0.25	\NN
\sbc 			& 11		& 86				& 2.1					& 5.7 					& 1.4						& 7.9					& 0.58 			& 1.3 			& 2.6 			& 320	& 0.35	\NN
\hiz 			& 3.5		& 230			& 13.6				& 30						& 70	 					& 70		 			& 0.7 			& 1.6 			& 3.2 			& 130	& 1.2		\LL
}

\subsection{Hydrodynamics and gravity}

As noted above, we use two different codes, \gadgetthree~and \arepo,
because we wish to investigate differences in the outcome that are driven by
variations in the method used to solve the hydrodynamics.
\gadgetthree~uses SPH \citep{Lucy:1977,Gingold:1977,Springel:2010}, a
pseudo-Lagrangian method.  In SPH, the gas is discretised into
particles, which are typically of fixed mass.  The density field and other
continuous quantities are calculated by taking the mean of the values
of some number of nearest particles (we use 32) weighted by the
smoothing kernel.  To derive the equations of motion, the Lagrangian
can be discretised and then the variational principle used
\citep{Gingold:1982}.  The formulation of SPH used in \gadgetthree~is
explicitly conservative even if the smoothing lengths vary
\citep{Springel:2002}.  The advantages of modern SPH include explicit
conservation of mass, energy, entropy, and linear and angular
momentum; Galilean invariance; resolution that naturally becomes finer
in regions of high gas density; and accurate treatment of
self-gravity.

\arepo~adopts a novel version of the other primary technique used in
astrophysical hydrodynamics, the finite-volume (i.e., Eulerian)
grid-based approach.  In traditional AMR codes, cubic cells at fixed
spatial locations are employed, and the cells are refined and
de-refined according to some criteria (e.g., the mass of cells can be
kept approximately constant). In \arepo, the grid cells are
not fixed in space; rather, mesh-generating points are advected with
the flow and a Voronoi tesselation is used to generate an unstructured
grid from the points. The Euler equations are solved using a
finite-volume approach. Specifically, \arepo~uses a second-order
unsplit Godunov scheme with an exact Riemann solver.  Advantages of
this moving-mesh approach compared with SPH include the following: it
is better at resolving shocks and contact discontinuities; it does not
suppress fluid instabilities; and the density field across a
resolution element (a cell) can be reconstructed to first order
(unlike in SPH, for which the density can only be constructed to
zeroth order).

Both codes use the same tree-based gravity solver, which is a modified
version of that used in \gadgettwo \citep{Springel:2005gadget}.
Collisionless particles are used to represent dark matter and stars;
these particles are assigned fixed gravitational softening lengths.
In \gadgetthree, the gas particles also have fixed gravitational
softening. This treatment guarantees that, by using suitably small
softening, one can have sufficient force resolution at all times.  In
contrast, the gravity solvers traditionally used in AMR codes have
force resolution that depends on the cell size and thus varies in time
and space. Even in collisionless simulations, this treatment can lead
to the suppression of small-scale structure (i.e., dwarf galaxies) if
cells around forming haloes are not refined sufficiently early
\citep{OShea:2005,Heitmann:2008}. Because \arepo~treats the
collisionless component using a tree-based method, it does not suffer
from this problem. The quasi-Lagrangian nature of the moving mesh
cells also enables a superior treatment of gas
self-gravity. \arepo~treats each cell as if the mass were concentrated
at the cell centre and calculates the softened gravitational force
using a softening that is of order the cell radius. For all components,
gravitational interactions are softened using a cubic spline that has
compact support \citep[e.g.,][]{Hernquist:1989treesph}. Full details of the
treatment of self-gravity used in \arepo~can be found in section 5 of
\citet{Springel:2010arepo}.

\ctable[
	caption = {Resolutions \label{tab:resolutions}},
	center,
	star,
	notespar,
	doinside=\small,
]{llcccc}{
	\tnote[a]{Runs that were performed with this resolution.
	For all but resolution R4, the isolated disc simulations and merger simulations for both orbits, both with and without BH accretion and AGN feedback, were performed.
	Only the $\smce$ simulation with BH accretion and AGN feedback was performed at resolution R4.}
	\tnote[b]{Dark matter particle mass.}
	\tnote[c]{Baryonic particle mass (and target gas cell mass in \arepo).}
	\tnote[d]{Gravitational softening for dark matter particles.}
	\tnote[e]{Gravitational softening for baryonic particles/cells.}
	
}{
																																	\FL
			&						& $m_{\rm dm}$\tmark[b]	& $m_{\rm bar}$\tmark[c]	& $\epsilon_{\rm dm}$\tmark[d]		& $\epsilon_{\rm bar}$\tmark[e]		\NN
Designation	&	Runs\tmark[a]			& ($10^5 \msun$)		& ($10^4 \msun$)		& (pc)						& (pc)						\ML
R1			&	\mw, \hiz				& 64					& 64					& 240						& 120						\NN
R2			& 	\mw, \hiz, \smc, \sbc		& 8					&8					& 120						& 60							\NN
R3			& 	\smc, \sbc				& 1					& 1					& 60							& 30							\NN
R4			& 	\smce with BHs			& 0.125				& 0.125				& 30							& 15							\LL
}

\subsection{Star formation and supernova feedback}

In both \gadgetthree~and \arepo, star formation and supernova feedback
are implemented via the effective equation of state (EOS) method of
\citet{Springel:2003}.  Only gas particles with density greater than a
low-density cutoff ($n \sim 0.1$ cm$^{-3}$) are assumed to
have an EOS governed by the sub-resolution model.  In the
\citet{Springel:2003} model, the ISM is considered to consist of two
phases in pressure equilibrium: cold dense clouds and a hot diffuse
medium in which the cold clouds are embedded. The instantaneous SFR
for each particle is calculated using a volume density-dependent
Kennicutt--Schmidt \citep{Kennicutt:1998,Schmidt:1959} prescription,
$\rho_{\rm SFR} \propto \rho_{\rm gas}^N$, with $N = 1.5$. Star
particles are spawned from gas particles or cells probabilistically
according to their SFRs. Feedback from supernovae is included as an
effective pressurization of the ISM such that the equation of state of
the gas is stiffer than an isothermal EOS.\footnote{In this work,
explicit stellar winds are not included. First, much of the previous work did
not include such winds, and one of our goals is to examine the reliability
of previous merger simulations. Second, in contrast with the
sub-resolution models for star formation, BH accretion, and AGN feedback,
the sub-resolution models for stellar winds differ significantly
enough that it would be difficult to disentangle the effects of the different
wind implementations from those of the different hydrodynamical
solvers. If stellar winds were included, the results
would likely differ more significantly for the same reasons that the properties
of the AGN outflows can differ significantly (these reasons are discussed below).
Indeed, \citet{Hopkins:2013merger_winds} found that the cold clumps at large
radii that are present in stellar-driven outflows in simulations run with the
traditional formulation of SPH are not present when the simulations are
run with the pressure-entropy formulation of SPH (see their fig. A3).
We would likely find similar results if we compared stellar-driven
outflows in \gadgetthree and \arepo simulations.} See \citet{Springel:2003} and
\citet{Springel:2005feedback} for further details. We stress that the
sub-resolution models for star formation and feedback in
\gadgetthree~and \arepo~are as similar as possible, which enables us
to investigate differences caused solely by variations in the
calculation of the hydrodynamics.

\subsection{BH accretion and AGN feedback}

As for the star formation and stellar feedback sub-resolution models,
we have attempted to keep the sub-resolution models for BH accretion
and AGN feedback in the two codes as similar as possible. However, for
reasons we discuss in Section \ref{S:tests}, the default sub-resolution models for BH
accretion and feedback used in \gadgetthree~ and \arepo~ differ
slightly. Because we shall explore different numerical implementations
of the BH accretion and feedback model, we describe the model in some
detail here, but see \citet{Springel:2005feedback} for a more thorough
presentation.

\subsubsection{BH accretion} \label{S:BH_accretion_method}

Each disc galaxy is initialised with a $10^5 \msun$ central sink
particle that undergoes modified Eddington-limited
Bondi--Hoyle--Lyttleton accretion
\citep{Hoyle:1939,Bondi:1944,Bondi:1952}.
The accretion rate is the Bondi--Hoyle--Lyttleton accretion rate multiplied by a dimensionless 
factor $\alpha$:
\begin{equation} \label{eq:bondi}
\dot{M}_{\rm B} = \frac{4 \pi \alpha G^2 M_{\rm BH}^2 \rho}{(c_{\rm s}^2 + v^2)^{3/2}},
\end{equation}
where $G$ is the gravitational constant, $M_{\rm BH}$ is the BH mass,
$\rho$ is the local gas density, $\cs$ is the local sound speed, and
$v$ is the velocity of the BH relative to the gas. Because we
typically do not resolve the Bondi radius, the parameter $\alpha$,
which we set equal to 100, is used to account for the fact that we
underestimate the density at the Bondi radius.  In practice, the
choice of $\alpha$ matters only at early times because most of the
mass growth occurs during times of Eddington-limited accretion, and
the mass of the final BH is insensitive to the value of $\alpha$.

The Eddington-limited accretion rate is
\begin{equation} \label{eq:edd_lim}
\dot{M}_{\rm Edd} = \frac{4 \pi G M_{\rm BH} m_{\rm p}}{\epsilon_{\rm r} \sigma_{\rm T} c},
\end{equation}
where $m_p$ is the proton mass, $\sigma_{\rm T}$ is the Thomson cross-section, and $\epsilon_{\rm r}$ is the radiative efficiency,
which is defined by
\begin{equation}
L_{\rm BH} = \epsilon_{\rm r} \dot{M}_{\rm BH} c^2,
\end{equation}
where $L_{\rm BH}$ is the luminosity of the BH. We assume
$\epsilon_{\rm r} = 0.1$ \citep{Shakura:1973}.

The BH particles accrete mass at a rate
\begin{equation}
\dot{M}_{\rm BH} = (1-\epsilon_{\rm r}) \min\left(\dot{M}_{\rm B}, \dot{M}_{\rm Edd}\right),
\end{equation}
in which the $(1-\epsilon_{\rm r})$ 
factor accounts for the rest-mass of the energy radiated away by the AGN (this factor was not included in the
original \citealt{Springel:2005feedback} treatment). In \gadgetthree, gas particles are swallowed stochastically. In \arepo,
the more natural treatment is to continuously drain
gas from the cell in which the BH is located. We have compared this treatment with one analogous to that in \gadgetthree,
in which cells are swallowed stochastically; we found that for the mass and time resolutions of our simulations,
the differences in the results are negligible (but lower-resolution simulations can exhibit differences).

To calculate the accretion rate given by equation~(\ref{eq:bondi}), we
must determine $\rho$ and $c_{\rm s}$ near the BH.  In \gadgetthree,
the SPH estimates for these quantities are used.  In \arepo, we can
adopt analogous estimates, but we can instead also employ the quantities
for the cell in which the BH is located. In principle, the latter
should better represent the properties of the gas around the
BH. However, the individual cell values can also be more noisy, so it
is not clear a priori which is preferred.  We compare the results
obtained using the different treatments in Section
\ref{S:accretion_test}. Based on those tests, we chose to use the cell
density and SPH-like estimate of the sound speed in our default
treatment.

One potential issue when calculating the accretion rate using either
the SPH density estimate or the cell density is that as the BH
consumes or expels the nearby gas, the region used for the density
estimate can grow in size. Consequently, the BH can continue to
accrete gas from larger and larger scales, whereas in reality, BH
accretion should terminate once the gas near the BH is consumed or
expelled. In SPH, this problem is unavoidable unless one reduces the
number of neighbours used to estimate the density (which is
problematic because the noise in the density estimate will be
increased significantly) or a scheme that is more
complicated than the simple Bondi--Hoyle--Lyttleton approach is employed; one
consequence is that lower-resolution simulations can sometimes exhibit
more BH growth \citep[e.g.,][]{Newton:2013}. In \arepo, this problem
may be avoided by preventing cells near the BH from becoming too
large. We do this by forcing the cells within some radius of the BH to
be refined if they have a radius greater than some maximum value. We
explore the effects of this refinement in Section
\ref{S:refinement_test}. Based on our tests, we decided to force cells
within 500 pc of a BH to have a maximum size of 50 pc.\footnote{For
comparison, 50 pc is the Bondi radius
of a $10^7 \msun$ BH accreting from a gas with sound speed $c_s = 30$
km $\ps$. Thus, an even stricter refinement criterion could be
justified, but because of the resolution of the simulations, there is
no structure on smaller scales.}

Finally, because the initial BH particles are similar in mass to the
stellar and dark matter particles, two-body interactions can cause the
BH to stray from the centre of the potential well. However, in
reality, dynamical friction would cause the BHs to rapidly sink to the
potential minimum. Thus, we pin the BH to the halo potential
minimum. For this reason, we also neglect the $v$ term in the denominator
of equation (\ref{eq:bondi}).

\subsubsection{AGN feedback} \label{S:BH_FB_method}

We also include a simple model for thermal feedback from the AGN
\citep{Springel:2005feedback}. The BH particles deposit some fraction
$\epsilon_{\rm f}$ of their luminosity (as thermal energy) to the
surrounding gas. We use $\epsilon_{\rm f} = 0.05$ because in previous
\gadgettwo~simulations, this value yielded an $M_{\rm BH}-\sigma$
relation normalization consistent with that observed
\citep{DiMatteo:2005}.
We scale the number of gas particles over which we distribute the
feedback energy with resolution such that the total mass of the
particles is constant.  As we demonstrate in Section \ref{S:FB_test},
this scaling minimizes the resolution dependence of the sub-resolution
model.  It is desirable to have
sub-resolution models that do not depend significantly on numerical
resolution, at least as long as the physics that the sub-resolution
treatment is meant to represent remains inaccessible.  Otherwise, the
problem is not well-posed numerically and the interpretation of the
sub-resolution model becomes unclear.

\subsection{Initial conditions}

\ctable[
	caption = {Orbital parameters \label{tab:orb_params}},
	center,
	notespar
]{lcc}{
	\tnote[a]{Progenitor disc galaxy identifier.}
	\tnote[b]{Initial separation of the discs.}
	\tnote[c]{Pericentric passage distance.}
}{
																			\FL
Galaxy model\tmark[a]	&	$R_{\rm init}$\tmark[b]	&	$R_{\rm peri}$\tmark[c]			\NN
					&	(kpc)					&	(kpc)							\ML
\mw			& 	200			&	11					\NN
\smc			&	60			&	5					\NN
\sbc			&	100			&	5					\NN
\hiz			&	100			&	7					\LL
}

The initial disc galaxies are created following
the procedure described in \citet{Springel:2005feedback}.  The
galaxies consist of dark matter haloes described by a
\citet{Hernquist:1990} profile with virial velocity $V_{200}$ and
concentration $c$, an exponential stellar disc with scalelength
$h_{\rm star}$ and scaleheight $z_0$, an exponential gaseous disc with
scalelength $h_{\rm gas}$ and scaleheight determined by requiring the
disc to be rotationally supported, and a bulge described by a
\citet{Hernquist:1990} profile with scalelength $b$.  Note that,
unlike in \citet{Springel:2005feedback}, the gaseous and stellar discs
do not have the same scalelength; rather, the gaseous discs can be
significantly more extended than the stellar discs.

Unlike SPH, grid-based methods cannot treat empty space. Thus, for the \arepo~simulations, we must add a background grid
of low-density cells to the initial conditions used for the \gadgetthree~simulations such that the entire simulation volume has positive
gas density. See section 9.4 of \citet{Springel:2010arepo} for details of how the background mesh is added.

Our intention is not to present a comprehensive suite of merger
simulations that addresses the full parameter space but rather to directly compare the results obtained using
\arepo~and \gadgetthree~for a set of simulations based on galaxy
models that are representative of a variety of actual galaxies. For definiteness, we use galaxy
models that are similar to those of
\citet*{Hopkins:2011self-regulated_SF} and the same merger parameters as
\citet{Hopkins:2013mergers}. Two of the isolated disc galaxies are
intended to be Milky Way (\mw) and Small Magellanic Cloud (\smc)
analogues. The other two represent a dwarf starburst (\sbc) and a
redshift $z \sim 2$ disc galaxy (\hiz). Thus, the initial discs capture
much of the diversity of real disc galaxies, but the sampling is by no
means complete.  The properties of the disc galaxies are given in
Table \ref{tab:galaxy_models}. We simulate each disc galaxy
with two different resolutions, which are specified
in Table \ref{tab:resolutions}.

We evolve each disc in isolation for 3 Gyr.
For the mergers, two identical disc galaxies are
placed on parabolic orbits with initial separation and pericentric
passage distance as specified in Table \ref{tab:orb_params}.  As in
\citet{Hopkins:2013mergers}, two orbits, the \textsf{e} and \textsf{f}
orbits of \cite{Cox:2006}, are used. For the \textsf{e} orbit, the
directions of the spin axes of the discs given in spherical
coordinates are $(\theta_1,\phi_1,\theta_2,\phi_2) = (30^\circ,
60^\circ, -30^\circ, 45^\circ)$. For \textsf{f},
$(\theta_1,\phi_1,\theta_2,\phi_2) = (60^\circ, 60^\circ, 150^\circ,
0^\circ)$. Note that neither orbit is coplanar; thus, our general
conclusions should be relatively insensitive to orbit (mergers with
perfectly coplanar orbits, which are of course highly unlikely in
nature, can exhibit pathological behavior that is not characteristic
of the behaviour for other orbits). Because our focus is to compare
the results of otherwise-identical \gadgetthree and \arepo simulations,
two orbits is sufficient; a comprehensive suite of simulations would ideally
include a much wider variety of orbits \citep[see, e.g.,][]{Moreno:2012,Moreno2013}.
We run each merger simulation for
$3-5$ Gyr, depending on the simulation, which is sufficient time for
the galaxies to coalesce.
We typically simulate each merger at two different resolutions (see
Table \ref{tab:resolutions} for details). Furthermore, to strengthen our
conclusions regarding whether any resolution dependence is
systematic, we performed the \smce simulation
with BH accretion and AGN feedback included at a third, higher
resolution. Except where otherwise noted, we always plot the results of
the highest-resolution run.

\section{Results}

\subsection{Simulations without BH accretion and AGN feedback} \label{S:results_no_BHs}

We first present a comparison of \gadgetthree~and \arepo~simulations
for which we disabled the BH accretion and AGN feedback treatments in
the codes.\footnote{Animations that show comparisons of the SFHs, BH
accretion rate versus time (when applicable), gas surface densities, and gas phase
diagrams for the highest-resolution simulations performed are available at \url{\urlformovies}.}

\subsubsection{Star formation histories}

\begin{figure}
  \centering
    \includegraphics[width=\columnwidth]{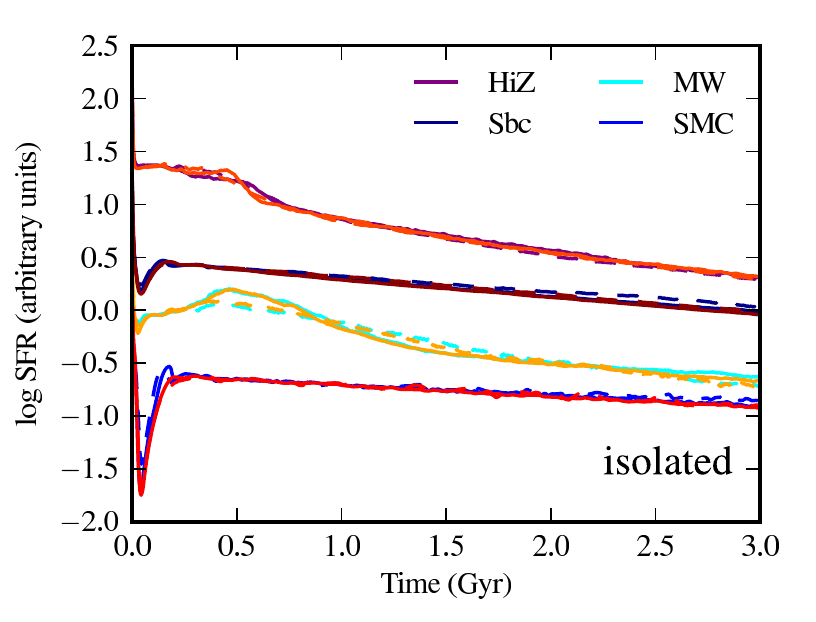} \\
    \includegraphics[width=\columnwidth]{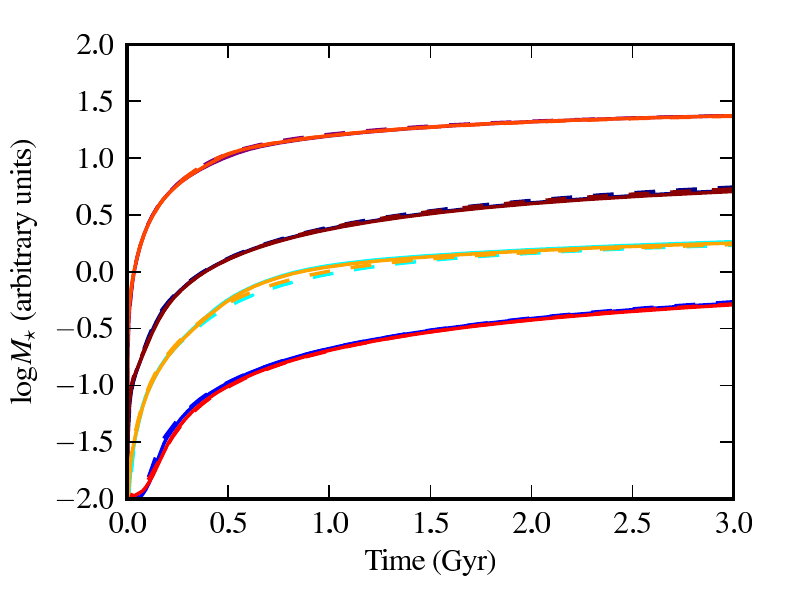}
  \caption{SFR (top) and cumulative stellar mass formed (bottom) versus time (Gyr) for the isolated disc simulations without BHs. The solid and dashed lines denote the
  higher- and lower-resolution runs, respectively.
  The blueish colours denote the \arepo~simulations and the reddish colours denote \gadgetthree~simulations. For each disc model (from top to bottom,
  \hiz, \sbc, \mw, and \smc), the legend shows the colour used for one of the simulations of that disc. The values for each simulation have been renormalised by an arbitrary
  constant (kept fixed for all simulations of a given initial disc) so that all runs can be shown clearly on the plot. At $t = 3.0$ Gyr, the absolute values of the SFRs for
  \hiz, \sbc, \mw, and \smc are $\sim 4$, $\sim 0.7$, $\sim 0.6$, and $\sim 0.07 \msunperyr$, respectively. The SFHs are almost identical regardless of the resolution or code used.}
  \label{fig:iso_NB_SFHs}
\end{figure}

\begin{figure*}
  \centering    
    \includegraphics[width=\columnwidth]{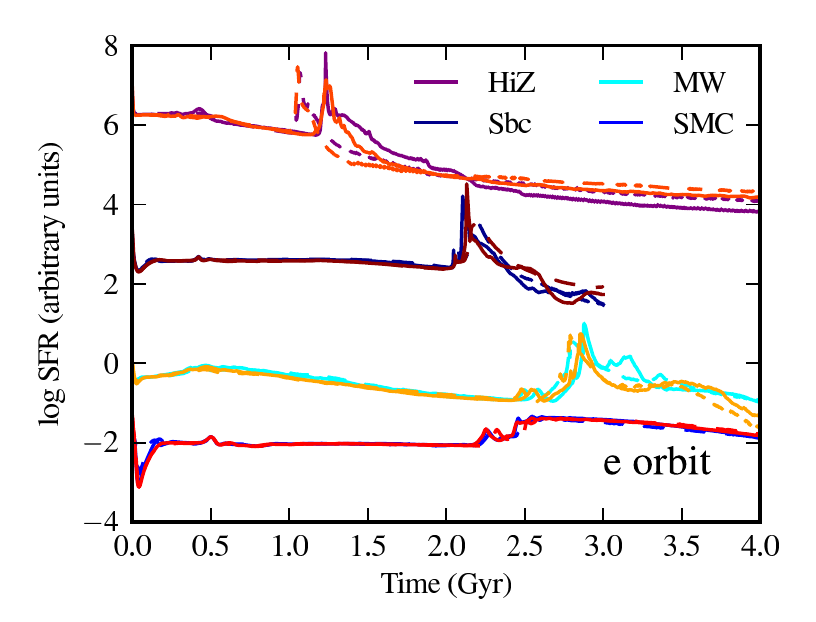}
    \includegraphics[width=\columnwidth]{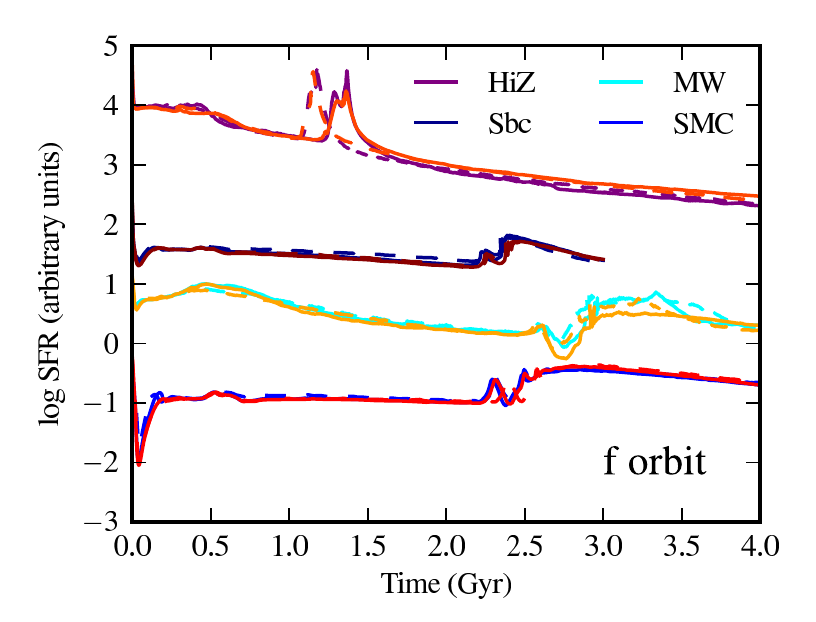} \\
    \includegraphics[width=\columnwidth]{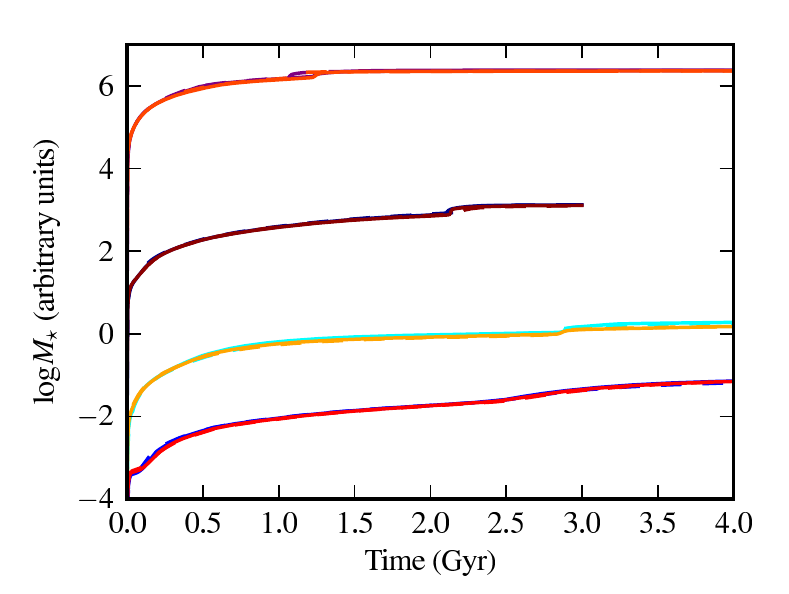}
    \includegraphics[width=\columnwidth]{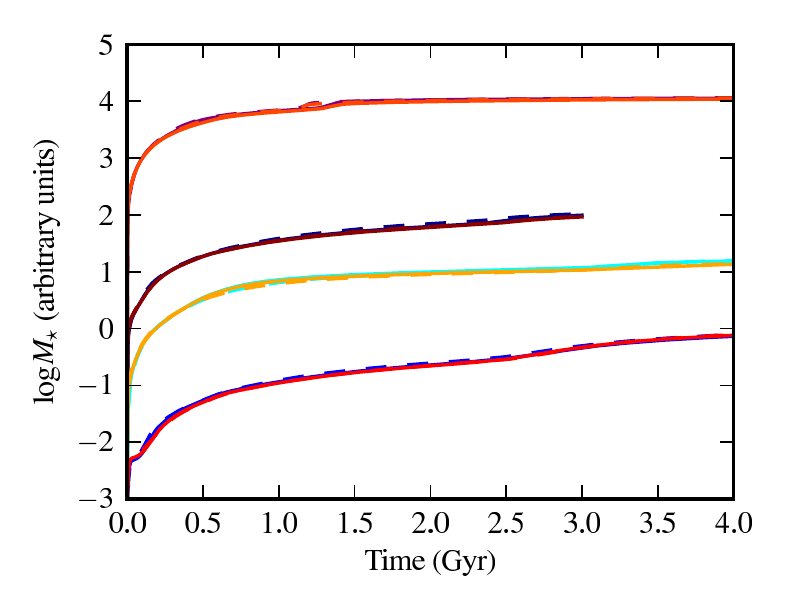}
  \caption{Similar to Fig. \ref{fig:iso_NB_SFHs}, but for the \eorbit~(left) and \forbit~(right) merger simulations without BH accretion and AGN feedback. At $t = 0.5$
  Gyr, the SFRs for the \hiz, \sbc, \mw, and \smc mergers are $\sim 75$, $\sim 4$, $\sim 10$, and $\sim 0.3 \msunperyr$, respectively, for both orbits.
  The SFHs agree less well for these simulations than for the isolated discs. However, the code-dependent differences are minor, and the
  cumulative stellar mass formed is almost indistinguishable.}
  \label{fig:merger_NB_SFHs}
\end{figure*}

A comparison of the star formation histories (SFHs) and cumulative
stellar mass formed versus time for the isolated disc simulations
is shown in Fig. \ref{fig:iso_NB_SFHs}.
The solid (dashed) lines indicate the higher (lower) resolution runs, and blueish (reddish) colours indicate \arepo~(\gadgetthree) simulations.

The isolated discs (Fig. \ref{fig:iso_NB_SFHs}) evolve in the expected
manner: after some initial settling into equilibrium, as the gas is
consumed (in these idealised simulations, no additional
gas is supplied during the simulations), the SFR decreases and the
stellar mass formed increases. For the \hiz simulations, there are
some minor resolution- and code-dependent differences in the SFHs at
$t \sim 0.5$ Gyr.  The different \sbc runs are almost
identical. The \mw simulations exhibit minor
resolution-dependent differences in the SFHs throughout the
simulation.  For the \smc model, there are minor differences in the
SFRs of the \gadgetthree~and \arepo~simulations at $t \sim 0.2$
Gyr. In all cases, the curves of cumulative stellar mass formed versus
time are almost indistinguishable. These results demonstrate that for
the isolated disc simulations, the two codes agree very well and the
simulations are converged with respect to particle number (at least in terms of their SFHs).

Fig. \ref{fig:merger_NB_SFHs} shows the SFHs for the merger
simulations without BH accretion and AGN feedback.  As expected from
much previous work, the mergers exhibit the following generic
evolution.  The SFR initially oscillates for a short time as the discs
settle into equilibrium.  Then, there is a slight elevation when the
discs are at first pericentric passage, but the bulges prevent the
discs from becoming very unstable. As the discs approach final
coalescence, in many cases, strong tidal torques drive gas into the
nucleus and fuel a starburst. However, the strength and shape of the
starburst depend on the progenitor properties and orbit. After the
strong starbursts (\hize, \hizf, \sbce,
and \mwe), the SFR decreases to the pre-merger level or
significantly below it.  Note that the decrease is driven solely by
gas consumption and shock-heating of the gas because these simulations
do not include AGN feedback. In the other merger simulations, in which
a strong starburst is not induced, the SFR can remain elevated for the
duration of the simulation (see especially \smce and
\smcf).

As can be inferred from Fig. \ref{fig:merger_NB_SFHs}, the agreement
for the merger simulations is also excellent, although there are more
noticeable differences than for the isolated disc cases.  The most
prominent difference is that for \hize, \hizf, \sbce, and \mwe, the
time at which the starburst occurs depends on resolution.  This is
likely caused by resolution-dependent variations in the hydrodynamical
drag experienced by the galaxies as they collide, which can slightly alter the
merging time-scales.  However, for a given resolution, the SFHs during
the starburst are similar for the two codes. The most significant,
albeit still relatively minor, differences between the \arepo~and
\gadgetthree~simulations occur after the starbursts, e.g., at $t \ga
2.4$, 2.9, and 2.6 Gyr for \sbce, \mwe, and
\mwf, respectively. The SFHs do not vary systematically
depending on the code used, e.g., for the post-starburst phase of
\sbce, the \arepo~SFRs tend to be lower, whereas for
\mwe, the opposite is true. As for the isolated discs, the
differences in the cumulative stellar mass formed versus time are
negligible.

\subsubsection{Gas morphologies}

\begin{figure*}
  \centering
    \includegraphics[width=0.5\columnwidth]{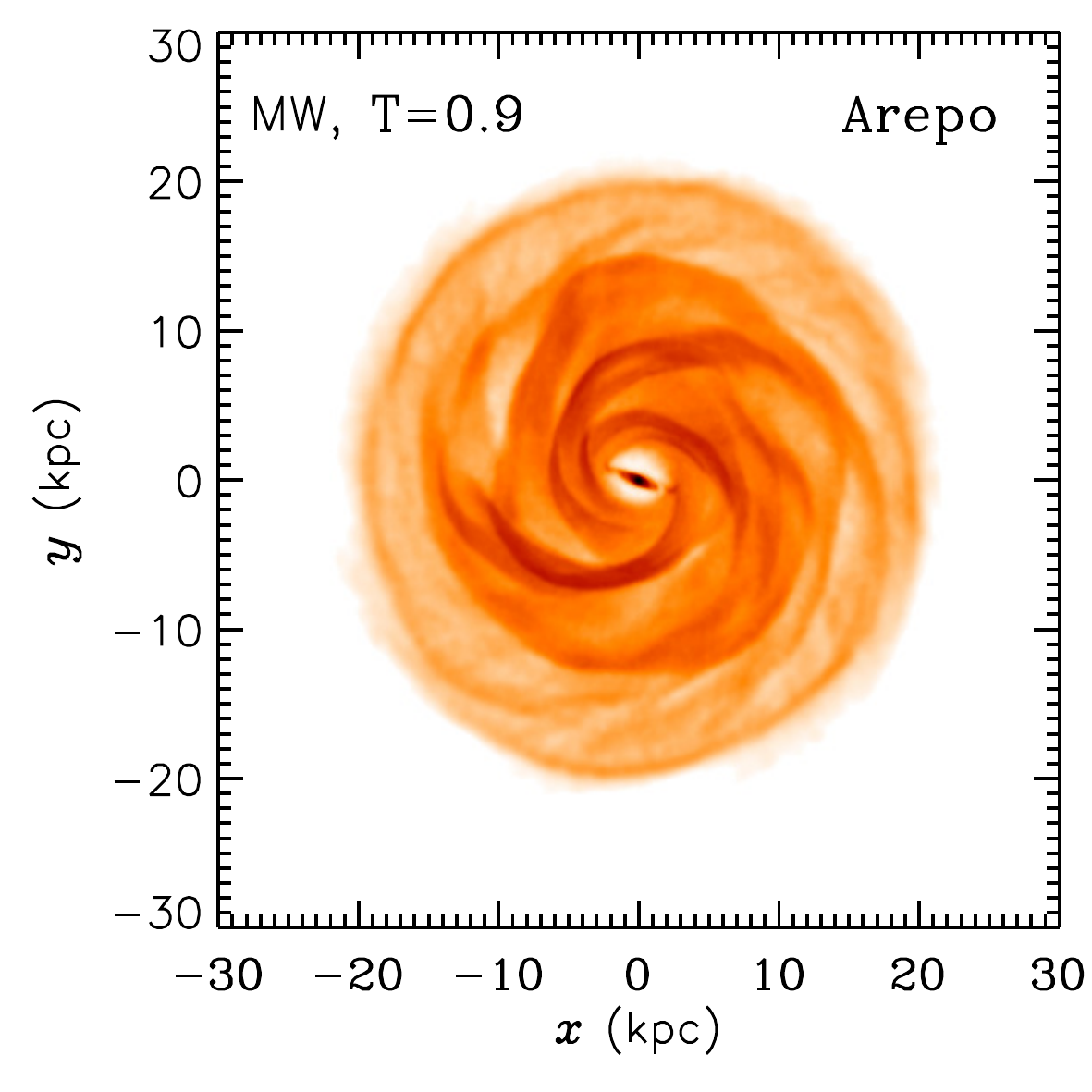}
    \includegraphics[width=0.5\columnwidth]{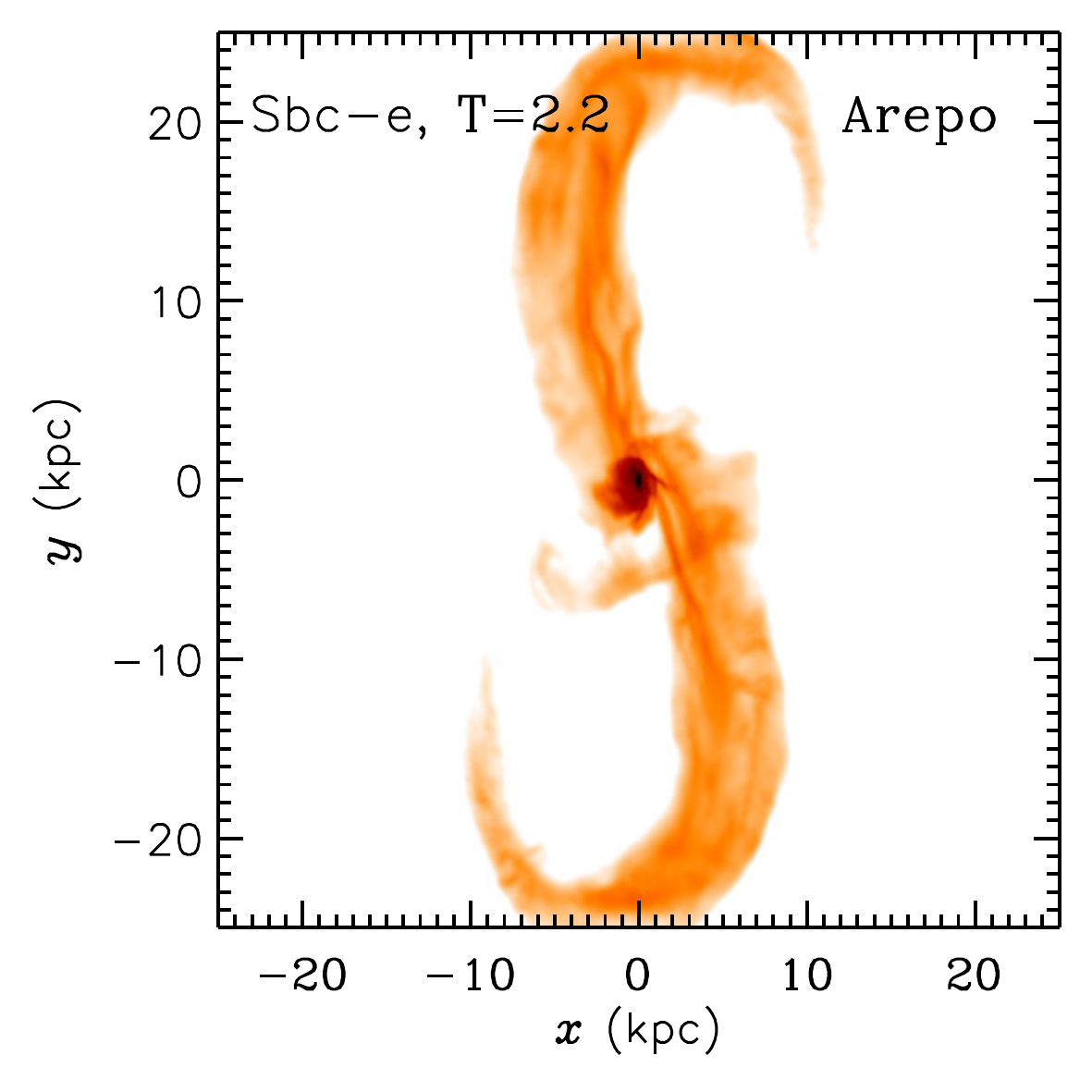}
    \includegraphics[width=0.5\columnwidth]{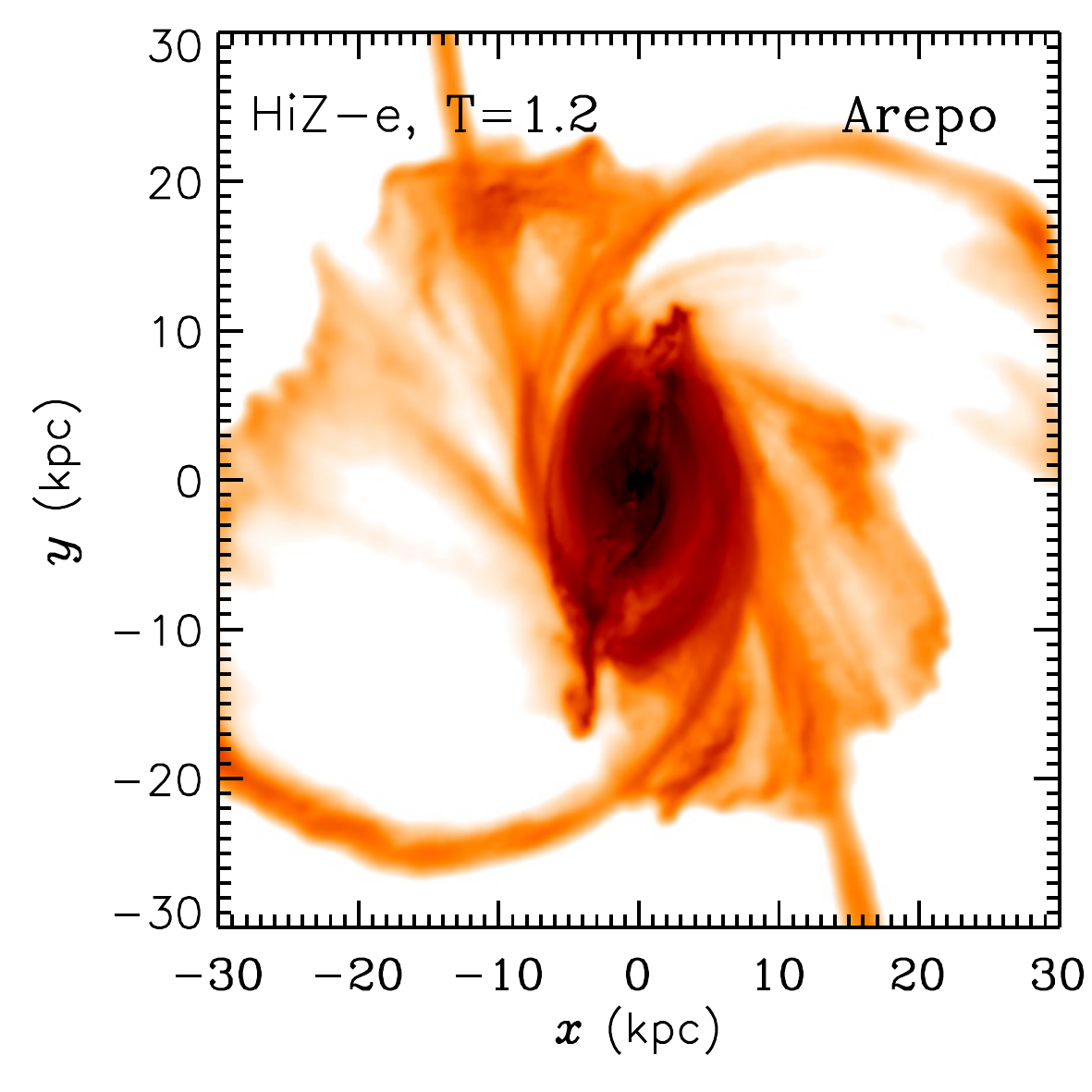}
    \includegraphics[width=0.5\columnwidth]{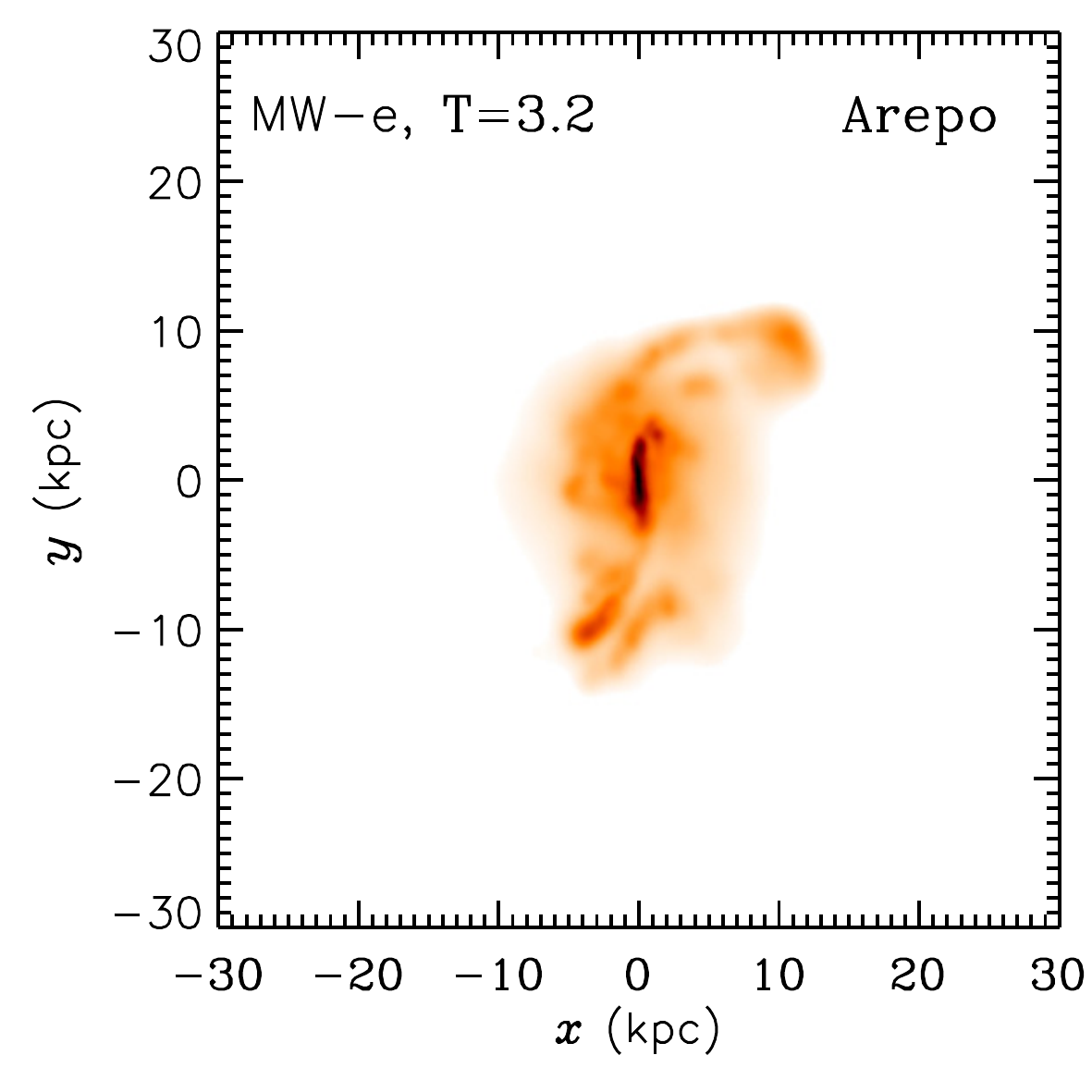} \\
    \includegraphics[width=0.5\columnwidth]{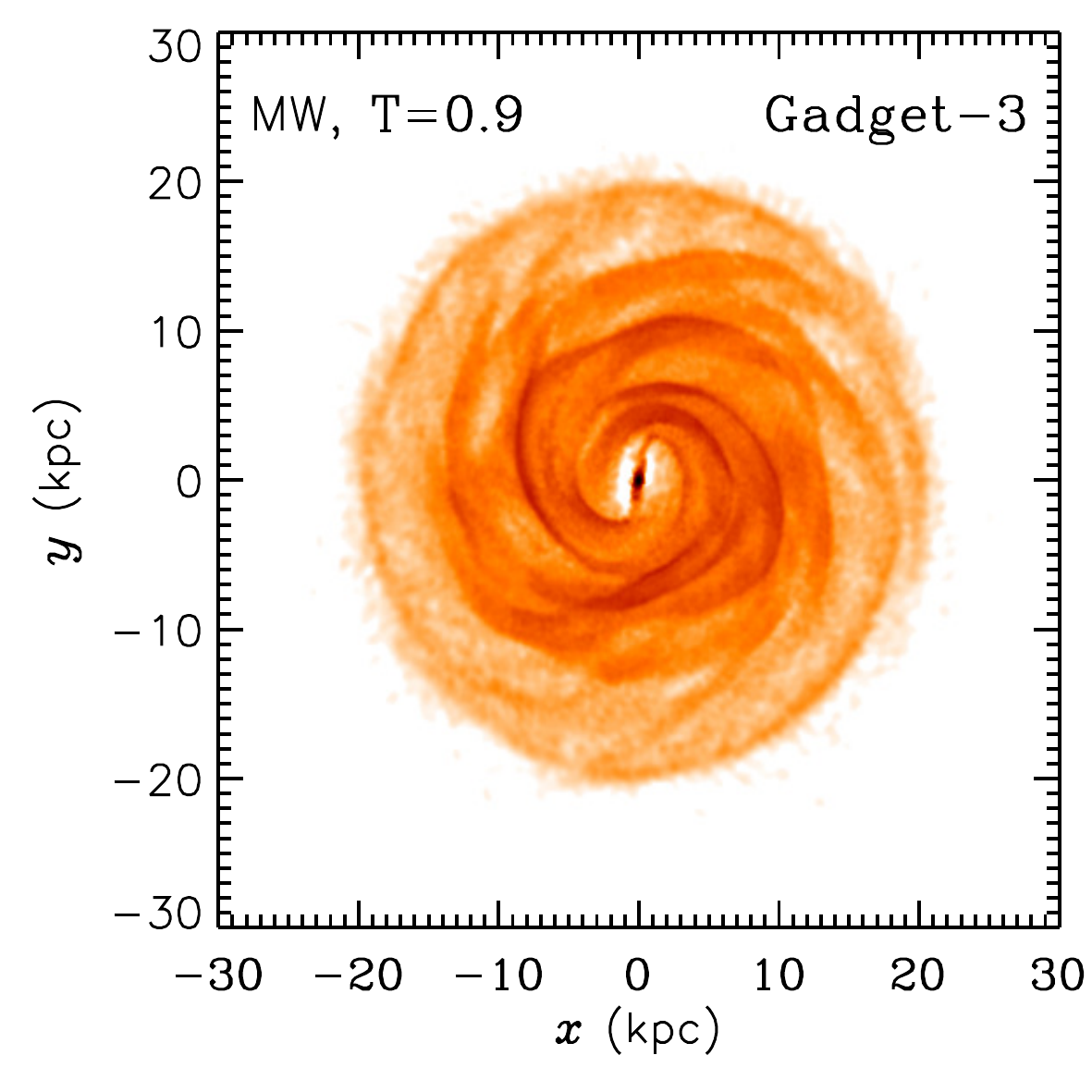}
    \includegraphics[width=0.5\columnwidth]{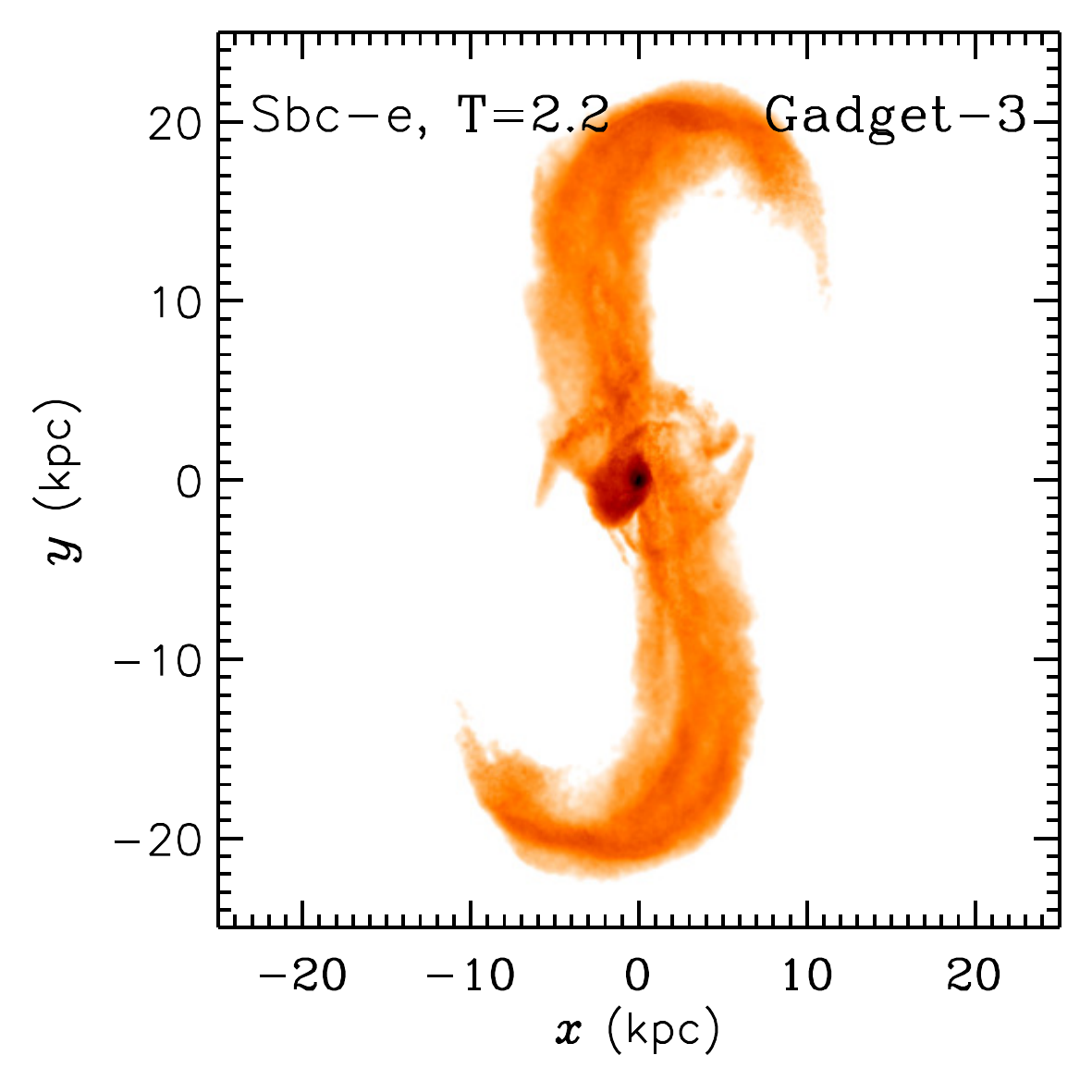}
    \includegraphics[width=0.5\columnwidth]{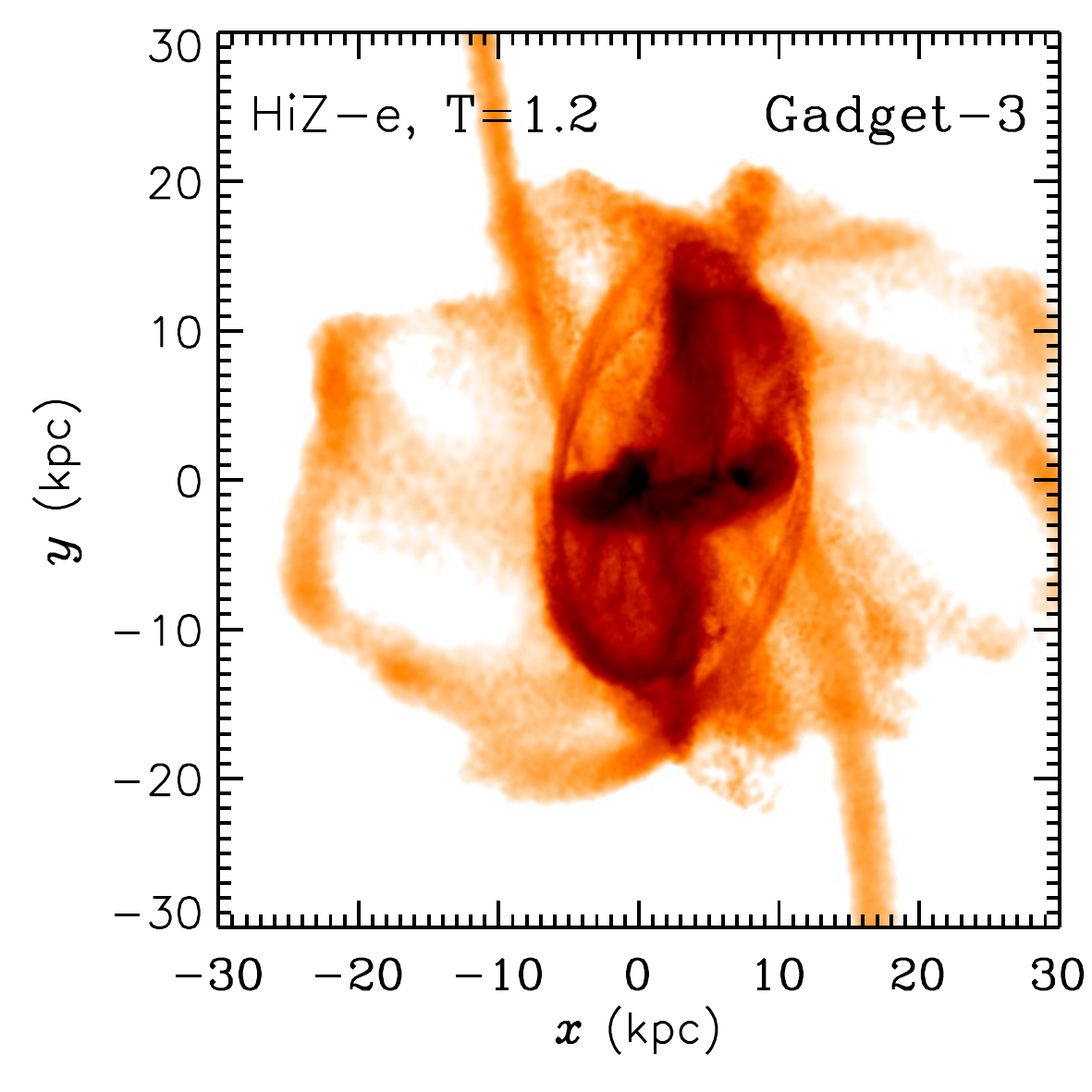}
    \includegraphics[width=0.5\columnwidth]{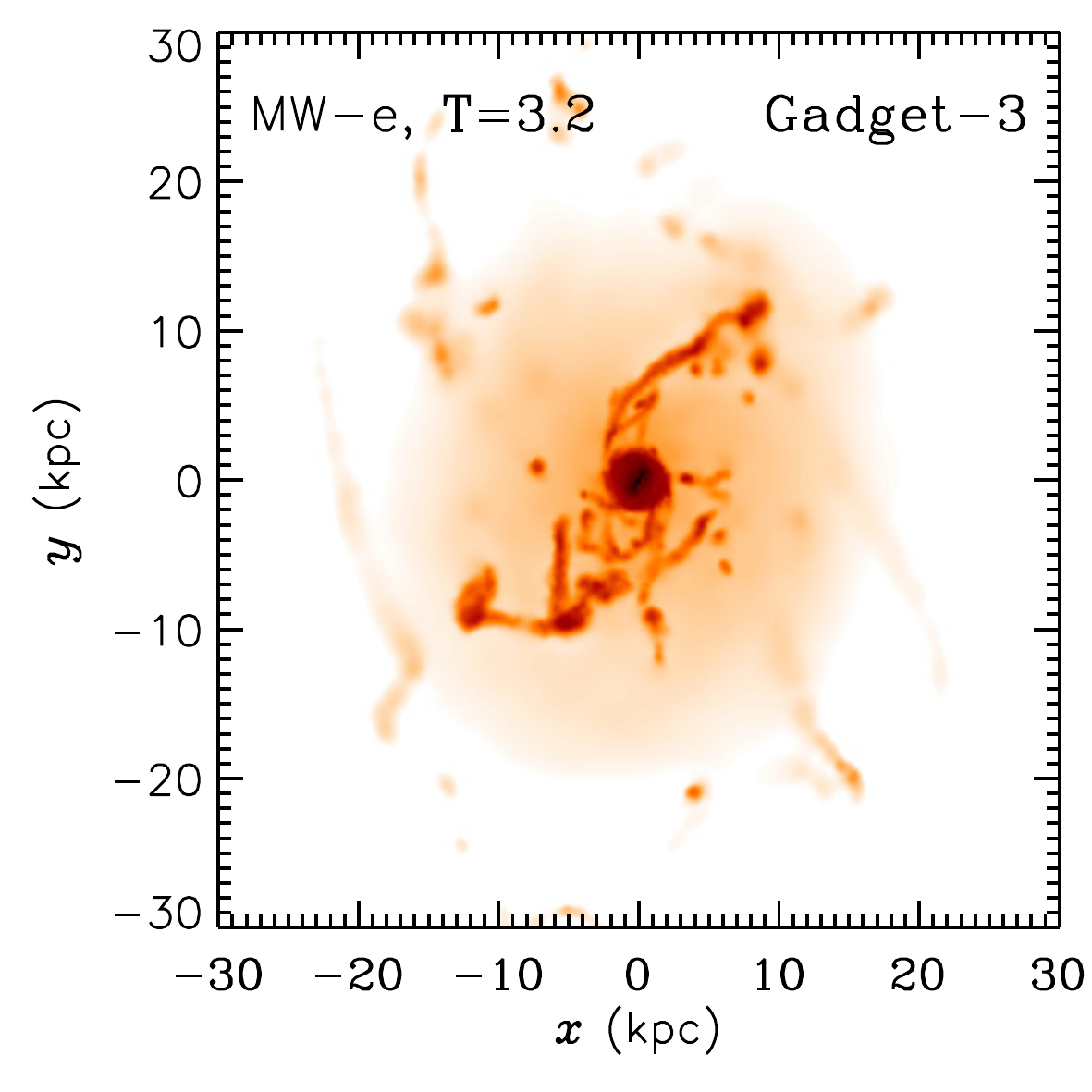}
  \caption{Example gas surface density plots for the following \arepo~(top) and \gadgetthree~(bottom) simulations that do not include BH accretion and AGN feedback:
  isolated \mw disc at $t = 0.9$ Gyr (first column), \sbce merger at $t = 2.2$ Gyr (second column), \hize merger at $t = 1.2$ Gyr (third column), and
  \mwe merger at $t = 3.2$ Gyr (fourth column). 
  Animations that show the time evolution of the gas surface density for all simulations are available at
  \url{\urlformovies}. Typically, the gas morphologies in the \gadgetthree~and \arepo~simulations are very similar, but
  in the post-starburst phase (e.g., column four), the gas morphologies are often clumpier in the \gadgetthree~simulations and the \arepo~simulations exhibit less
  hot halo gas.}
  \label{fig:NB_gas_comparison}
\end{figure*}

The agreement of the SFHs of the \gadgetthree~and \arepo~simulations
for both the isolated disc and merger simulations indicates that the
differences in the numerical schemes do not dramatically affect the
global evolution of the simulated systems. However, it is possible
that the detailed properties of the simulated mergers differ
despite the good agreement for the integrated quantities, and this is
indeed the case. Fig. \ref{fig:NB_gas_comparison} presents gas surface
density plots for four different simulation snapshots. (The interested
reader is invited to visit the aforementioned URL to view animations that compare
the time-evolution of the gas surface density for the highest-resolution
runs of all simulations
presented in this work.) The top row shows the \arepo~result, and the
bottom row shows the \gadgetthree~result for the same snapshot and
resolution. The first column shows the \mw isolated disc at $t = 0.9$
Gyr. The \arepo~and \gadgetthree~results are qualitatively similar,
but the gas distribution appears slightly smoother in the
\arepo~simulation and the orientation of the bar differs. The second
column shows the \sbce merger at $t = 2.2$ Gyr, which is
near final coalescence. Again, the morphologies are very similar, but
the spatial extent of the tidal tails differs slightly.  The third
column shows the \hize merger near the peak of the starburst
($t = 1.2$ Gyr). Here, the gas distribution is again smoother in the
\arepo~run, and the gas morphology of the nuclear region differs quite
dramatically. Finally, the fourth column shows the \mwe
merger at $t = 3.2$ Gyr ($\sim 0.4$ Gyr after the peak of the
starburst). Here, the differences are the most dramatic: the nuclear
disc that has re-formed is oriented edge-on in the \arepo~simulation
but face-on in the \gadgetthree~simulation.  This is likely driven by
the stochastic nature of the torques acting on the gas that
accumulates in the centre of the remnant
\citep[e.g.,][]{Hernquist:1991}.  Furthermore, the
\gadgetthree~simulation features a clumpy, extended hot halo that is
not present in the \arepo~simulation.

In general, the \arepo~morphologies tend to be smoother than those
yielded by \gadgetthree, and the clumps that are often observed in the
\gadgetthree~simulations (and are spurious; e.g.,
\citealt{Sijacki:2012}) are not present in the \arepo~simulations.
The differences between the two codes are most pronounced during the
starburst and post-starburst phases. We discuss the physical reasons
for these differences in detail in Section
\ref{S:reasons_for_agreement}.

\subsubsection{Gas phase structure}

\begin{figure*}
  \centering
    \includegraphics[width=\columnwidth]{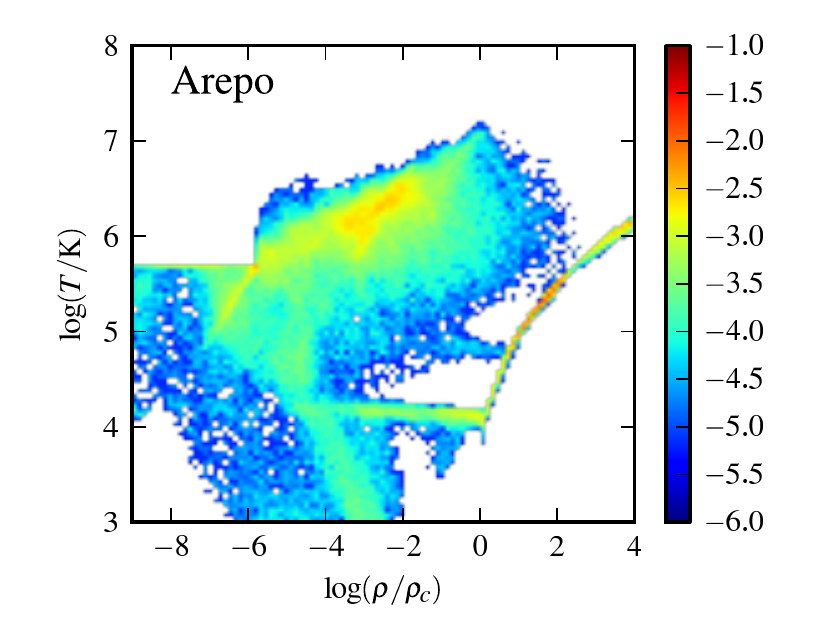}
    \includegraphics[width=\columnwidth]{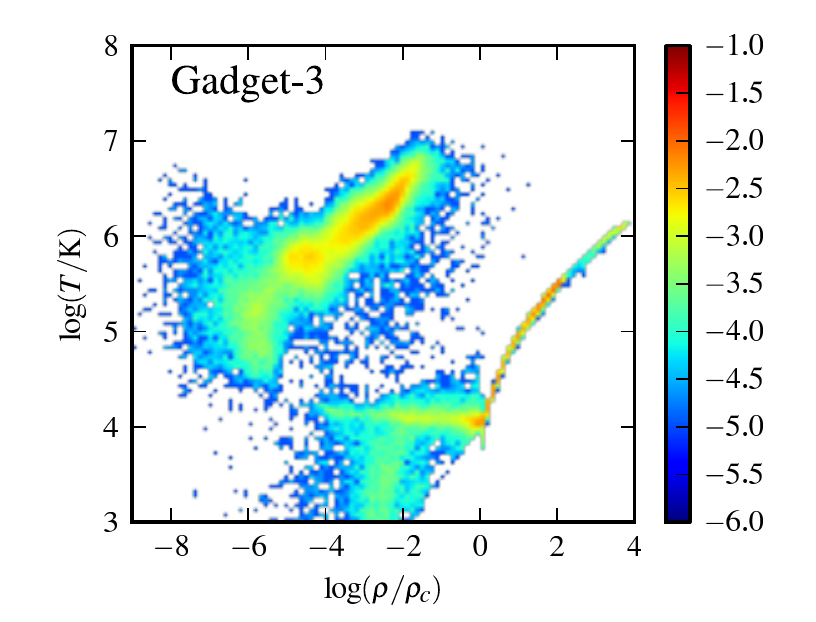}
  \caption{Example phase diagrams (the density is given in terms of the critical density above which gas can be on the \citealt{Springel:2003} EOS,
  which corresponds to $n \sim 0.1$ cm$^{-3}$)
  for the \mwe merger simulation without BHs $\sim 400$ Myr after the peak of the starburst ($t = 3.2$ Gyr) performed using \arepo~(left)
  and \gadgetthree~(right). (Note that the sharp cutoff in the upper-left corner of the \arepo~phase diagram is artificial: in \arepo, we limit the maximum temperature
  of the gas below a specified density threshold to avoid wasting computational effort on cells with very low densities and high temperatures,
  which would have small time steps because of the high sound speed but are unimportant for the hydrodynamics.)
  Typically, the phase structure of the gas at a fixed time in the \arepo~and \gadgetthree~simulations is similar. The largest differences,
  which are still relatively minor, arise during the starburst and post-starburst periods. During these
times, the \arepo~simulations often contain somewhat less hot gas, as this
  example demonstrates. The reason is that in \arepo, the hot halo gas cools more efficiently, which is why the fourth column of Fig. \ref{fig:NB_gas_comparison} demonstrates
  that after the starburst the hot halo is less prominent in the \arepo~simulations.}
  \label{fig:NB_pd_comparison}
\end{figure*}

We shall now compare the gas phase structure in the \gadgetthree~and
\arepo~simulations.  Animations showing the evolution of the gas phase
structure for all simulations are available at the aforementioned
URL.  For brevity, we will only discuss the general trends and
present an illustrative example.

Throughout the simulations, the evolution of the gas phase structure
is very similar in the \arepo~and \gadgetthree~simulations when BH
accretion and AGN feedback are not included. Given the results
presented above, this result is not surprising: if the phase structure
differed significantly, then the SFHs would not agree so well. As for
the gas morphology, the differences are more pronounced in the
starburst and post-starburst periods of the simulations. Specifically,
the \arepo~simulations tend to exhibit less low-density, hot halo
gas. In both the \arepo~and \gadgetthree~simulations, a hot halo forms
when gas is shock-heated during final coalescence of the discs. In the
\arepo~simulations, however, the hot halo gas cools more effectively
for the reasons discussed in Section \ref{S:reasons_for_agreement}.

Fig. \ref{fig:NB_pd_comparison} demonstrates this difference. This figure shows example phase diagrams for the post-starburst phase
($t = 3.2$ Gyr, $\sim 400$ Myr after the starburst) of the \mwe merger simulation performed using \arepo~(left) and \gadgetthree~(right).
Note that these correspond to the same simulation and output time as the gas surface density plots shown in the fourth column of Fig. \ref{fig:NB_gas_comparison}.
In the \arepo~simulation, there is less hot halo gas. The enhanced cooling of hot halo gas in the \arepo~simulation explains why the SFR
is somewhat higher in the \arepo~simulation at this time (see Fig. \ref{fig:merger_NB_SFHs}), which is also the case in some of the other merger
simulations (e.g., \hize).

\subsection{Simulations with BH accretion and AGN feedback} \label{S:results_BHs}

We now compare simulations that include BH accretion and AGN feedback.
Here, we use the default accretion and feedback schemes discussed
above, but we explore the implications of different choices in Section \ref{S:tests}.

\subsubsection{Star formation histories}

\begin{figure*}
  \centering
    \includegraphics[width=0.88\columnwidth]{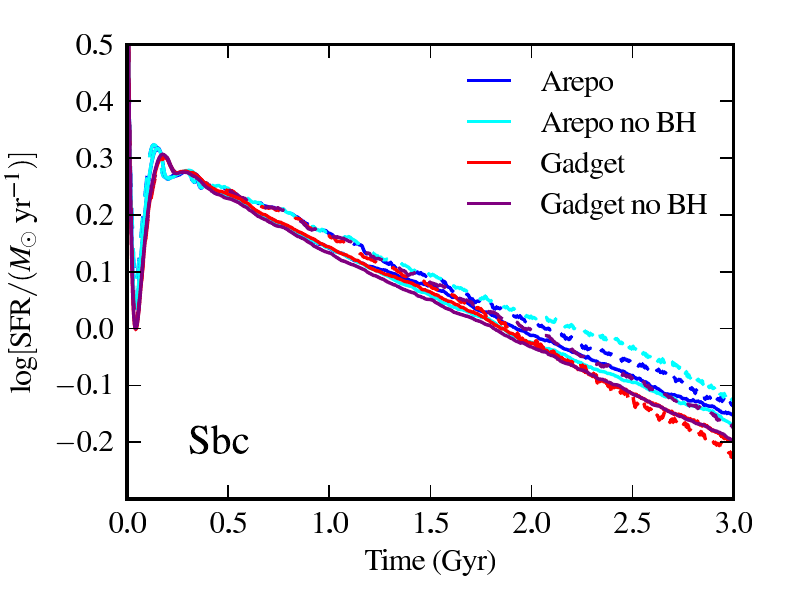}
    \includegraphics[width=0.88\columnwidth]{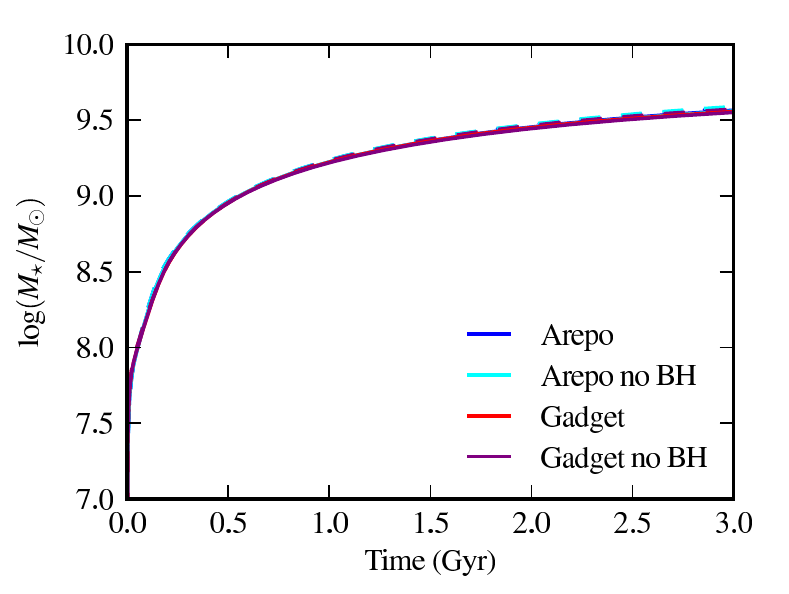} \\
    \includegraphics[width=0.88\columnwidth]{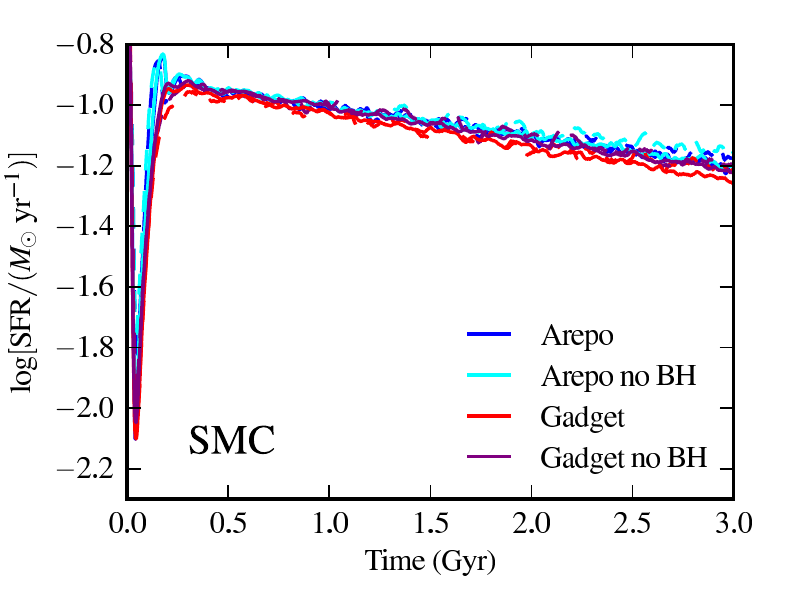}
    \includegraphics[width=0.88\columnwidth]{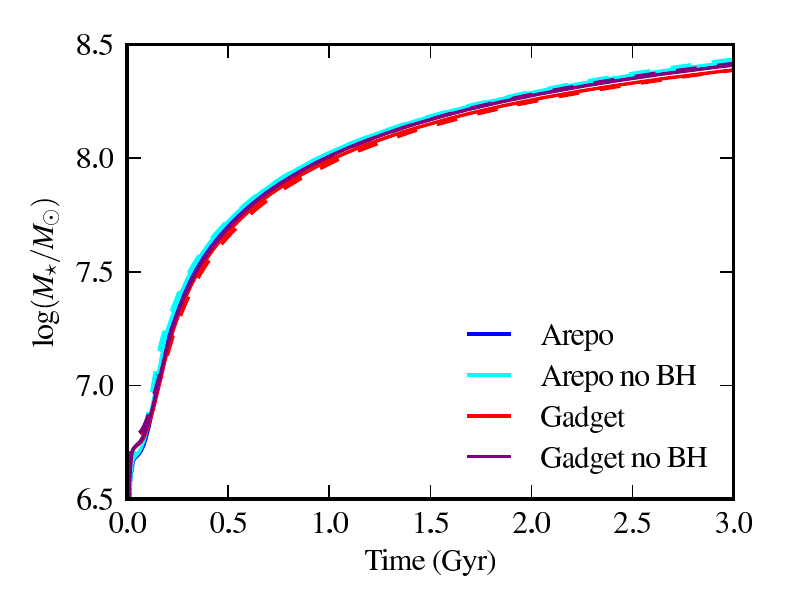} \\
    \includegraphics[width=0.88\columnwidth]{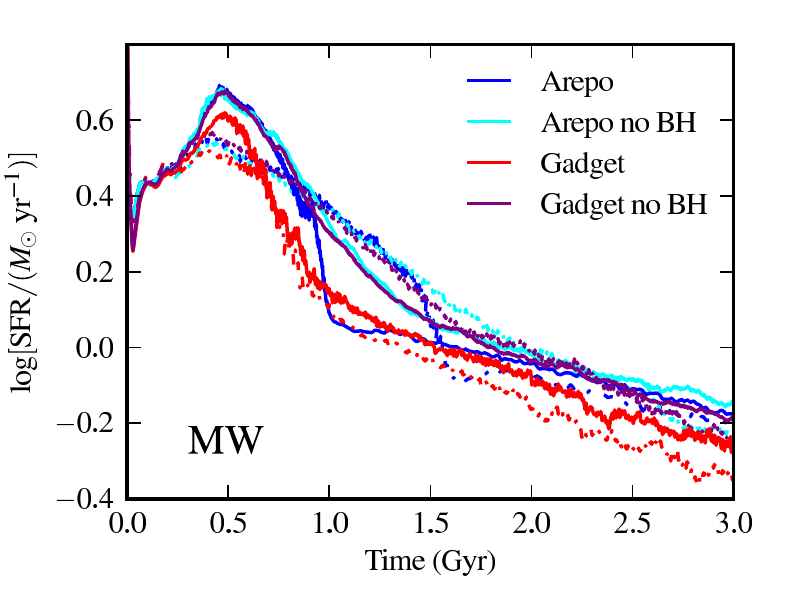}
    \includegraphics[width=0.88\columnwidth]{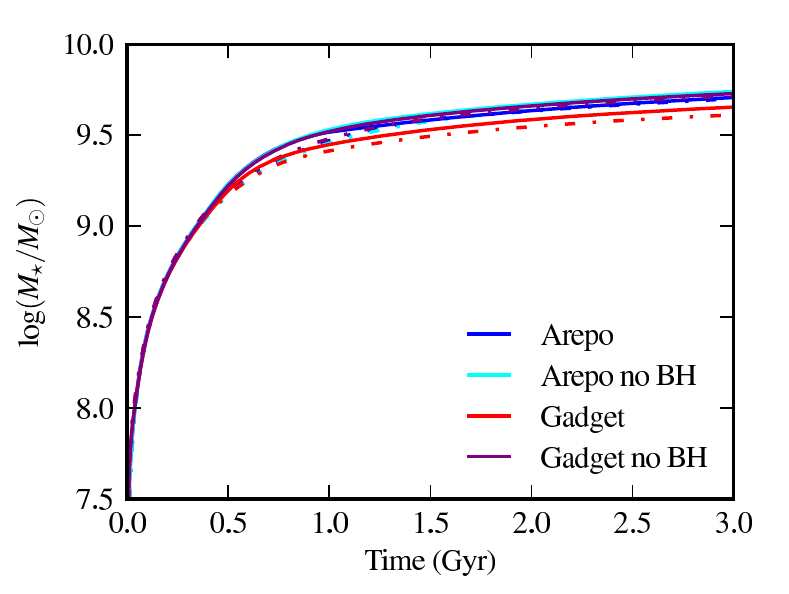} \\
    \includegraphics[width=0.88\columnwidth]{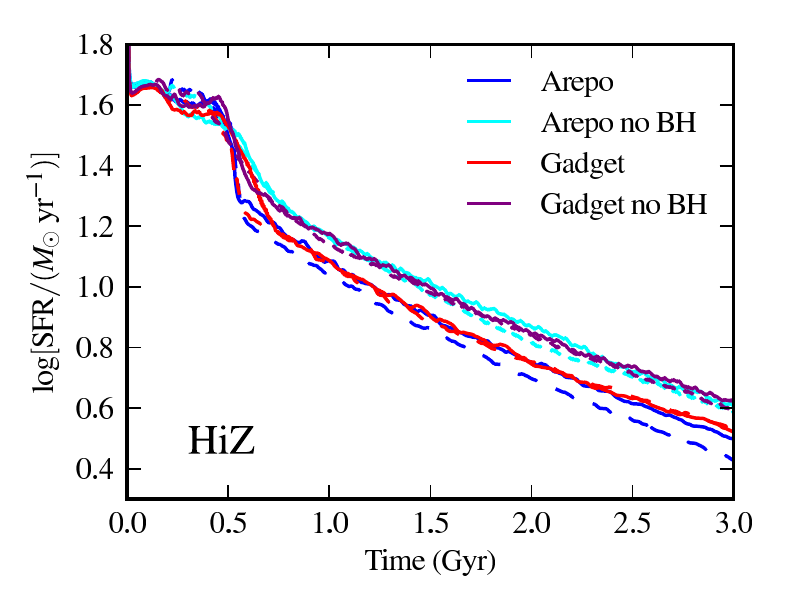}
    \includegraphics[width=0.88\columnwidth]{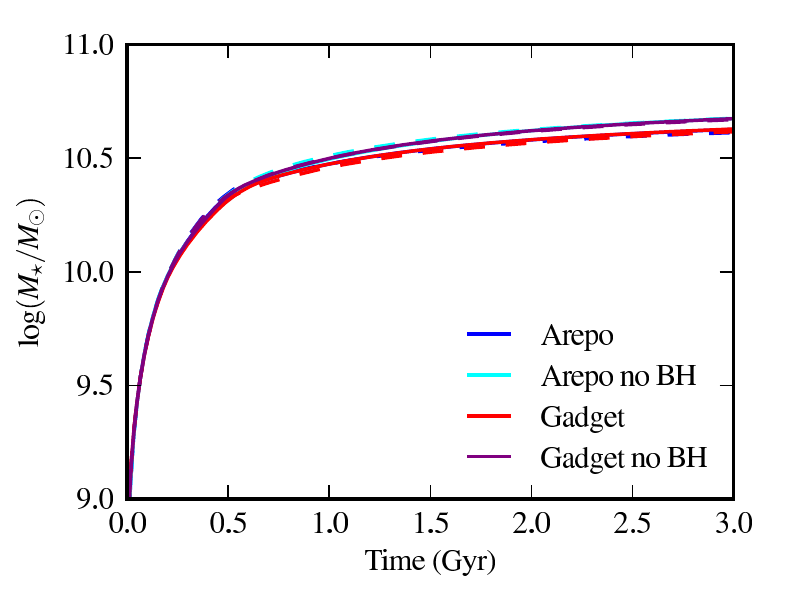}
  \caption{SFHs (left) and cumulative stellar mass formed versus time (right) for the isolated disc galaxy simulations: \sbc (first row),
  \smc (second row), \mw (third row), and \hiz (fourth row). Blue (red) indicates \arepo~(\gadgetthree) simulations with BH accretion and AGN feedback;
  the corresponding simulations without BH accretion and AGN feedback are indicated in cyan (magenta). Solid (dashed) lines indicate higher (lower) resolution simulations.
  For the \mw and \hiz simulations, inclusion of BH accretion and AGN feedback tends to decrease the SFR. When BH accretion and AGN feedback are included, the SFHs agree
  somewhat less well than when those sub-resolution models are disabled, but the agreement is still quite good.}
  \label{fig:iso_SFHs}
\end{figure*}

\begin{figure*}
  \centering
    \includegraphics[width=0.88\columnwidth]{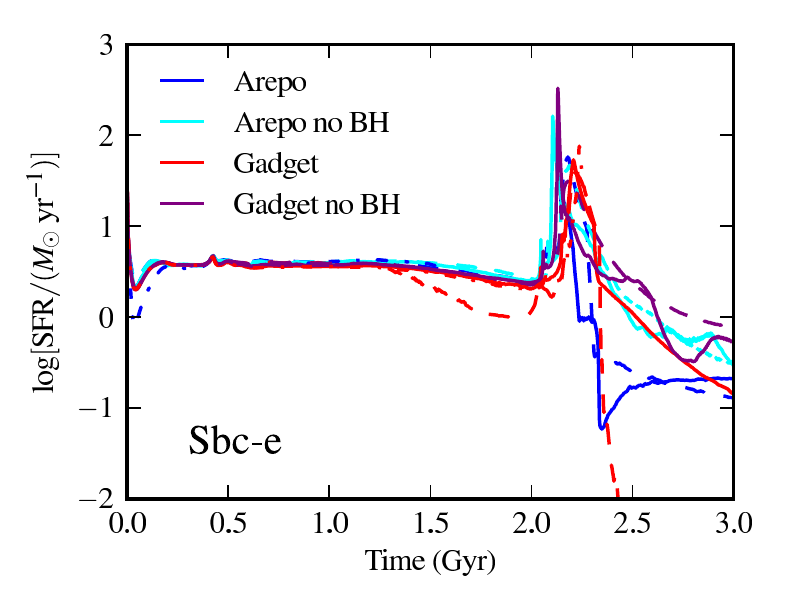}
    \includegraphics[width=0.88\columnwidth]{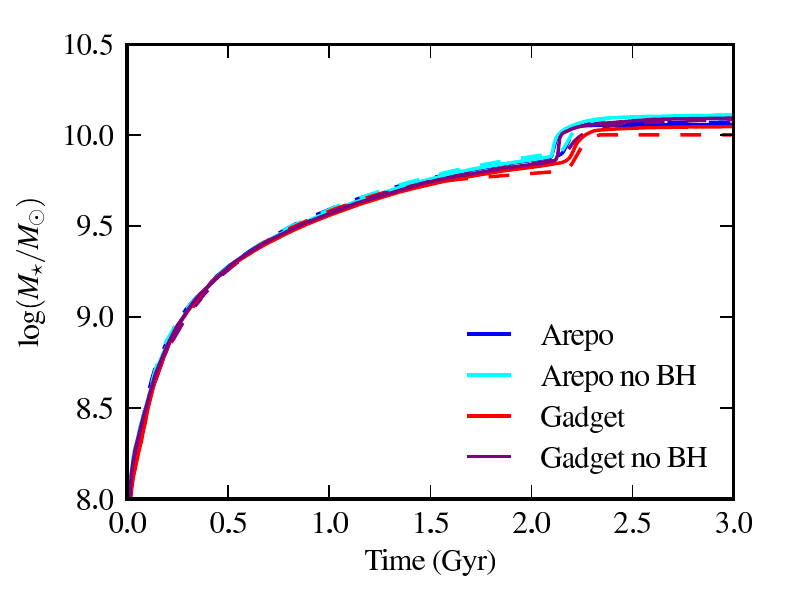} \\
    \includegraphics[width=0.88\columnwidth]{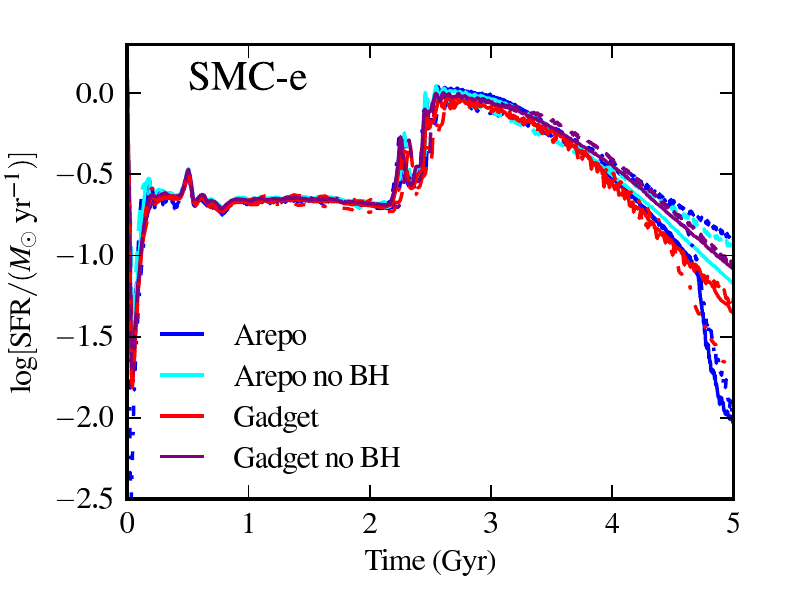}
    \includegraphics[width=0.88\columnwidth]{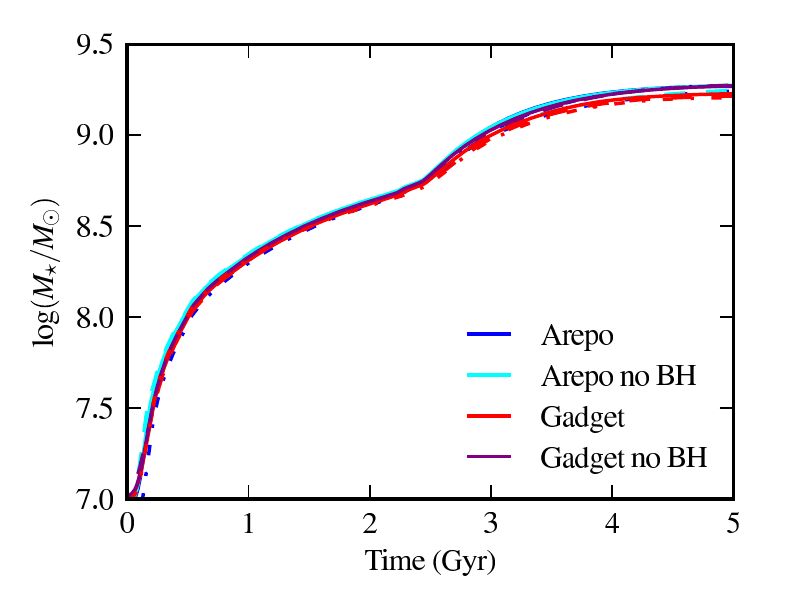} \\
    \includegraphics[width=0.88\columnwidth]{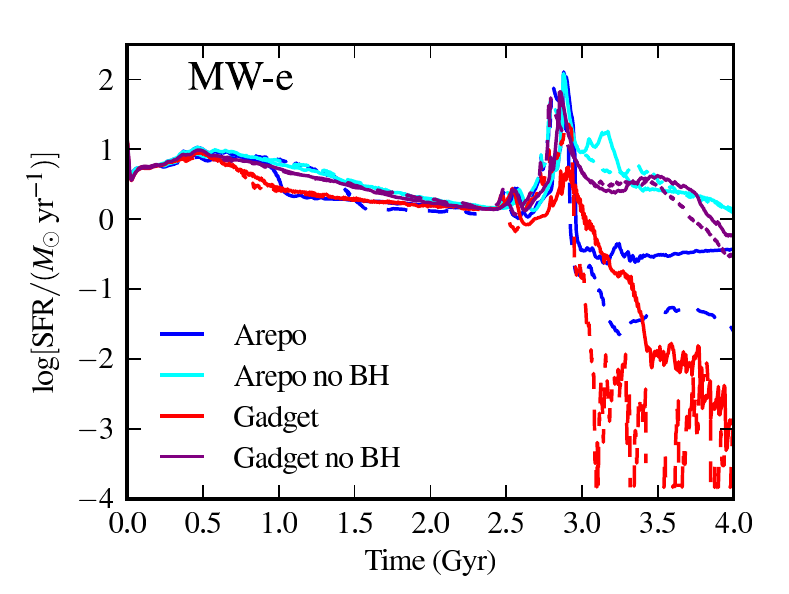}
    \includegraphics[width=0.88\columnwidth]{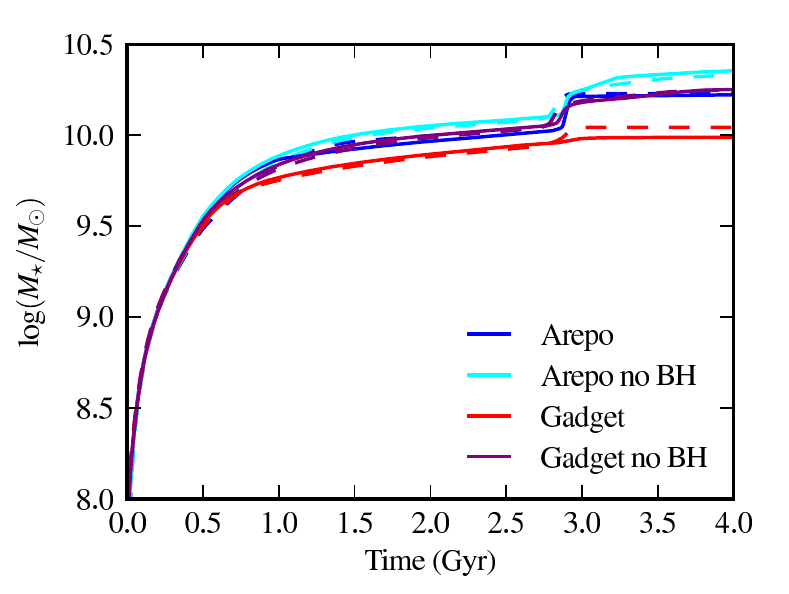} \\
    \includegraphics[width=0.88\columnwidth]{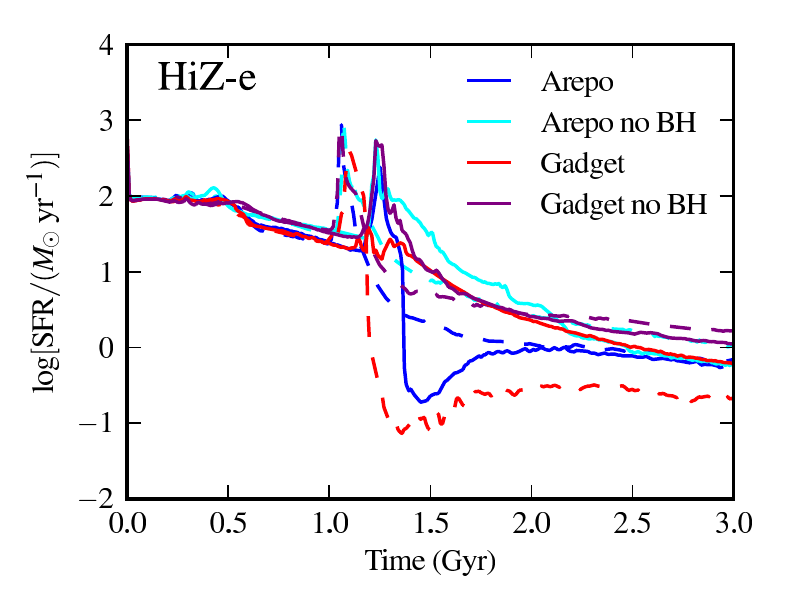}
    \includegraphics[width=0.88\columnwidth]{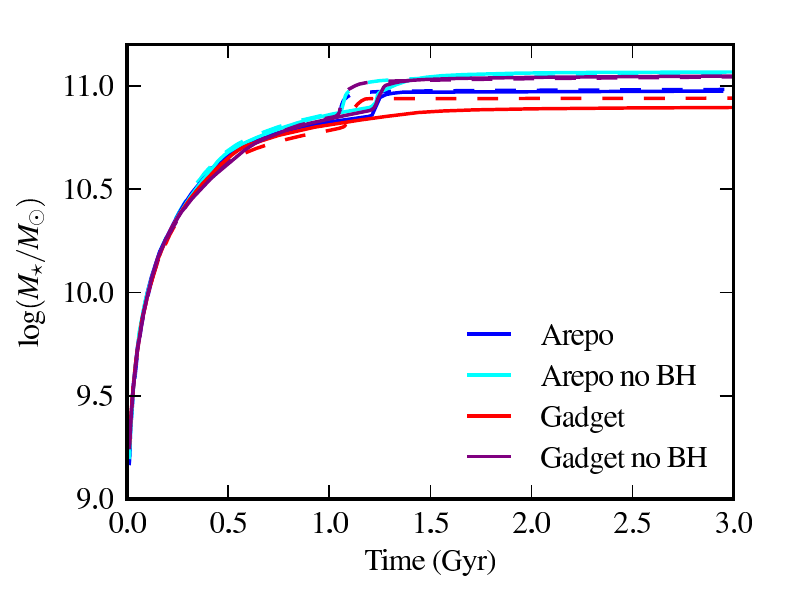} \\
  \caption{Similar to Fig. \ref{fig:iso_SFHs}, but for the \eorbit~merger simulations. The $\smce$ simulation was performed at a third, higher resolution (R4); the results for this resolution
  are denoted with dot-dashed lines. The agreement between the \arepo~and \gadgetthree~results is less good than for the isolated
  discs, but the differences are comparable with those between different resolutions. In many cases, AGN feedback decreases the post-starburst SFR, but the strength of the effect depends
  on both the code used and the resolution.}
  \label{fig:e_SFHs}
\end{figure*}

\begin{figure*}
  \centering
    \includegraphics[width=0.88\columnwidth]{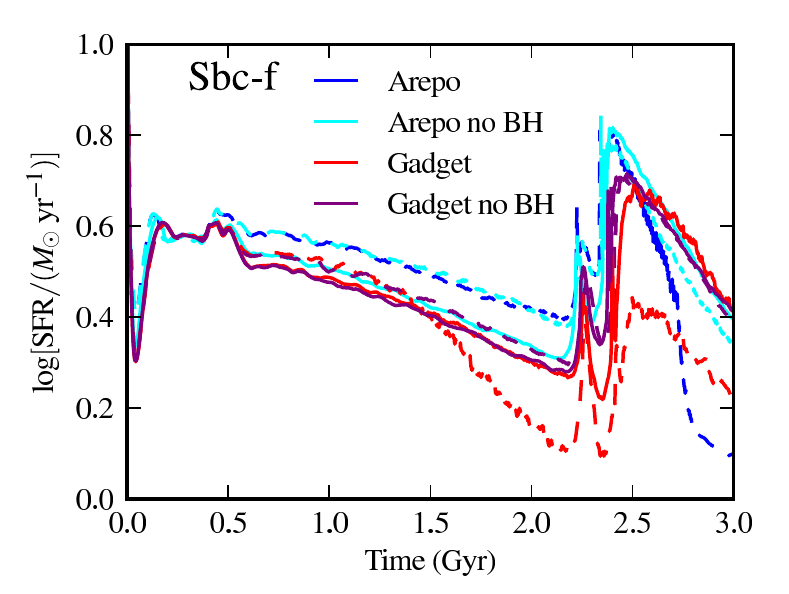}
    \includegraphics[width=0.88\columnwidth]{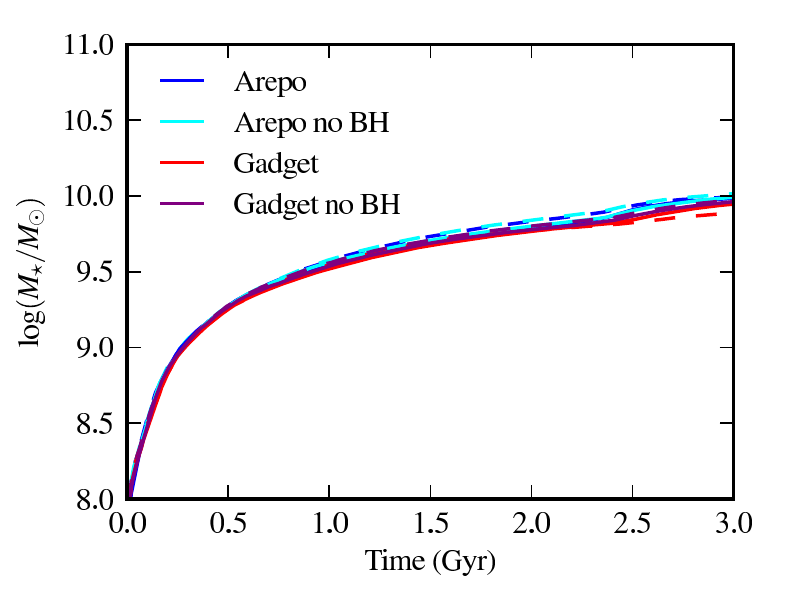} \\
    \includegraphics[width=0.88\columnwidth]{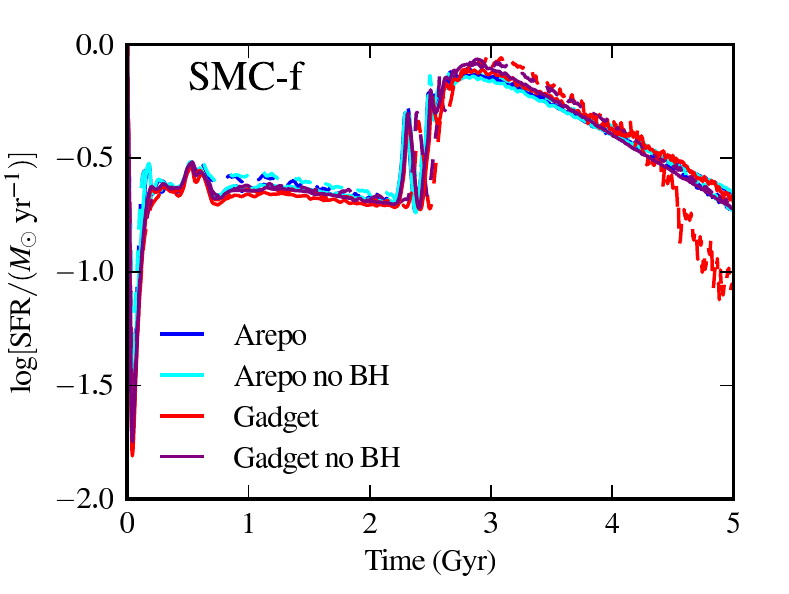}
    \includegraphics[width=0.88\columnwidth]{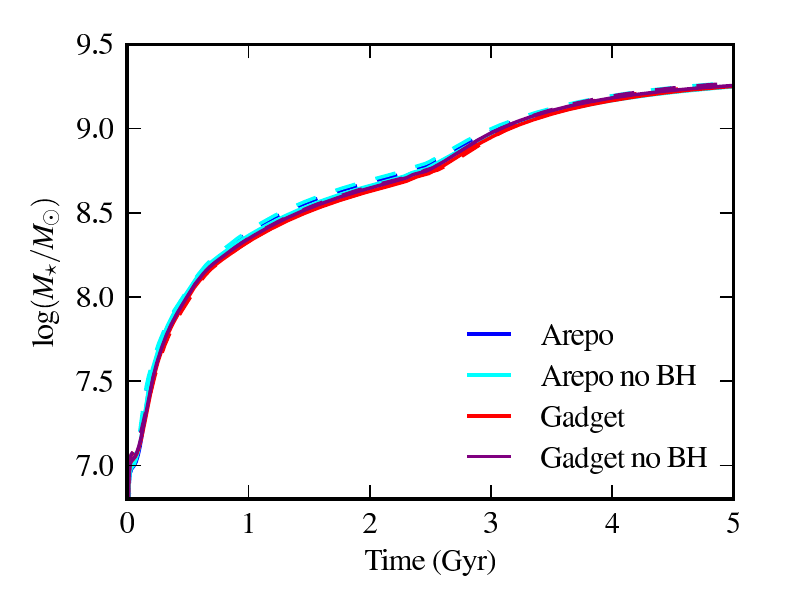} \\
    \includegraphics[width=0.88\columnwidth]{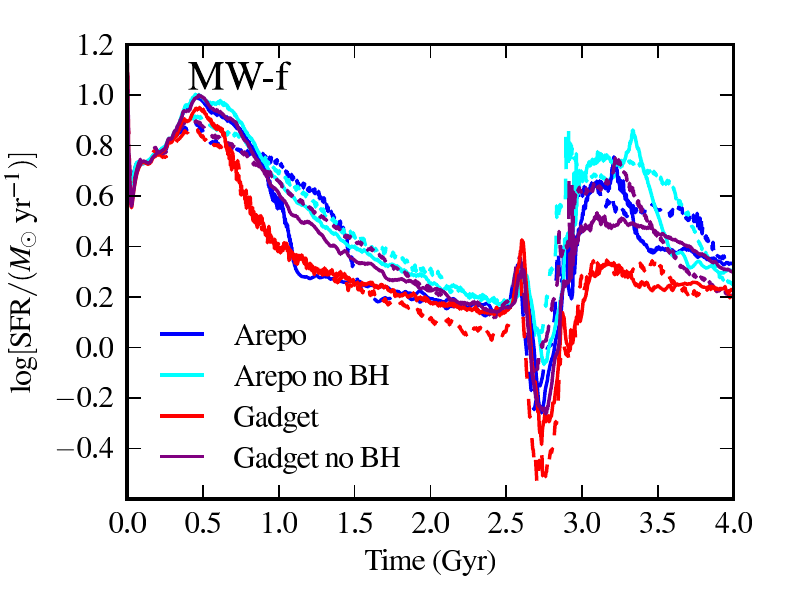}
    \includegraphics[width=0.88\columnwidth]{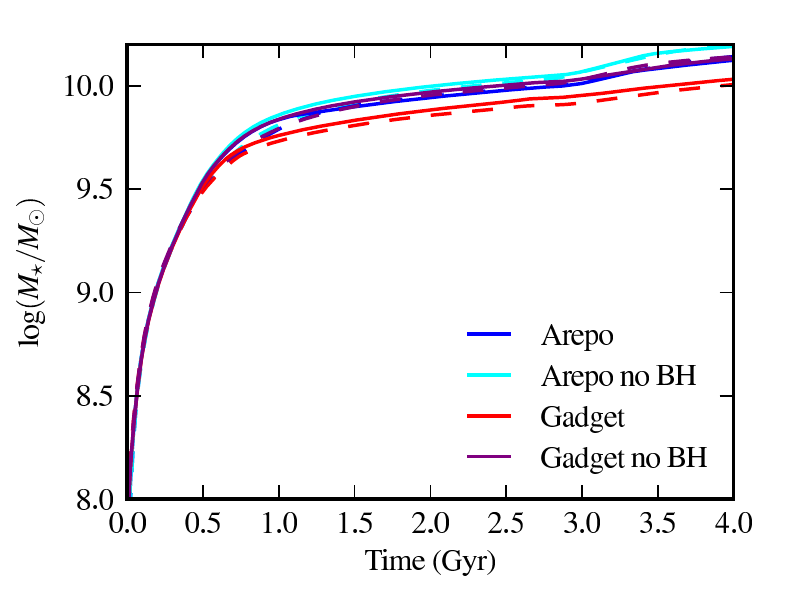} \\
    \includegraphics[width=0.88\columnwidth]{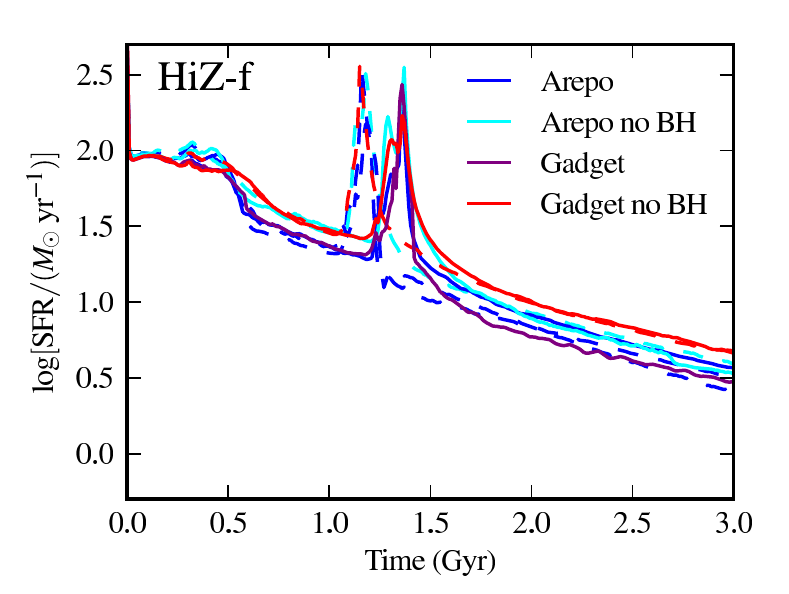}
    \includegraphics[width=0.88\columnwidth]{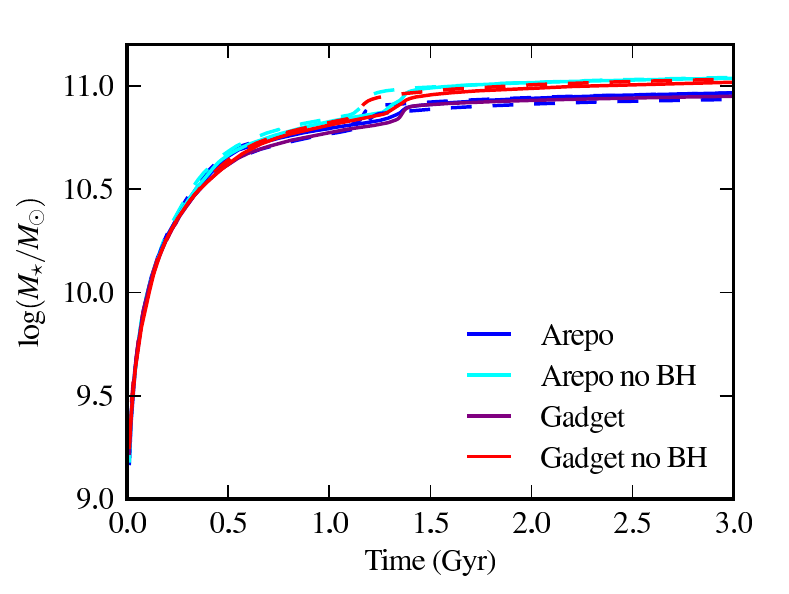} \\
  \caption{Similar to Fig. \ref{fig:iso_SFHs}, but for the \forbit~merger simulations.}
  \label{fig:f_SFHs}
\end{figure*}

Fig. \ref{fig:iso_SFHs} shows the SFHs and cumulative stellar mass formed versus time for the isolated disc simulations.
Blue (red) indicates \arepo~(\gadgetthree) simulations with BH accretion and AGN feedback;
the corresponding simulations without BH accretion and AGN feedback are indicated in cyan (magenta) for comparison.
Solid (dashed) lines indicate higher (lower) resolution simulations.

In general, for the isolated disc simulations, the agreement among the
SFHs for the two codes and different resolutions is still relatively
good, but the differences are clearly more significant than when BH
accretion and AGN feedback are disabled. For the \sbc case
(first row of Fig. \ref{fig:iso_SFHs}), the \arepo~simulations tend to
have slightly higher SFRs, but the differences between the simulations'
SFRs are less than $\sim 0.1$ dex at all times. The cumulative stellar mass
formed is indistinguishable.  The SFHs for the \smc simulations
(second row) agree similarly well, and the cumulative stellar mass
formed is nearly the same. For the \mw simulations (third
row), the SFRs differ by as much as $\sim 0.3$ dex, but the
resolution-dependent variations are as significant as those between
the codes and the differences for simulations that
vary only in whether they include BH accretion and AGN feedback. 
In this case, the total stellar mass formed over the course of the
simulations (3 Gyr) is $\sim 0.1$ dex less in the
\gadgetthree~simulations with AGN feedback, but all other simulations
agree very well. Finally, the SFRs of the \hiz isolated disc
simulations (fourth row) can vary by as much as $\sim 0.2$ dex, and
the primary cause of the difference is whether BH accretion and AGN
feedback are included (in such simulations, the SFRs are
systematically lower, as expected).  However, the stellar mass
formed is the same to within $\la 0.1$ dex.

Figs. \ref{fig:e_SFHs} and \ref{fig:f_SFHs} show the SFHs and
cumulative stellar mass formed versus time for the \eorbit~and
\forbit~merger simulations, respectively.  For brevity, we will not
discuss each panel individually but rather highlight general
trends. For a given progenitor combination and orbit, the shapes of
the SFHs are qualitatively similar for both codes and all resolutions
(except in the cases for which AGN feedback significantly impacts the
post-starburst SFR). However, there are significant quantitative code-
and resolution-dependent differences. For many of the simulations
(i.e., \sbce, \smce, \hize,
\smcf, and \hizf), the SFHs are almost identical
up to final coalescence. If a strong starburst is triggered, the
precise time and amplitude of the starburst can vary depending on the
code and resolution. In some -- but not all -- cases (\sbce
and \mwe, in particular), inclusion of AGN feedback causes
the post-starburst SFR to decrease more rapidly compared with the
corresponding simulations without AGN feedback. Although the SFHs
differ significantly at some times, the cumulative stellar mass formed
versus time is often very similar for the different codes and
resolutions, as we saw above for the simulations without AGN feedback
and the isolated disc simulations with AGN feedback.  In most cases,
the cumulative stellar mass formed differs by less than $\sim 0.1$ dex
at all times. The \mwe merger simulations exhibit the most
significant differences in the cumulative stellar mass formed: when
AGN feedback is included, the cumulative stellar mass formed in the
\arepo~simulations is $\sim 0.25$ dex greater than in the
\gadgetthree~simulations.

\subsubsection{BH masses}

\begin{figure*}
  \centering
    \includegraphics[width=0.88\columnwidth]{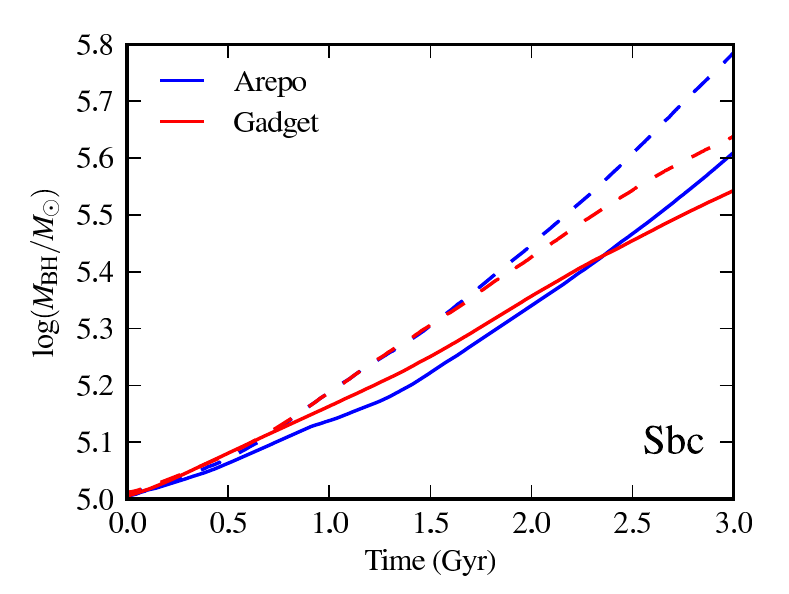}
    \includegraphics[width=0.88\columnwidth]{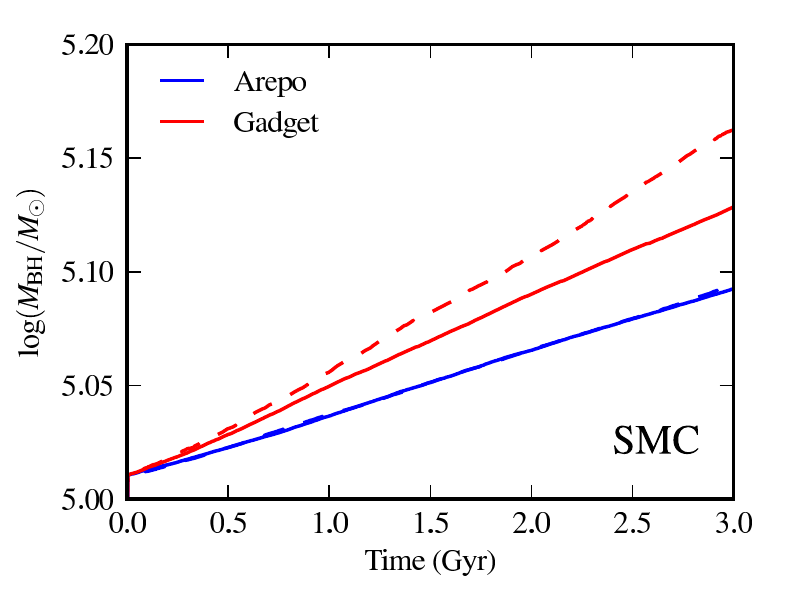} \\
    \includegraphics[width=0.88\columnwidth]{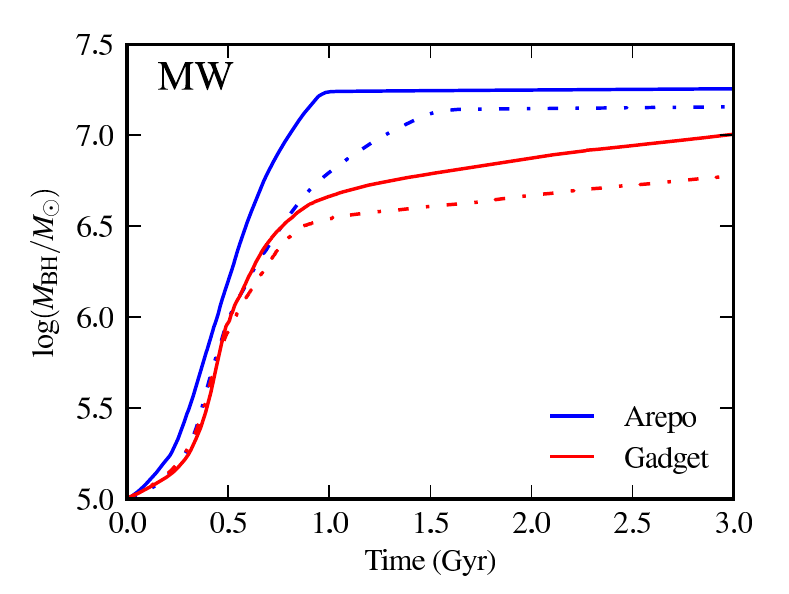} 
    \includegraphics[width=0.88\columnwidth]{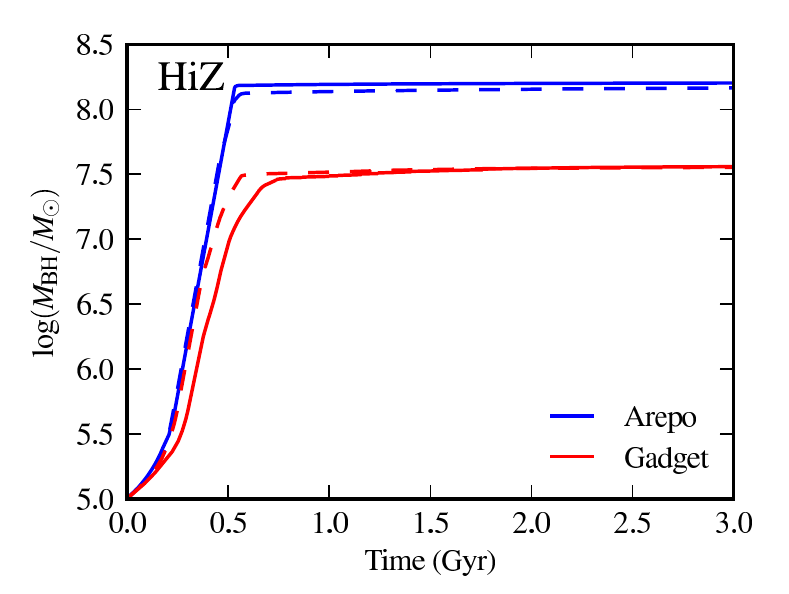} 
  \caption{BH mass versus time for the isolated disc galaxy simulations: \sbc (top left), \smc (top right), \mw (bottom left), and \hiz (bottom right).
  Blue (red) indicates \arepo~(\gadgetthree) simulations with BH accretion and AGN feedback. Solid (dashed) lines indicate higher- 
 (lower-) resolution simulations.
  For these simulations, the BH masses are reasonably well converged with respect to particle number, and the code-dependent 
  effects are more significant
  than the differences between the two resolutions. For all but the \smc simulation, for a given particle number, \arepo~yields more massive BHs.}
  \label{fig:iso_BH_mass}
\end{figure*}

\begin{figure*}
  \centering
    \includegraphics[width=0.88\columnwidth]{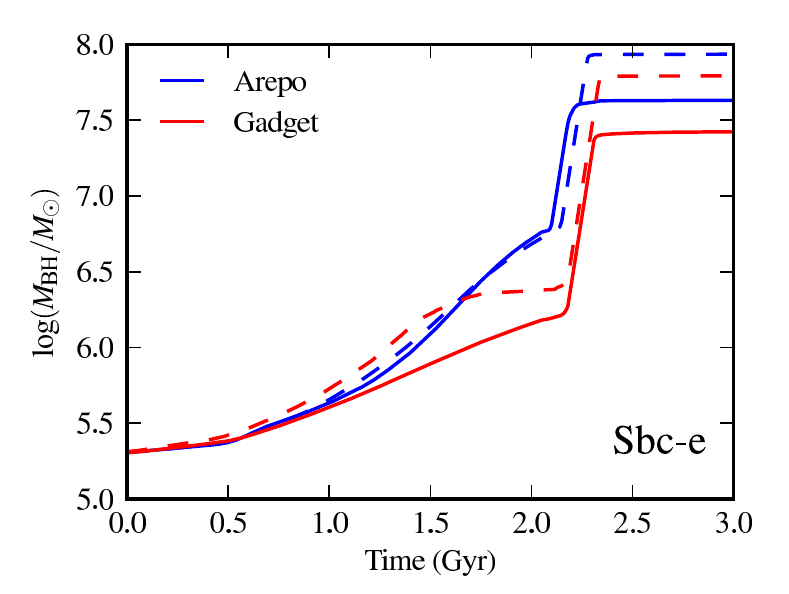}
    \includegraphics[width=0.88\columnwidth]{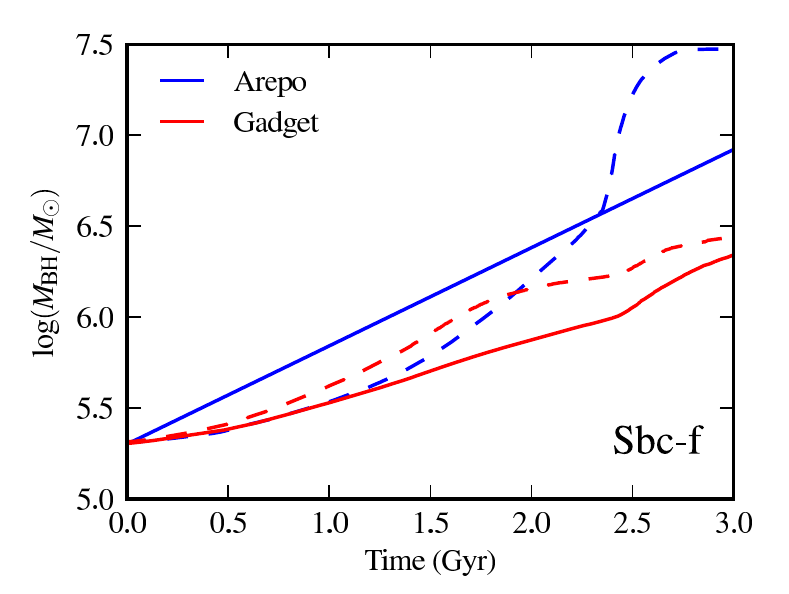} \\
    \includegraphics[width=0.88\columnwidth]{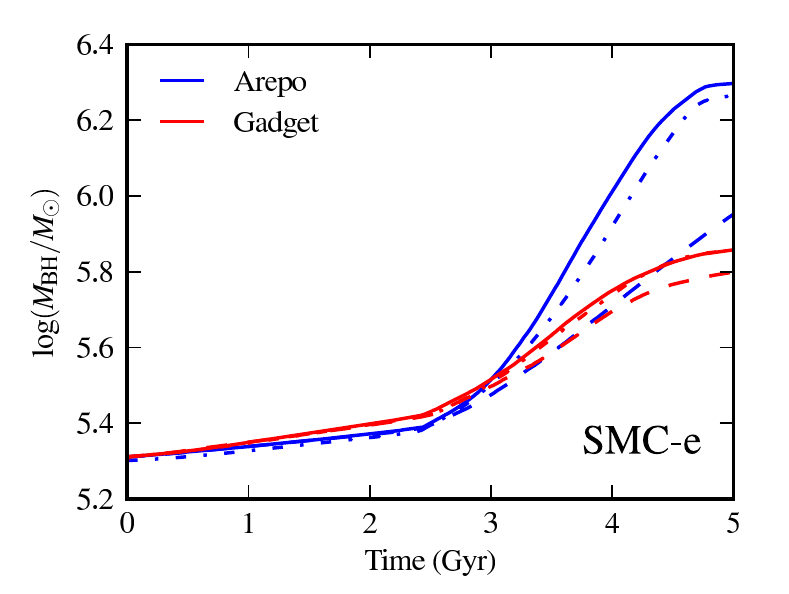}
    \includegraphics[width=0.88\columnwidth]{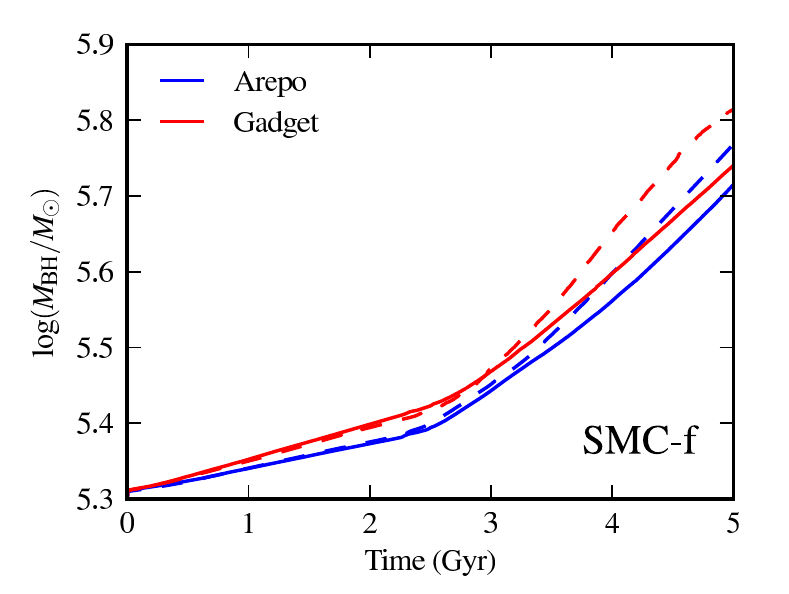} \\
    \includegraphics[width=0.88\columnwidth]{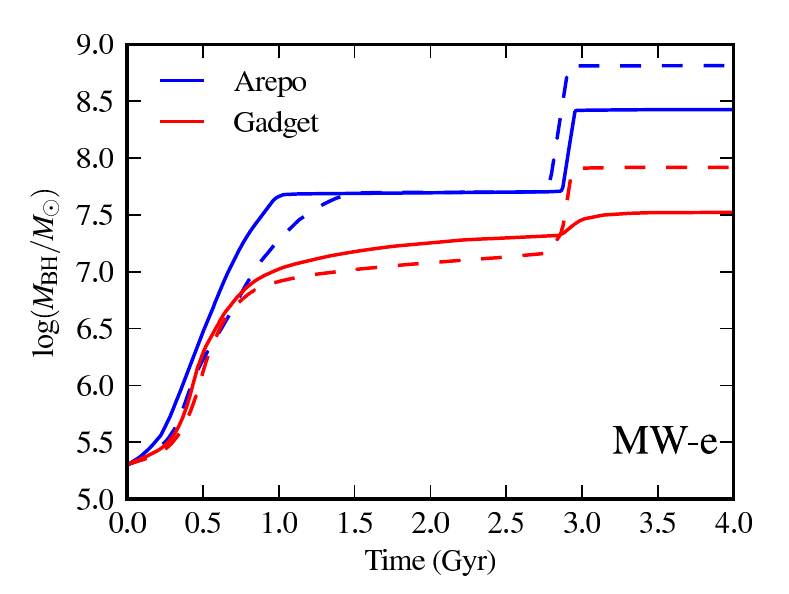}
    \includegraphics[width=0.88\columnwidth]{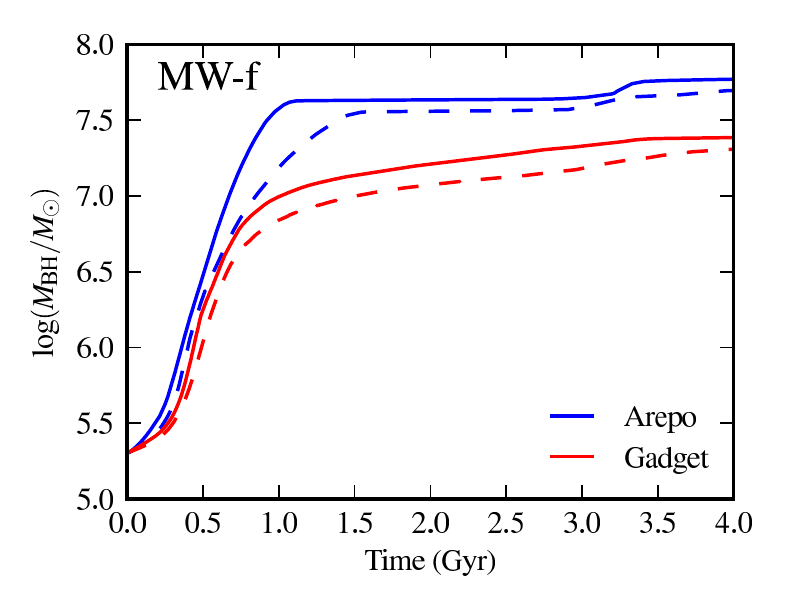} \\
    \includegraphics[width=0.88\columnwidth]{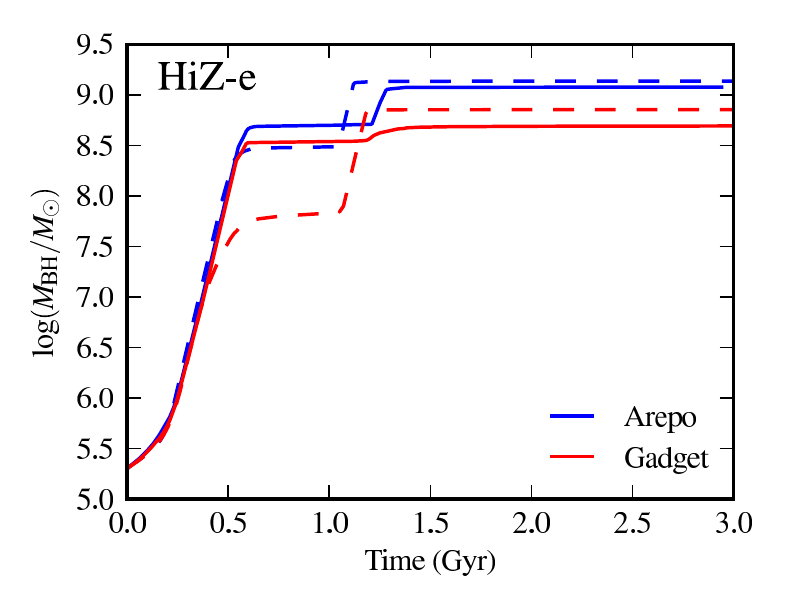}
    \includegraphics[width=0.88\columnwidth]{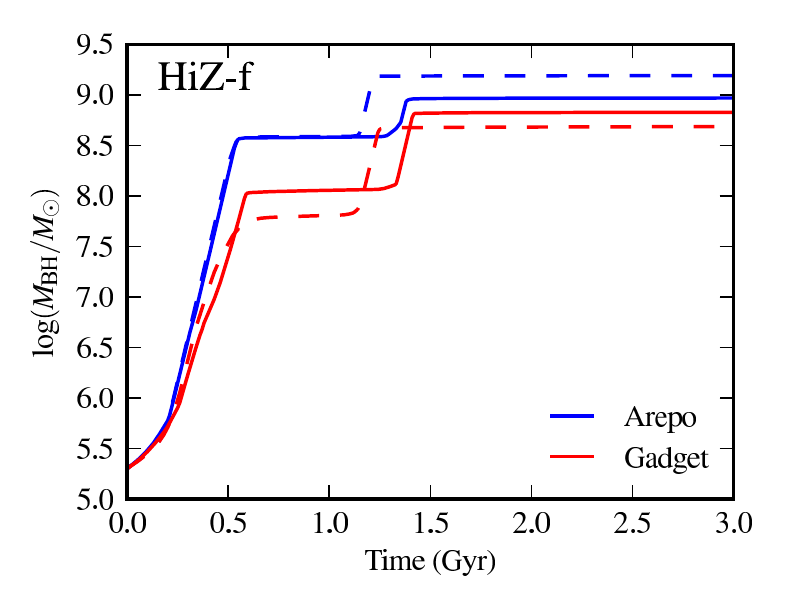}
  \caption{Similar to Fig. \ref{fig:iso_BH_mass}, but for the merger simulations: \sbc (first row), \smc (second), \mw (third), and \hiz (fourth).
  The left (right) column shows the results for the \eorbit~(\forbit) merger simulations. Note that the \smce merger was run at a third, higher
  resolution (R4); the results are denoted with dot-dashed lines. During the mergers, the BH masses can increase by multiple orders of magnitude.
  The code- and resolution-dependent differences in the BH masses can be as large as a factor of a few. The differences with resolution are not systematic.
  For all but the \smcf simulations, for a given resolution, the BH masses yielded by \arepo~are greater than those yielded by \gadgetthree.}
  \label{fig:merger_BH_mass}
\end{figure*}

Another quantity of interest is the BH mass versus time because
differences in the BH masses would alter the strength of the AGN
feedback and thereby influence whether or not the merger remnants lie
on the $M_{\rm BH}-\sigma$ relation \citep{Ferrarese2000,Gebhardt2000}.
It is natural to expect that the
BH mass evolution is more code- and resolution-dependent than the SFH
because the BH growth depends sensitively on the gas conditions in the
nuclear region(s) and could thus be affected by small-scale
variations that would not significantly alter the integrated SFH.

Fig. \ref{fig:iso_BH_mass} shows the BH mass versus time for the
isolated disc simulations. In the \sbc and \smc simulations, the BHs
grow by only a modest amount ($\la 0.8$ and $\la 0.15$ dex,
respectively) over the 3.0 Gyr of the simulations. In the \mw and \hiz
simulations, the BHs rapidly increase in mass by 2-3 orders of
magnitude. This strong BH growth in the absence of a merger is driven
by bar instabilities, which is clear from examination of the gas
morphologies. The BH growth terminates once the gas near the BH is consumed or expelled.

For a given code, the final BH masses in the isolated disc simulations
differ with resolution by $\la 0.2$ dex. However, the code-dependent
variations can be more significant.  In particular, note that the
final BH masses in the \arepo~\hiz simulations are a factor of $\sim
5$ greater than those in the \gadgetthree~simulations.

The BH mass evolution for the merger simulations is shown in
Fig. \ref{fig:merger_BH_mass}. In these plots, the total mass of all
BHs -- recall that each progenitor disc is seeded with a BH -- is
plotted. For a given progenitor, the BHs grow significantly more in
the merger simulations than in the isolated disc simulations. The \smcf
merger exhibits the weakest growth ($\sim 0.5$ dex), and the \hiz
merger simulations exhibit the strongest BH growth (almost four orders
of magnitude).  As for the SFHs, the code- and resolution-dependent
differences amongst the BH masses in the merger simulations are more
significant than for the isolated disc simulations.  In many -- but not all
-- examples, the BH masses are greater in the \arepo~simulations than
in the \gadgetthree~runs. The resolution-dependent variations (which
are at most $\sim 0.6$ dex and usually significantly less) are
typically less than the code-dependent differences (for a given
resolution, these can be as great as an order of magnitude), and the
resolution dependence is not systematic. Consequently, for a given
code, the final BH masses should be considered uncertain by as much as
a factor of a few. This reflects the high degree of non-linearity in
the feedback-regulated BH growth. Any small variation in the local gas
conditions at the BH's position can influence its exponential growth
rate and hence become strongly amplified with time.  Note also that the
BH accretion histories can differ significantly depending on the
resolution and code; thus, during the merger, the BH masses at a given
time can differ more significantly than the final BH masses (i.e., the
BH masses after the BH growth has terminated, which can occur at
different times for different resolutions and codes).

Interestingly, for the \smce merger, which was simulated at three resolutions,
the BH mass evolution in the two higher-resolution
(resolutions R3 and R4) simulations performed with a given code is almost identical,
but the $\sim 0.4$ dex difference in the final BH masses yielded by the two
different codes persists. Thus, it is possible that the BH masses
yielded by a given code would converge if all simulations were performed
at even higher resolution, but we have not performed such simulations
because of the computational expense and because the code-dependent
differences, which are the focus of this work, remain even for the highest-resolution
\smce simulations.

\subsubsection{Gas morphologies}

\begin{figure*}
  \centering
    \includegraphics[width=0.5\columnwidth]{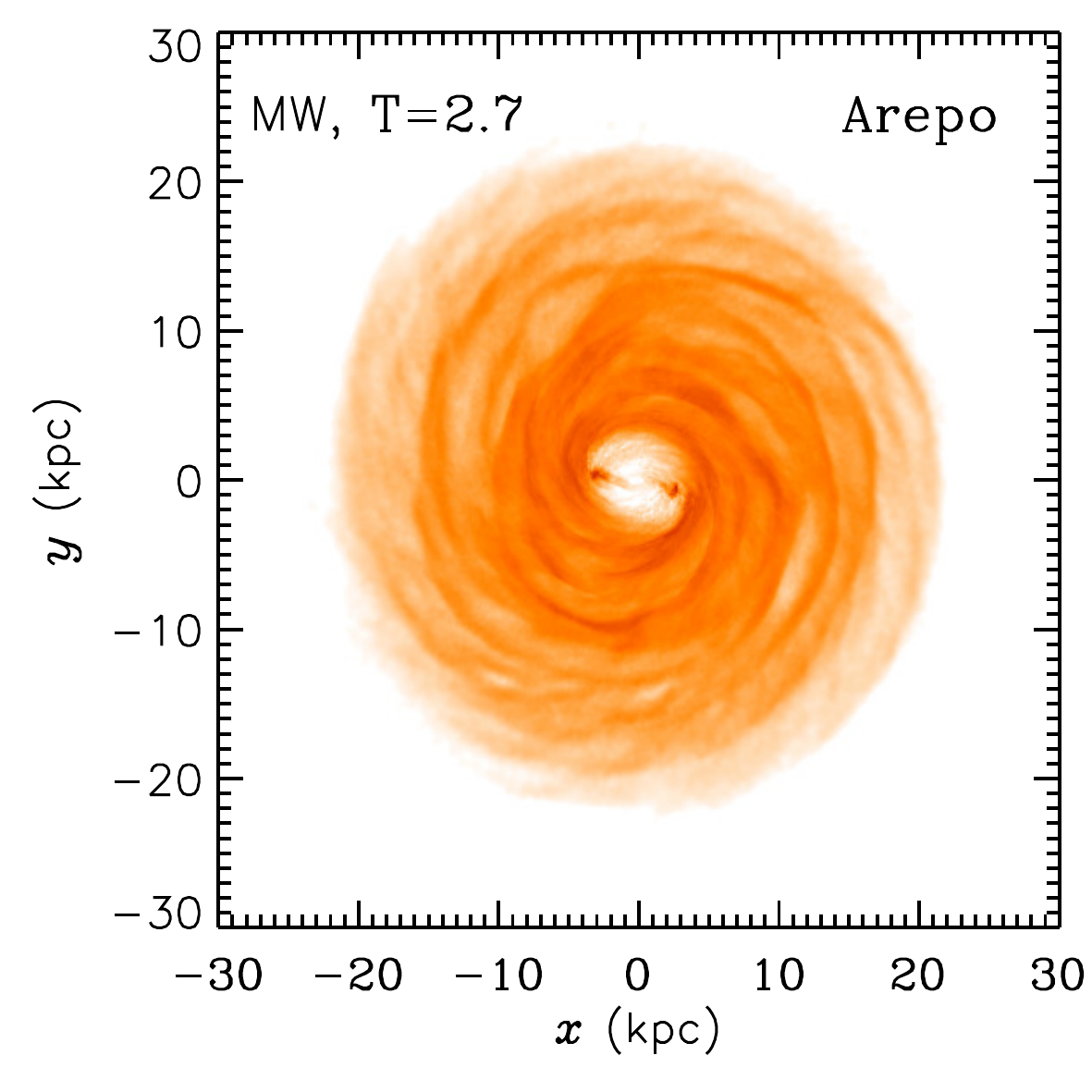}
    \includegraphics[width=0.5\columnwidth]{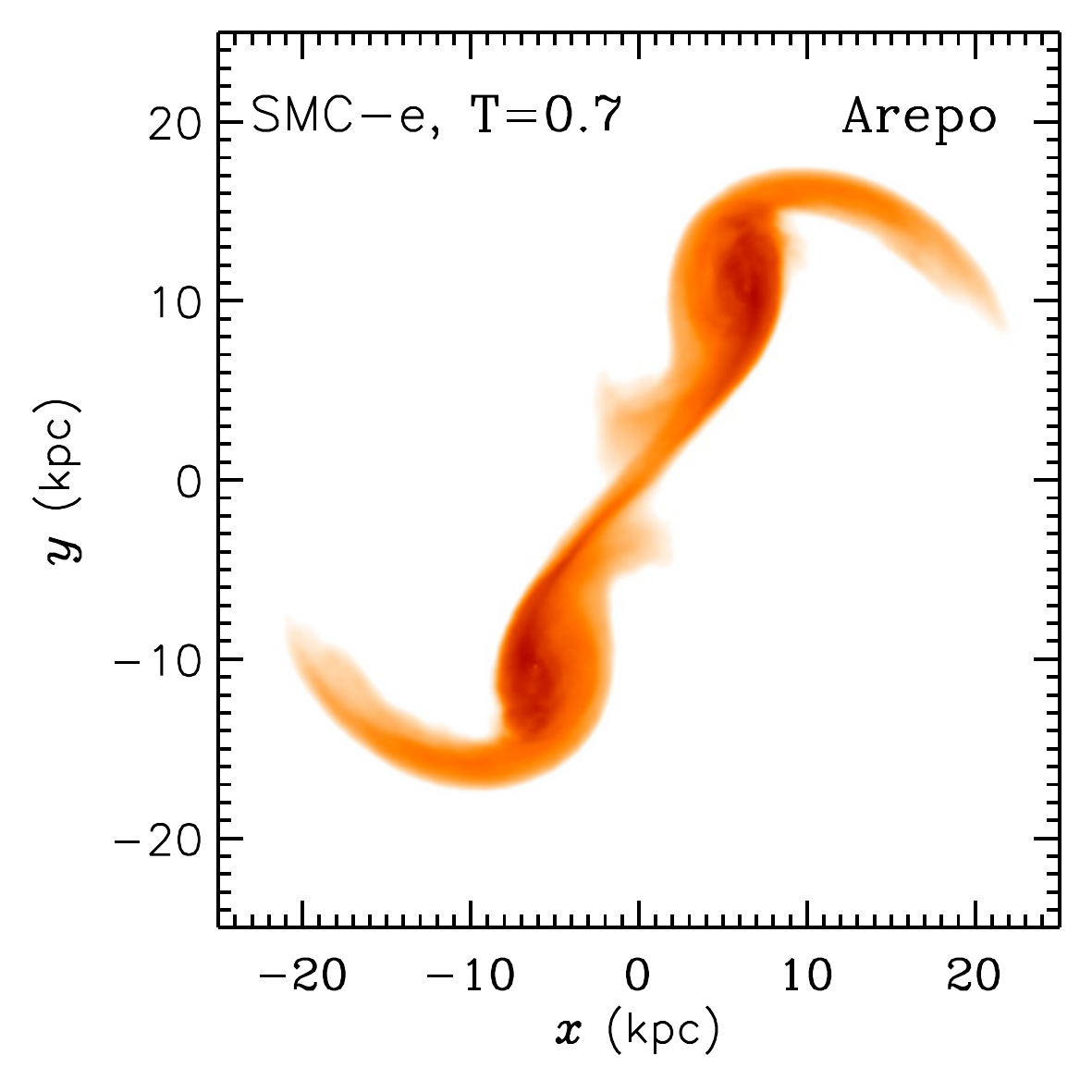}
    \includegraphics[width=0.5\columnwidth]{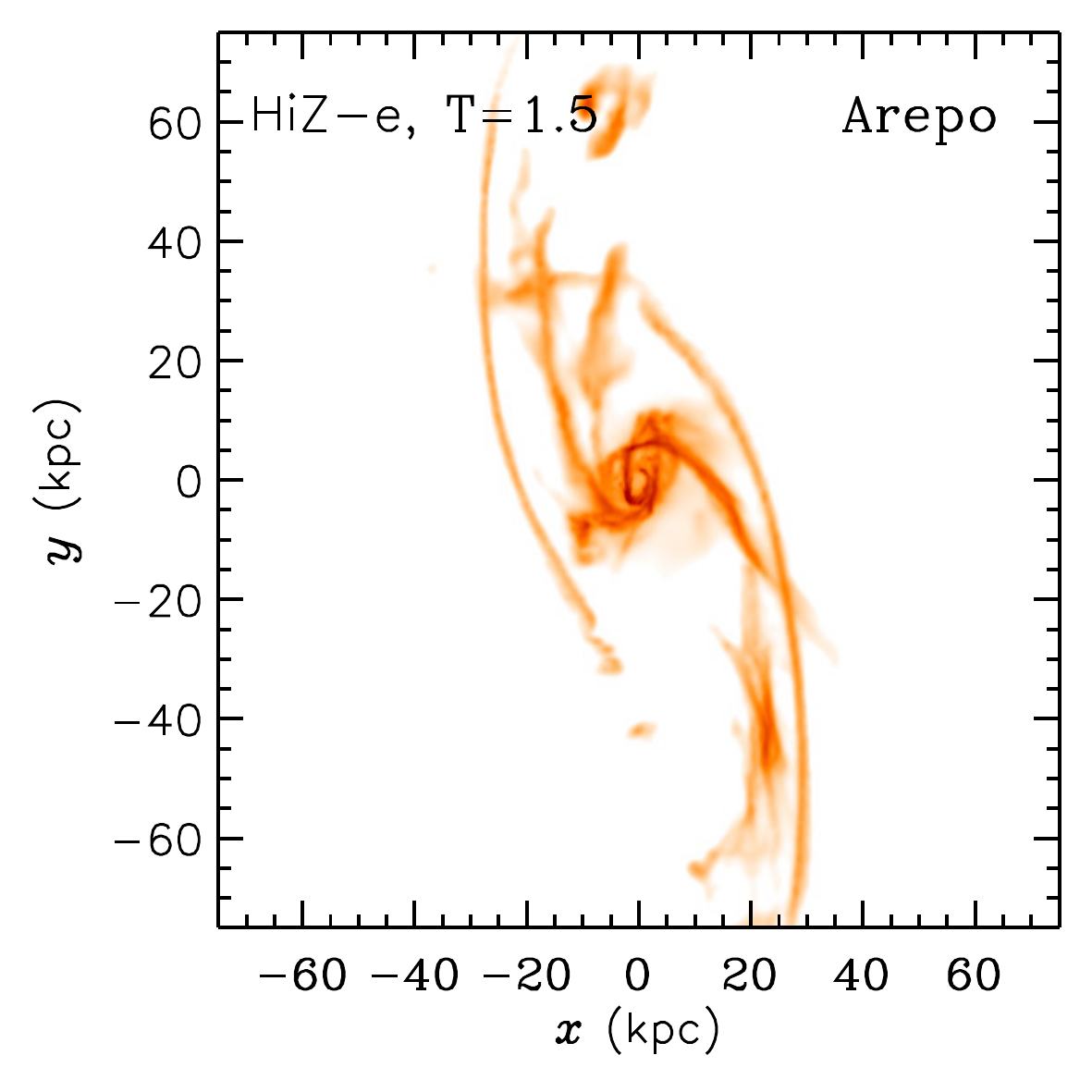}
    \includegraphics[width=0.5\columnwidth]{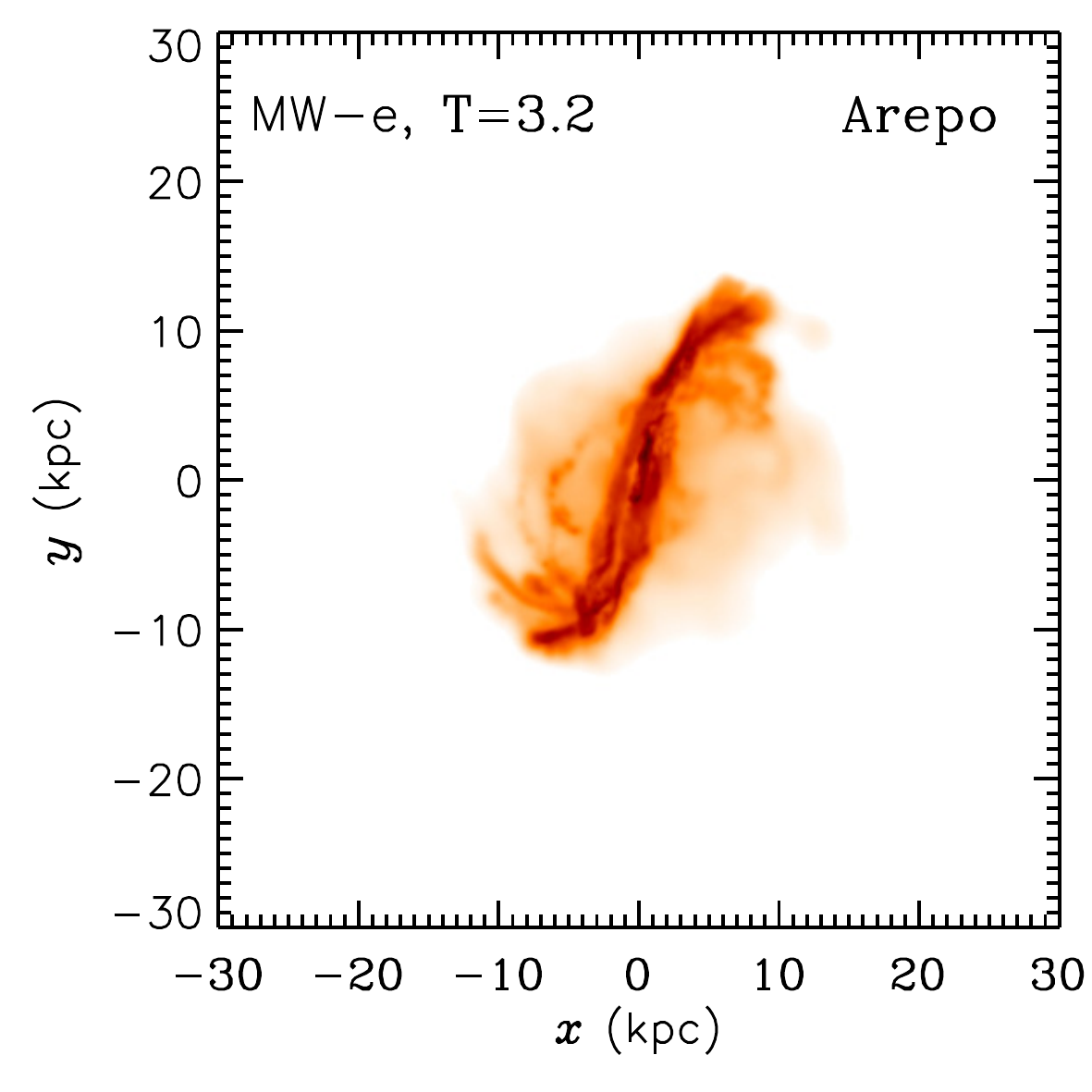} \\
    \includegraphics[width=0.5\columnwidth]{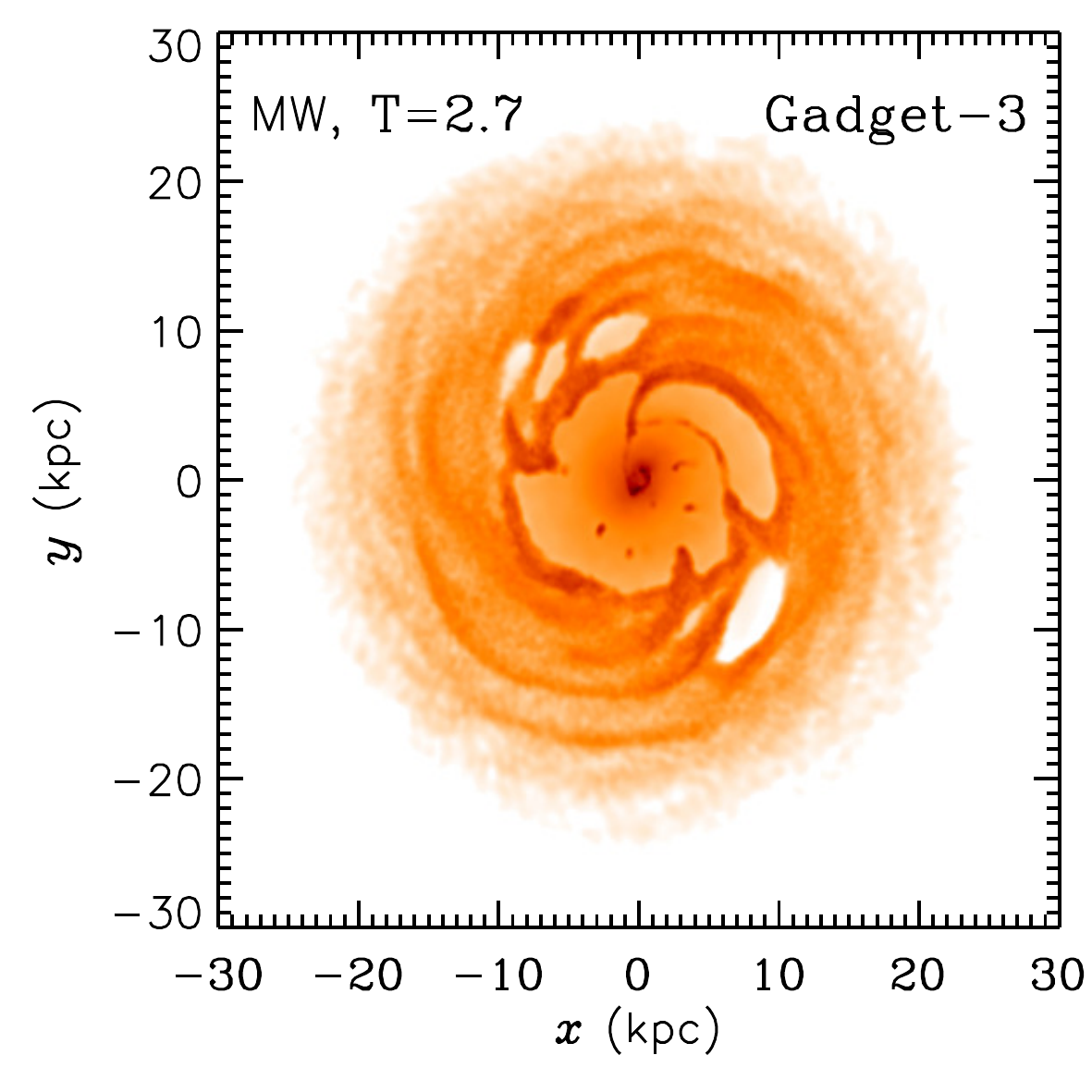}
    \includegraphics[width=0.5\columnwidth]{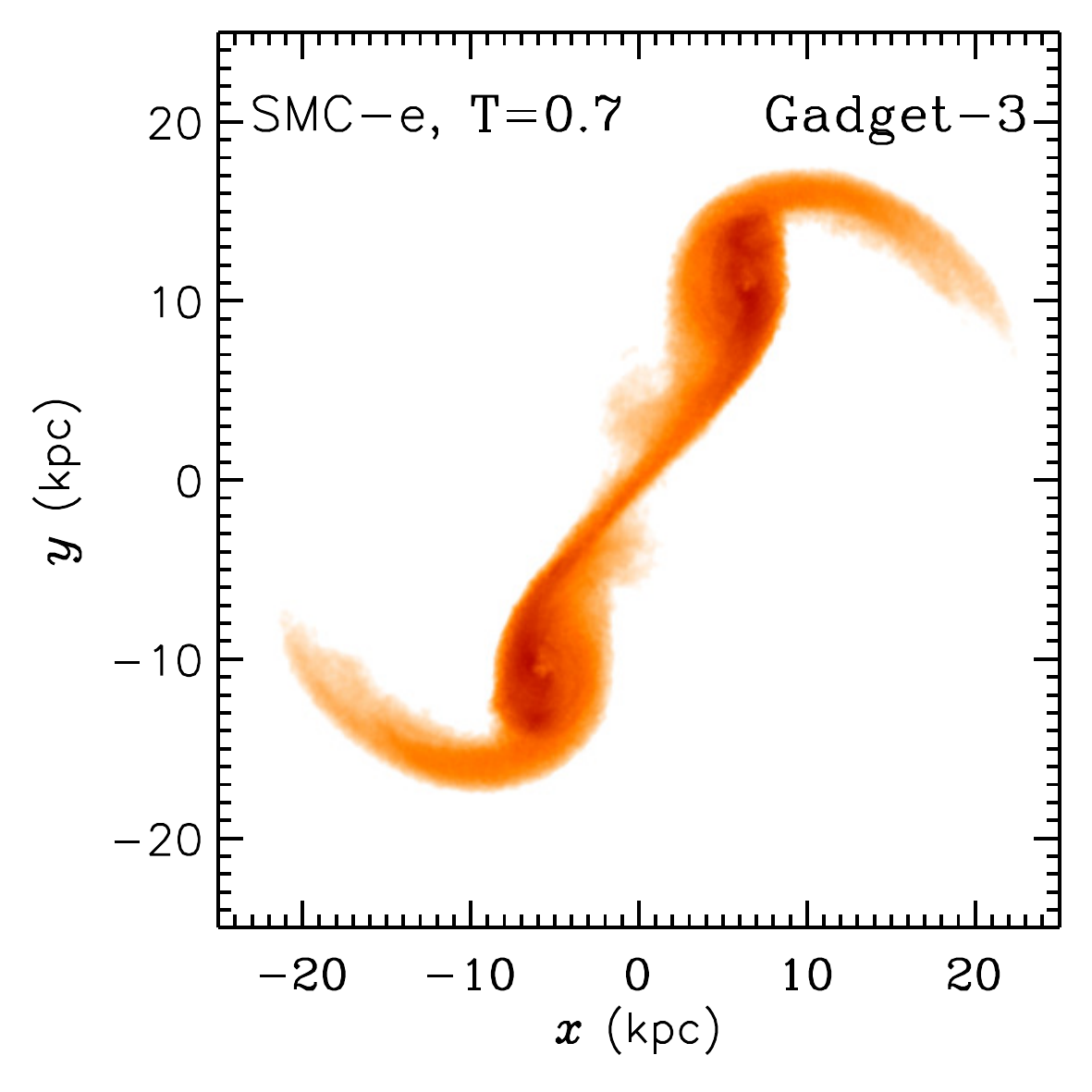}
    \includegraphics[width=0.5\columnwidth]{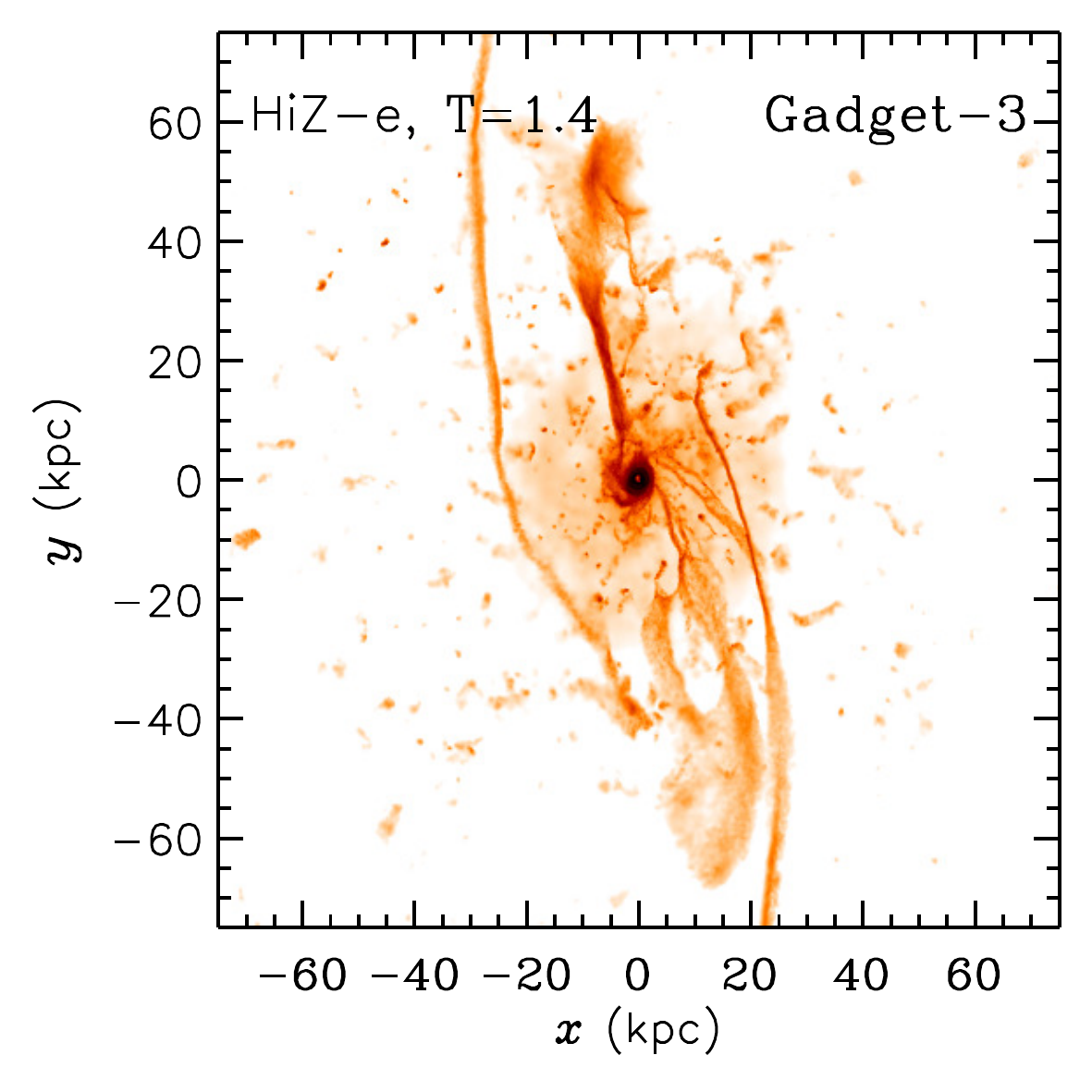}
    \includegraphics[width=0.5\columnwidth]{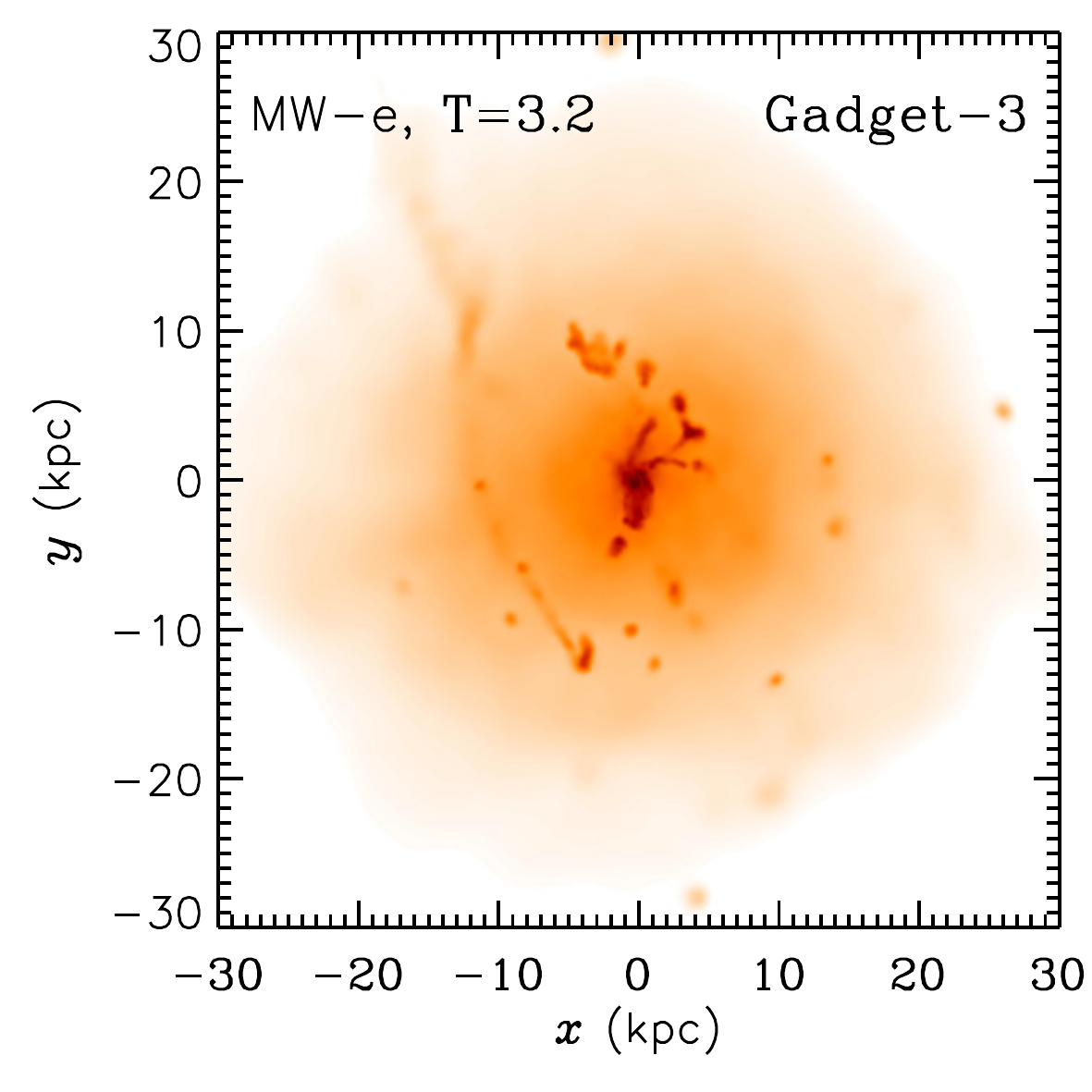}
  \caption{Example gas surface density plots for the following \arepo~(top) and \gadgetthree~(bottom) simulations that include BH accretion and AGN feedback:
  isolated \mw disc at $t = 2.7$ Gyr (first column), \smce merger at $t = 0.7$ Gyr (second column), \hize merger at $t = 1.4$ Gyr (third column), and
  \mwe merger at $t = 3.2$ Gyr (fourth column). As for the simulations without BH accretion and AGN feedback, in the starburst and post-starburst phases, the \gadgetthree~results
  feature more prominent hot haloes and spurious clumps.}
  \label{fig:gas_comparison}
\end{figure*}

Fig. \ref{fig:gas_comparison} shows example gas surface density plots
for the \arepo~(top row) and \gadgetthree~(bottom row) simulations
with BH accretion and AGN feedback. As for the simulations without BH
accretion and AGN feedback, the morphologies are qualitatively
similar, but the details differ.  Furthermore, as for the SFHs, the
code-dependent differences in the gas morphologies are more
significant when BH accretion and AGN feedback are included. The first
column shows the \mw isolated disc at $t = 2.7$ Gyr. At this time, a
bar is evident in the \arepo~simulation and much of the gas in the
central region has been consumed because of the bar instability. In
the \gadgetthree~simulation, a bar is not evident; rather, a large,
irregularly shaped cavity, which contains some small clumps of dense
gas, has formed. Part of the reason for the significant differences
between the \arepo~and \gadgetthree~results is that in the
\arepo~simulations, the BH can only consume gas that is within 50 pc,
whereas in \gadgetthree, the region from which gas is accreted grows
as the gas near the BH is depleted because a fixed number of
neighbours is used to calculate the SPH density estimate. Note, however,
that generically, holes tends to form around the BHs for the following
physical reason: in the
quiescent state that is eventually reached in isolation or at the end of a
merger (when the BH growth has effectively shut off),
a small bubble of hot, low-density gas around the BH is created by
the pressure that is sustained in our feedback model by the
residual accretion.

The second column shows the \smce merger simulation at $t =
0.7$ Gyr (after first pericentric passage). The results of both codes
exhibit extended, smooth tidal features, and the morphologies are
almost indistinguishable. This column exemplifies the good agreement
that is characteristic of the pre-starburst phase of the merger
simulations.

The third column of Fig. \ref{fig:gas_comparison} shows the
\hize merger near the peak of the starburst ($t = 1.4$
Gyr). Some of the filamentary structure is similar in the \arepo~and
\gadgetthree~simulations, but the detailed morphologies differ
significantly. As noted above, the \gadgetthree~result exhibits many
spurious clumps of gas that are not present in the \arepo~simulation.

Finally, the fourth column shows the \mwe merger at $t = 3.2$ Gyr,
$\sim 400$ Myr after the starburst. Note that for comparison, this is
the same simulation and time as shown in the fourth column of
Fig.~\ref{fig:NB_gas_comparison}, except that BH accretion and AGN
feedback are included here. For a given code, the gas morphologies are
similar to those of the simulations for which BH accretion and AGN
feedback were not included (Fig. \ref{fig:NB_gas_comparison}).  One
notable difference is that in the \gadgetthree~simulation, the gas
disc is less pronounced and the hot halo is more prominent. Once
again, the \gadgetthree~simulation features an extended halo of hot
gas and spurious clumps, both of which are not present in the
\arepo~simulation. Note that in this example, the code-dependent
differences in the gas morphologies are more significant than those
caused by the inclusion of AGN feedback. This result is a
counterexample to the conclusion of \citet{Scannapieco:2012} regarding
the effects of various star formation and stellar feedback models
compared with differences between codes and implies that it is highly
desirable to use the most accurate hydrodynamical solver possible.

\subsubsection{Gas phase structure}

\begin{figure*}
  \centering
    \includegraphics[width=\columnwidth]{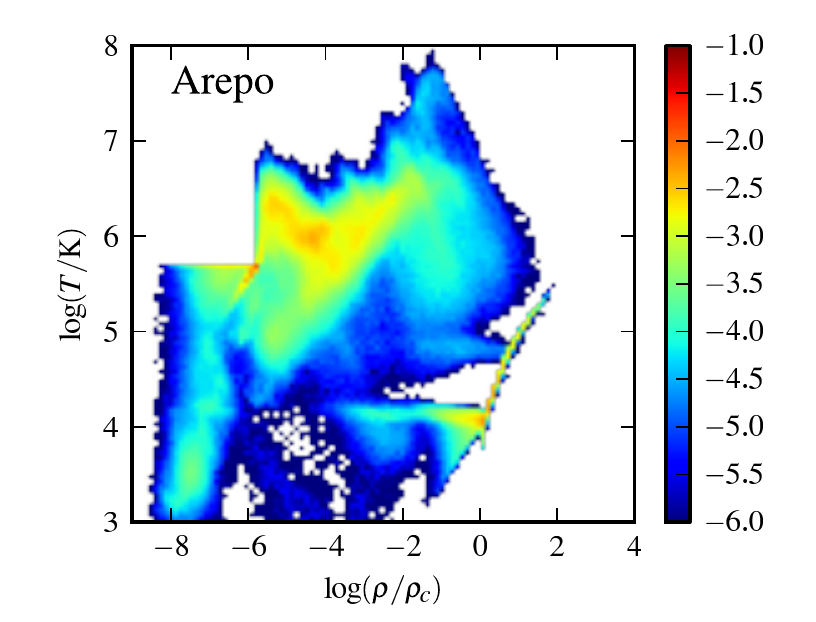}
    \includegraphics[width=\columnwidth]{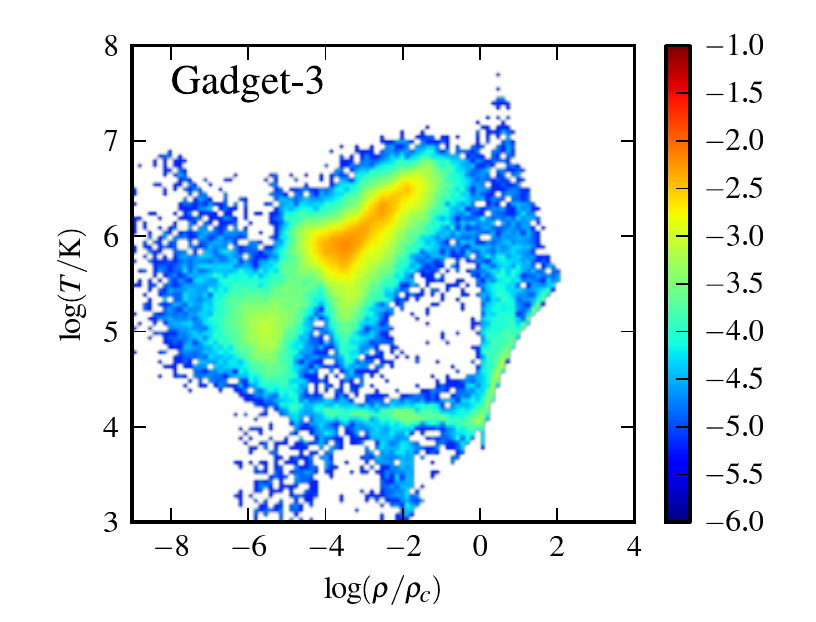}
  \caption{Similar to Fig. \ref{fig:NB_pd_comparison}, but for the \mwe simulations with BH accretion and AGN feedback at the same time as in Fig. \ref{fig:NB_pd_comparison}
  ($t = 3.2$ Gyr, $\sim 400$ Myr after the peak of the starburst and AGN activity). For both codes, AGN feedback results in more hot halo gas. As for the simulations without
  AGN feedback, the \gadgetthree~simulation features hot halo gas than the \arepo~simulation; in the latter, the hot halo gas cools more efficiently on to the EOS.}
  \label{fig:pd_comparison}
\end{figure*}

As for the simulations without BH accretion and AGN feedback, the gas
phase structure in the \gadgetthree~and \arepo~simulations is similar
in the pre-starburst phase but can differ significantly during and
after the starburst. Again, we only present one example to illustrate
the characteristic differences here, but the interested reader can
visit the aforementioned URL to examine the evolution of the gas phase
structure for all simulations.

Fig. \ref{fig:pd_comparison} shows gas phase diagrams for the
\mwe simulations with BH accretion and AGN feedback at the
same time as in Fig. \ref{fig:NB_pd_comparison} ($t = 3.2$ Gyr, $\sim
400$ Myr after the peak of the starburst and AGN activity). For a
given code, the inclusion of AGN feedback causes there to be more gas
in the hot halo and correspondingly less gas on the EOS (the thin line
in the lower-right corner). As for the simulations that did not
include AGN feedback, the \gadgetthree~simulation features more hot
halo gas than the \arepo~simulation, in which the hot halo gas cools
more efficiently.

\subsection{Tests of the BH accretion and AGN feedback models} \label{S:tests}

Here, we present various tests that demonstrate the effects of the different treatments of BH accretion and AGN feedback discussed above.
We use the \smc analogue as one test case. Whereas for this simulation,
the differences in the results are small in an absolute sense, they are
systematic and can have more significant effects in other simulations.
We chose to use the \smc analogue as a test case because
the SFH and BH growth are comparatively simple; thus, differences can be more easily understood. Furthermore, the BH
grows very little ($M_{\mathrm{BH}} \sim 10^5 \msun$, the seed mass, throughout the simulation) and should have a negligible effect on the
SFH of the galaxy. Thus, the `no BH' case can be used as the baseline with which to compare the other runs; ideally, the SFHs should
be the same for the BH and `no BH' cases. We also investigate the effects
of different BH accretion and AGN feedback for the \mwe merger simulation. In this significantly more complicated case, the interpretation
of the comparisons among the different treatments of BH accretion and AGN feedback is less straightforward, but many of the effects
observed for the \smc case are also observed here. The tests presented here justify our fiducial choices for the BH accretion
and feedback implementations and demonstrate some important numerical effects that are not always appreciated in the literature.

\subsubsection{Refinement near the BH} \label{S:refinement_test}

\begin{figure}
  \centering
    \includegraphics[width=\columnwidth]{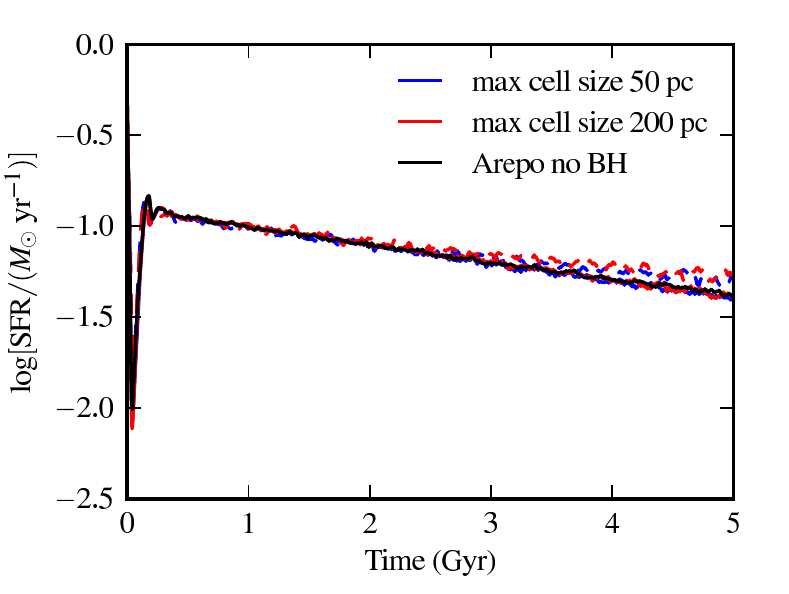} \\
    \includegraphics[width=\columnwidth]{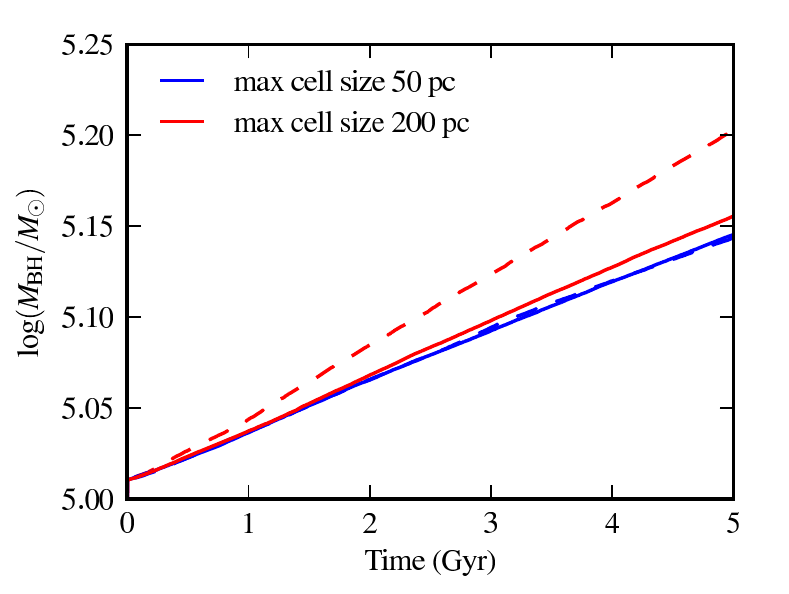}
  \caption{Demonstration of the effect of limiting the cell size near the BH for the isolated \smc disc simulation. The top (bottom) panel shows
  the SFH (BH mass versus time). The blue (red) lines show the results when cells that are less than 500 pc from the BH are forced to have size
  less than 50 (200) pc. The higher-resolution simulation with BH accretion and AGN feedback disabled is shown in black. When the maximum
  cell size near the BH is 50 pc, the agreement between the two resolutions is excellent.}
  \label{fig:refinement_test}
\end{figure}

\begin{figure}
  \centering
    \includegraphics[width=\columnwidth]{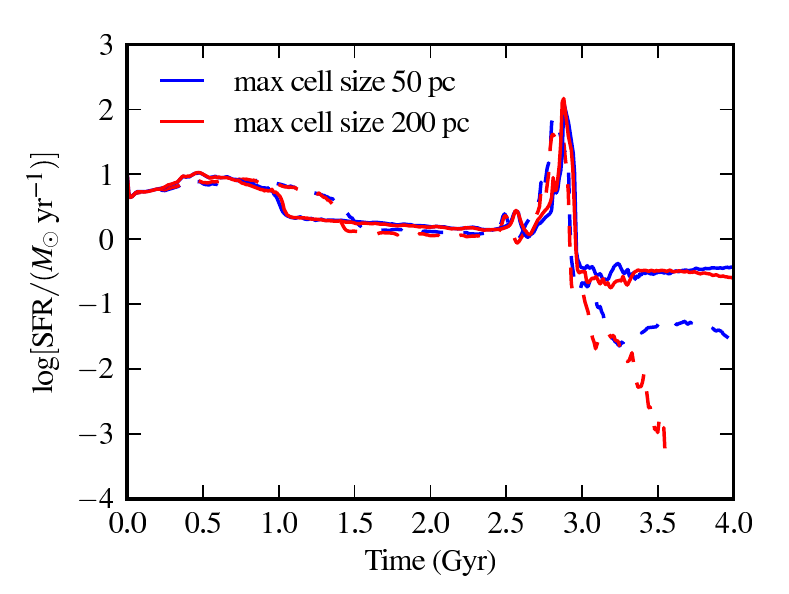} \\
    \includegraphics[width=\columnwidth]{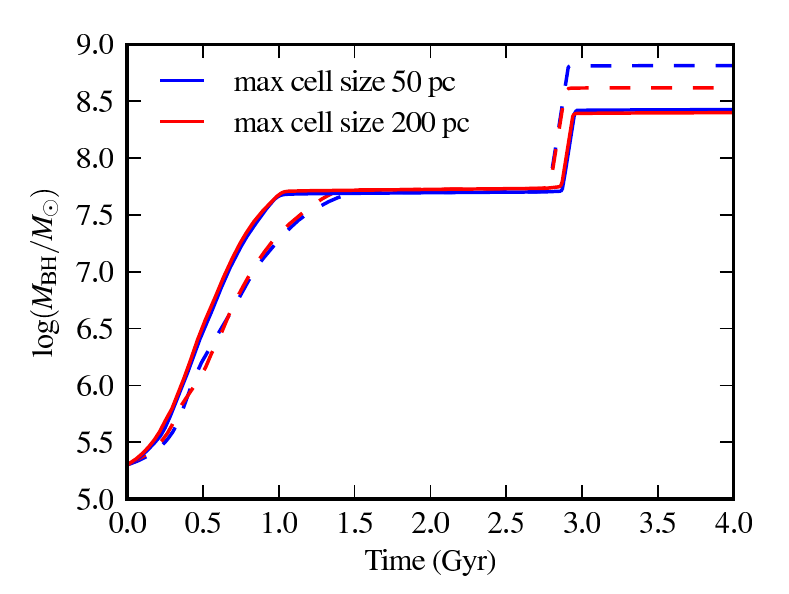}
  \caption{Similar to Fig. \ref{fig:refinement_test}, but for the \mwe merger. The results
  are almost independent of the $\rmax$ value for the higher resolution because the
  resolution around the BH is sufficiently high without the forced refinement. The final BH
  masses for the two resolutions agree better when $\rmax = 200$ pc, but this agreement
  does not indicate better convergence. See the text for details.}
  \label{fig:refinement_test_merger}
\end{figure}

In SPH, as the gas density in the immediate vicinity of a BH decreases
because of gas consumption and expulsion, the radius over which the
density is calculated increases because the number of neighbours used
for the density estimate and the particle masses are fixed.  Thus, in
some situations, the BH accretion rate can be overestimated and gas
fuels the BH from unphysically large scales; this is especially
problematic in cosmological simulations, for which the resolution is
often not better than a kiloparsec. In \arepo, the standard cell refinement
scheme attempts to keep cell masses comparable; thus, a similar effect,
in which cells near the BH grow large, can occur. However, unlike in
SPH, we can overcome this potential problem by preventing cells near
the BH from becoming too large and using the gas density of the cell
that contains the BH to calculate the accretion rate. Consequently, if
the gas density in the vicinity of the BH decreases, the accretion
rate decreases concomitantly.

Fig. \ref{fig:refinement_test} demonstrates the effects of limiting
the maximum cell size near the BH particle for the \smc
isolated disc case. We force cells located
within 500 pc of the BH to be refined if their size is greater than
some value $\rmax$. We show the results for $\rmax = 50$
and 200 pc in Fig. \ref{fig:refinement_test}. The top panel indicates
that the choice of $\rmax$ has no effect on the SFH, and the slight
difference between the different resolutions at late times is
independent of $\rmax$.

However, the growth of the BH differs systematically. For both
resolutions, the BH grows more when $\rmax = 200$ pc because of the
effect described above, and the consequences are more severe for the
lower-resolution simulation. When $\rmax = 50$ pc, the final BH mass
is slightly less and the two different resolutions agree perfectly.

Fig.~\ref{fig:refinement_test_merger} shows the results of changing
$\rmax$ for the \mwe merger simulation. In this case, the SFH is better converged
when a maximum cell when $\rmax = 50$ pc. For the higher-resolution simulations,
the BH growth history is unaffected by the choice of $\rmax$. Thus, for this resolution
($\epsilon_{\mathrm{bar}} = 60$ pc), the refinement near the BH is sufficiently high even without
forcing refinement. Interestingly, the final BH mass of the lower-resolution simulation
with $\rmax = 200$ pc agrees better with that of the higher-resolution simulations
than when $\rmax = 50$ pc. In this case, allowing larger cells near the BH by setting
$\rmax = 200$ pc serves to mitigate some of the resolution dependence: in the
lower-resolution simulation with $\rmax = 50$ pc, during the starburst at final coalescence,
the central gas density is greater than in the higher-resolution runs. Consequently, the BH
grows more rapidly during that time. Using $\rmax = 200$ pc results in a decreased
central gas density and thus less-massive BH.
However, this result should not be taken as an indication
that the BH mass is better-converged when $\rmax = 200$ pc:
if the BH mass were truly converged,
the $\rmax = 50$ pc simulation should agree at least as well because in this simulation,
the number of resolution elements is greater than or equal to that of the
lower-resolution simulation.

\subsubsection{BH accretion rate calculation} \label{S:accretion_test}

\begin{figure*}
  \centering
    \includegraphics[width=\columnwidth]{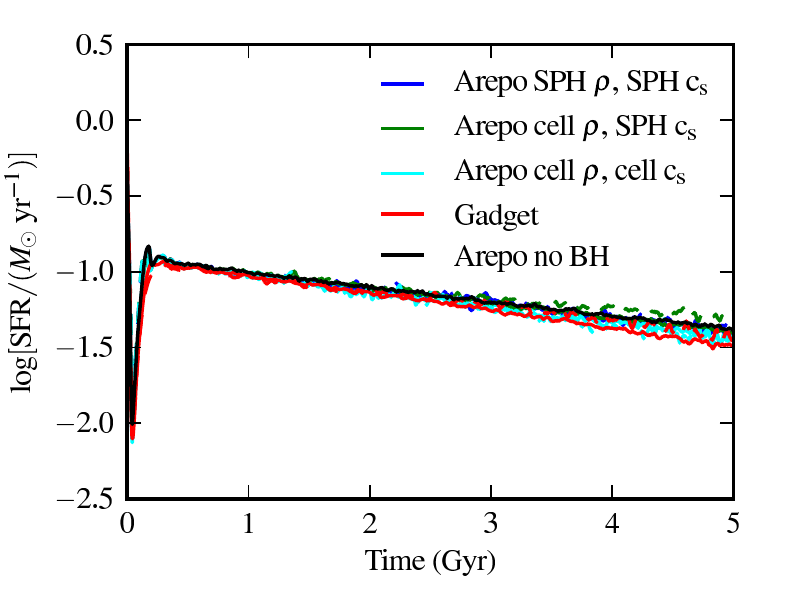}
    \includegraphics[width=\columnwidth]{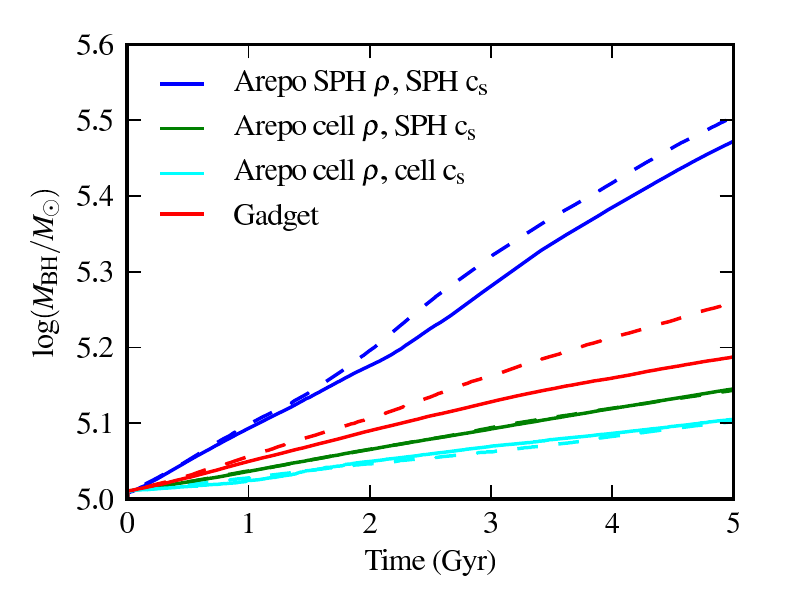} \\
    \includegraphics[width=\columnwidth]{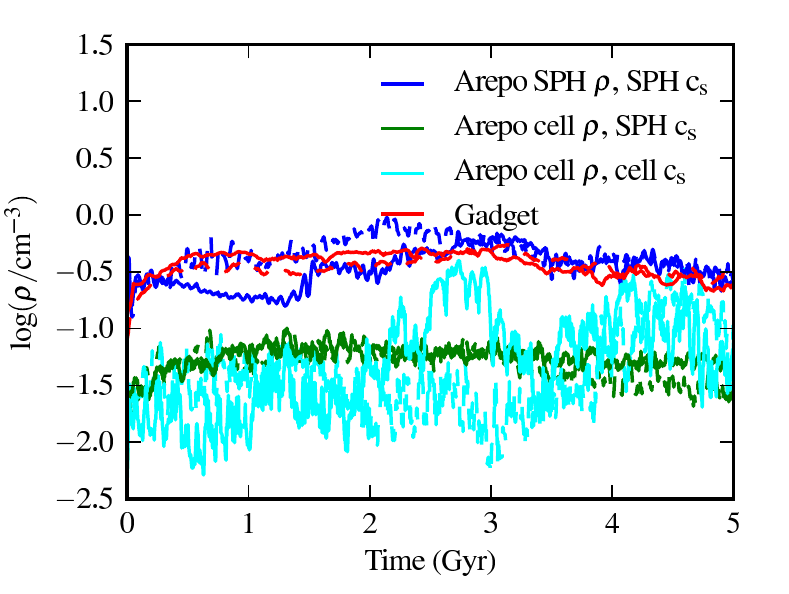}
    \includegraphics[width=\columnwidth]{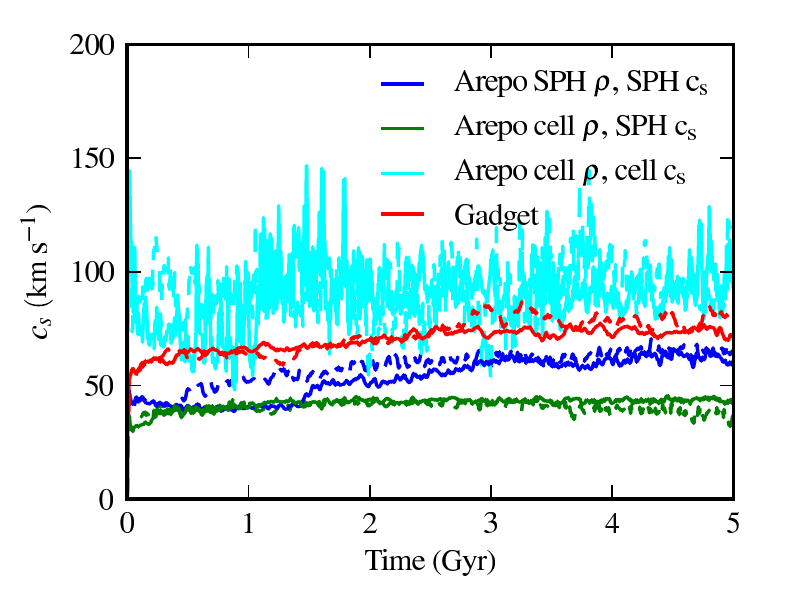} \\
  \caption{Comparison of different methods for calculating the BH accretion rate in \arepo. Either the cell value or the SPH density estimate
  of both the gas density and sound speed near the BH can be used. Different combinations are shown (see the legend). For comparison,
  the \arepo~simulation with no BH and the \gadgetthree~simulation with BH accretion and AGN feedback are also shown.
  Solid (dashed) lines indicate higher- (lower-) resolution simulations.
  The panels show the following quantities versus time: the SFR (upper left), the BH mass (upper right), and the gas density (lower left) and
  sound speed (lower right) estimates used to calculate the accretion rate (the last two quantities have been smoothed with a median filter of size 1001 time steps
  for plotting purposes). Use of the cell density and SPH estimate for the sound speed yields the
  best convergence, and the convergence is also good when the cell sound speed is used. Still, we have opted to use the SPH estimate for the
  sound speed in our fiducial model because unlike the cell density, the cell sound speed is very noisy.}
  \label{fig:accretion_test}
\end{figure*}

\begin{figure*}
  \centering
    \includegraphics[width=\columnwidth]{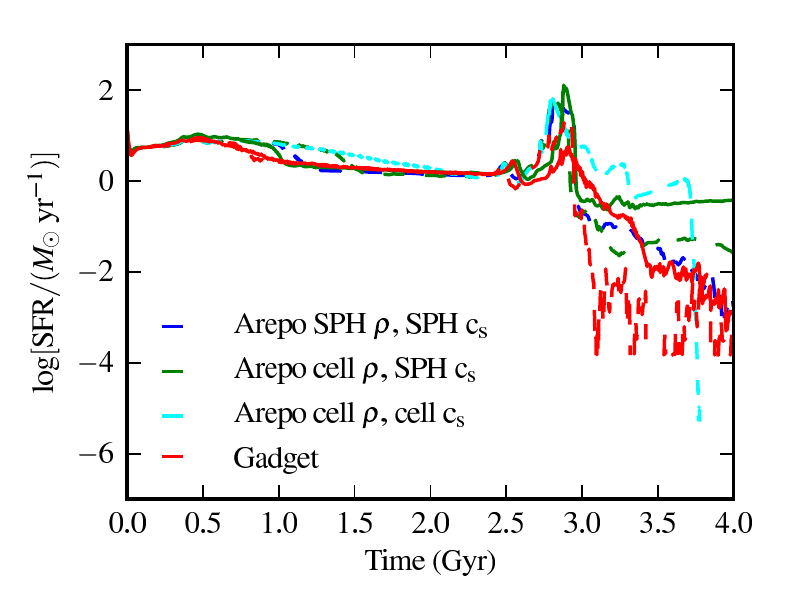}
    \includegraphics[width=\columnwidth]{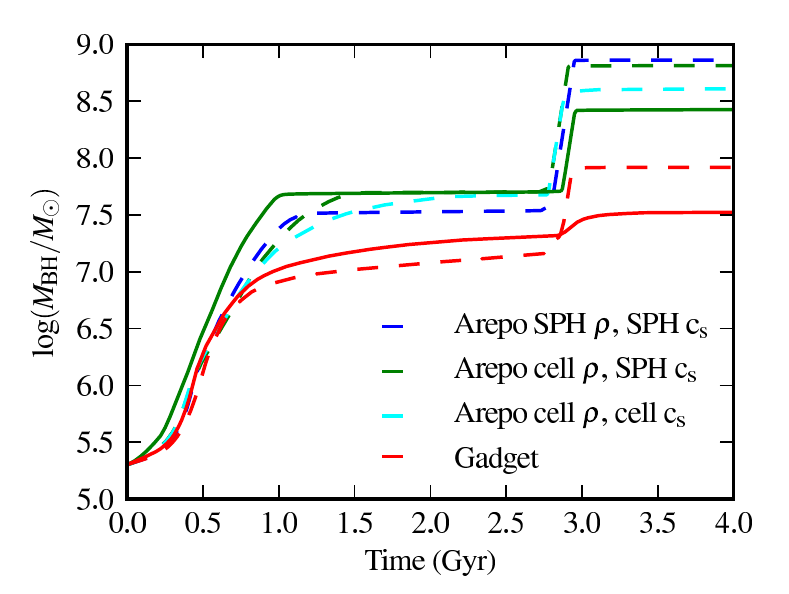} \\
    \includegraphics[width=\columnwidth]{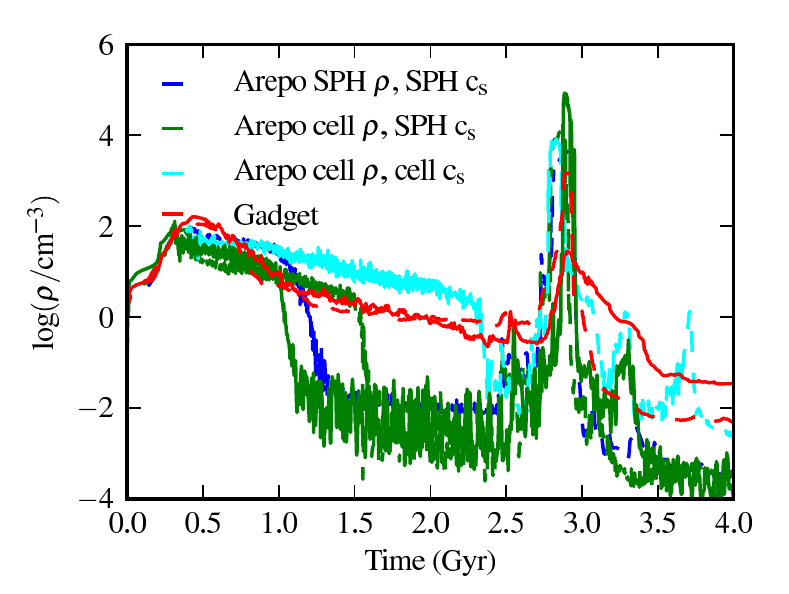}
    \includegraphics[width=\columnwidth]{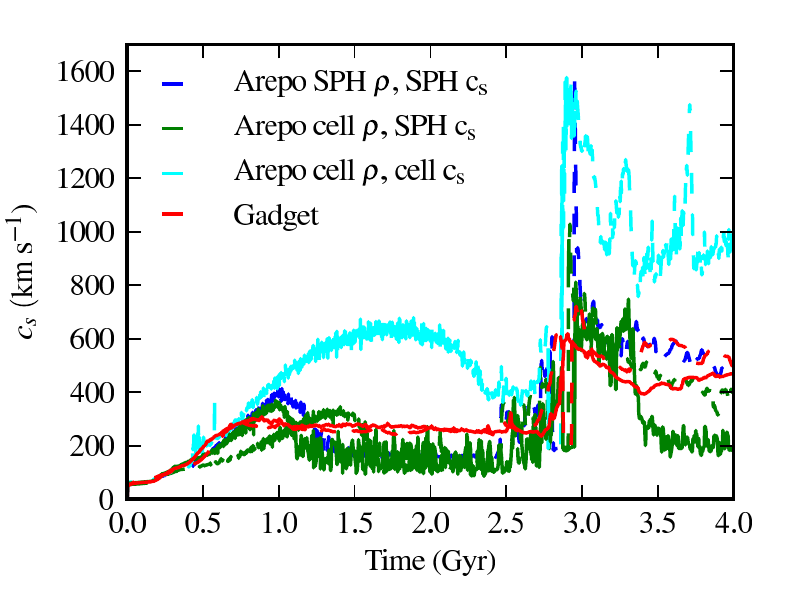} \\
  \caption{Similar to Fig. \ref{fig:accretion_test}, but for the \mwe merger. As for the \smc isolated disc, use of the cell sound speed to calculate
  the BH accretion rate results in a smaller final BH mass because the cell sound speed can be more than a factor of three greater than the
  SPH estimate.}
  \label{fig:accretion_test_merger}
\end{figure*}

To calculate the BH accretion rate using equation~(\ref{eq:bondi}), we require the gas density and sound speed near the BH. In \arepo, both quantities
can be estimated in a manner analogous to that in SPH, i.e., by averaging over some number of nearest neighbour cells (typically, 32). However,
it is also possible to use the gas density and sound speed for the cell in which the BH is located. In principle, the cell values should be more
representative of the conditions near the BH and thus ensure a more physical calculation of the accretion rate, but they may be too noisy to
be useful. We will explore the effects of these choices now.

Fig. \ref{fig:accretion_test} demonstrates the variations that arise from
using different methods to calculate the BH accretion rate in \arepo
for the \smc isolated disc case.
The \gadgetthree~results are shown for comparison. In all cases, the
mass over which the feedback energy is distributed is kept constant,
i.e., the number of cells or particles over which the energy is
distributed is increased as the cell mass is decreased. Furthermore,
cells within 500 pc of the BH are forced to have size less than $\rmax
= 50$ pc.

The differences in the SFH (upper-left panel) for the various
treatments are small and comparable with the resolution-dependent differences.
The BH growth history, in contrast, can be
affected significantly; the upper-right panel indicates that the final BH
mass can differ by as much as 0.4 dex, and the effect in merger
simulations can be even larger. The BH grows most in \arepo~when the
SPH density and sound speed are used (the blue curve).  The reason is
that, as shown in the lower-left panel, in \arepo, the SPH density
estimate is systematically greater than the cell density (and similar
to the \gadgetthree~density). The \gadgetthree~simulations (red curve)
exhibit the next highest BH growth; the reason that they feature less growth
compared with the analogous \arepo~simulations is that the SPH sound
speed estimate (lower-right panel) is systematically higher in
\gadgetthree than \arepo.

When the cell density is used, the BH grows less because the cell density is systematically lower than the SPH density.
The growth is most suppressed when the cell sound speed is used (cyan curve) because the cell sound speed is systematically
higher than the SPH sound speed. Furthermore, the cell sound speed is very noisy, with variations of $\ga 50$ km$\ps$ from the mean value,
and the cell density is significantly more noisy when the cell sound speed is used to calculate the BH accretion rate (compare the cyan and green
curves in the lower-left panel of Fig. \ref{fig:accretion_test}).

Fig.~\ref{fig:accretion_test_merger} compares the different BH accretion rate
treatments for the \mwe merger simulation. As for the \smc isolated disc case, the BH grows
most when the SPH estimates for the gas density and sound speed are used to calculate
the accretion rate. Using the cell density rather than the SPH estimate results in a very similar
final BH mass. Furthermore, using both the cell density and sound speed results in the smallest final BH
mass of the three lower-resolution simulations because the cell sound speed can be more than a factor
of three greater than the SPH estimate. Although the differences in the BH masses caused by varying
the sub-resolution model accretion rate calculation are significant ($\sim 0.3$ dex), they are less
than the differences between the two resolutions for a fixed code and those between the two codes
for a fixed number of particles/cells.

Based on these results, we chose to use the cell density for our
production runs because this is more representative of the density
near the BH than is the SPH density, yet the two have comparable
amounts of noise. However, we chose to use the SPH estimate for the
sound speed because the cell value is too noisy.

\subsubsection{AGN feedback} \label{S:FB_test}

\begin{figure*}
  \centering
    \includegraphics[width=\columnwidth]{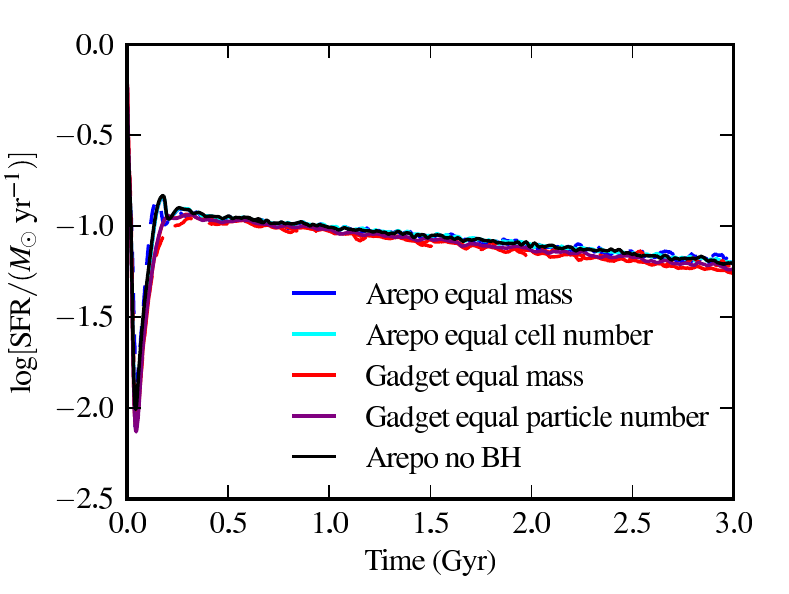}
    \includegraphics[width=\columnwidth]{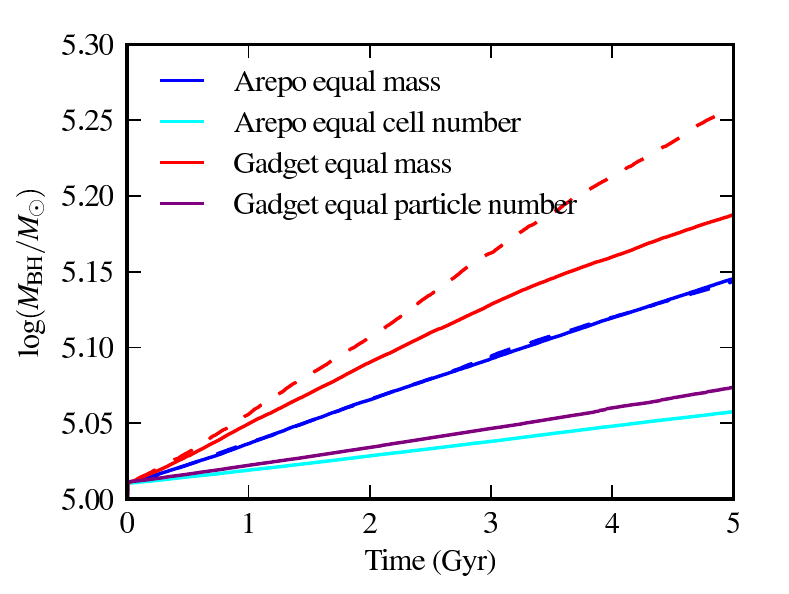} \\
    \includegraphics[width=\columnwidth]{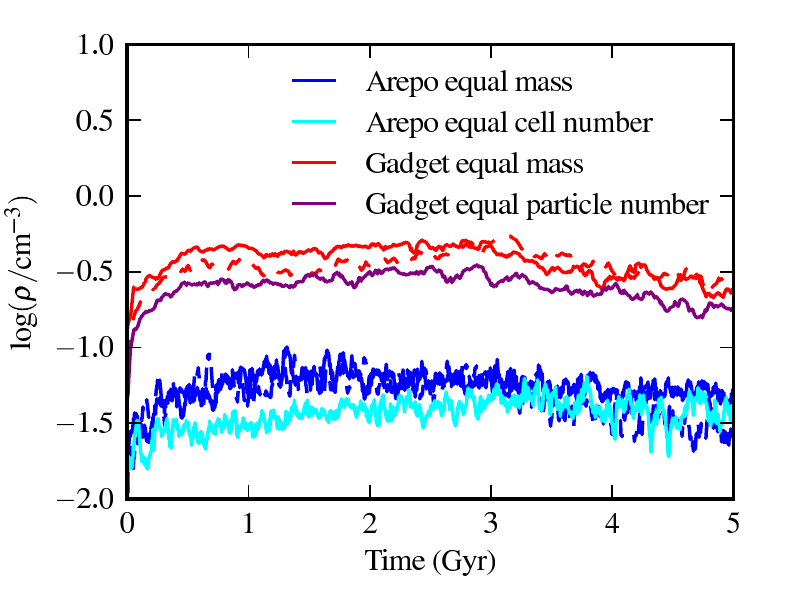}
    \includegraphics[width=\columnwidth]{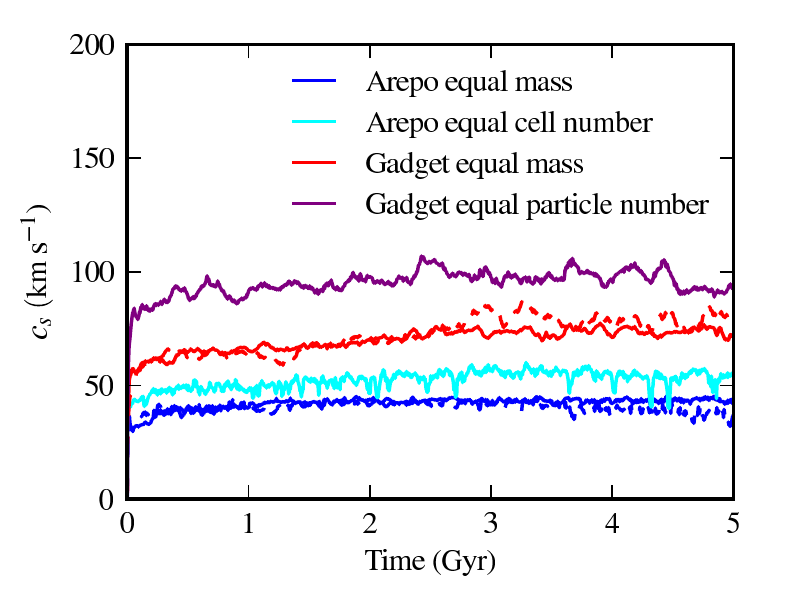} \\
  \caption{Similar to Fig. \ref{fig:accretion_test}, but for different methods of distributing the BH feedback energy. In both \gadgetthree~and \arepo, keeping
  the mass -- rather than the number of particles or cells -- over which the AGN feedback energy is distributed fixed with resolution yields the best convergence.
  Thus, we have used this as our standard method. Note that the convergence of the BH masses in the \arepo~simulations is excellent when this treatment
  is used.}
  \label{fig:FB_test}
\end{figure*}

\begin{figure*}
  \centering
    \includegraphics[width=\columnwidth]{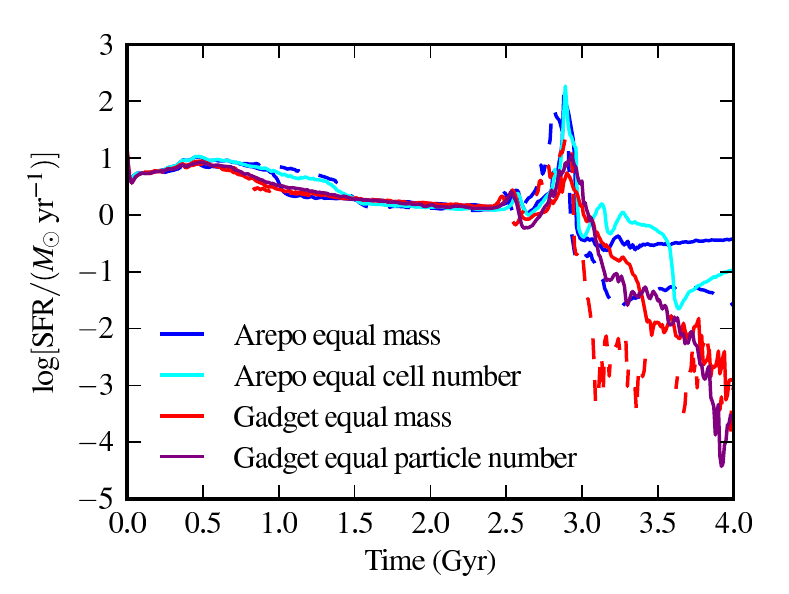}
    \includegraphics[width=\columnwidth]{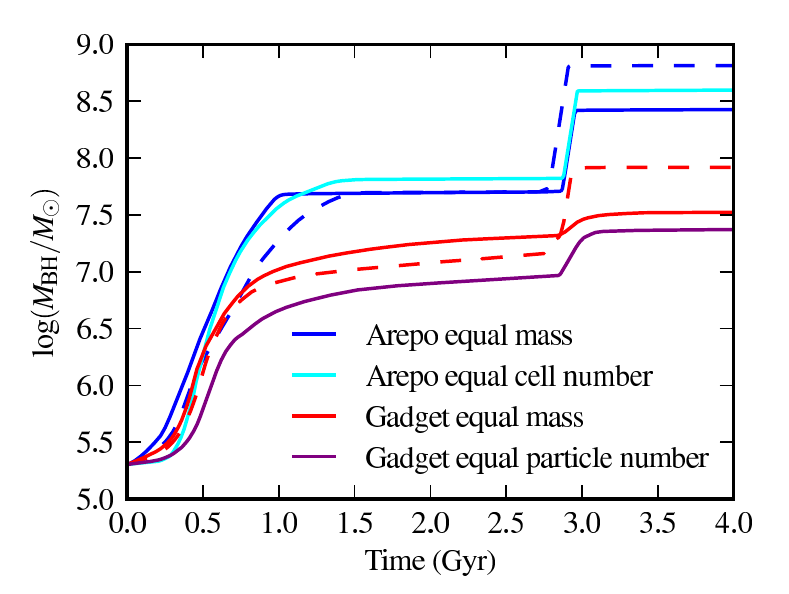} \\
    \includegraphics[width=\columnwidth]{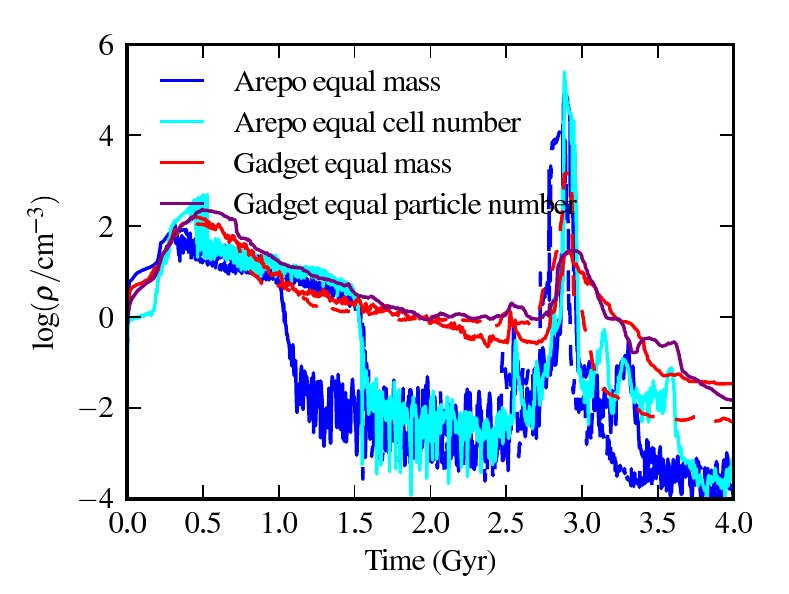}
    \includegraphics[width=\columnwidth]{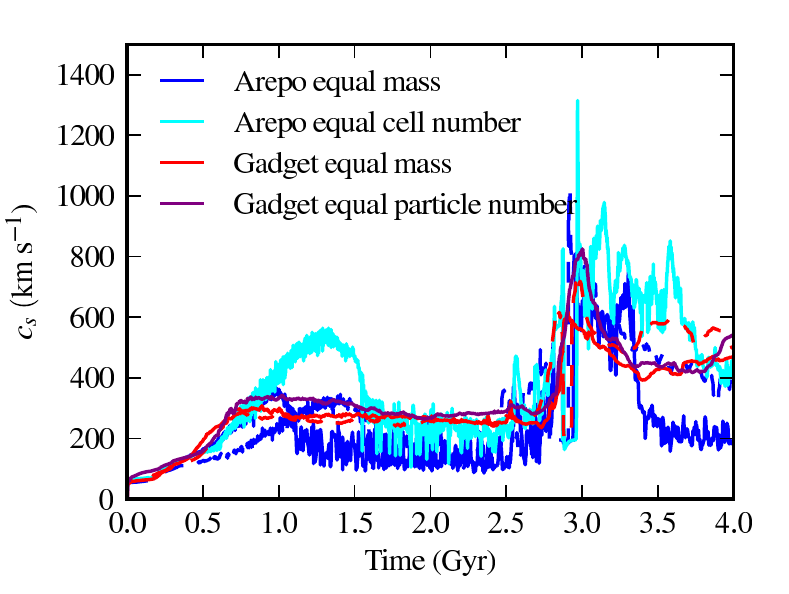} \\
  \caption{Similar to Fig. \ref{fig:FB_test}, but for the \mwe merger. In this case, keeping the mass over which the feedback energy is distributed
  constant yields a better-converged final BH mass for \gadgetthree but not \arepo.}
  \label{fig:FB_test_merger}
\end{figure*}

The manner in which the AGN feedback energy is distributed can also
affect the BH growth. As explained in Section \ref{S:BH_FB_method},
our preferred method is to keep the mass over which the feedback
energy is distributed (the `feedback mass') constant. Thus, for
higher-resolution runs, in which the particle/cell mass is lower,
the number of particles/cells over which the feedback energy is
distributed should be increased. However, this is not the only
possible approach; an alternative approach is to keep the number of
particles/cells over which the feedback energy is distributed fixed.
But, as we will show now, this causes a
stronger resolution dependence, which is undesirable for a
sub-resolution model.

Fig. \ref{fig:FB_test} demonstrates the differences between these two
choices for the \smc isolated disc case.
As in the previous test, there are minor variations in the
SFHs at late times, but the magnitude of the difference is similar to
that caused by varying the resolution. However, varying the method for distributing
the feedback energy causes significant and
systematic changes in the BH growth history. First, note that the
dashed and solid blue lines, which correspond to the lower-resolution
\arepo~simulation and the higher-resolution simulation in which the
feedback mass is kept constant, agree perfectly. The corresponding
curves for the density and sound speed are (necessarily) in good
agreement. In contrast, the \arepo~run in which the number of cells
over which the feedback is distributed is kept constant (the cyan
line) has systematically less BH growth. The reason for this
discrepancy is that the latter simulation has systematically lower gas
density (bottom left) and higher sound speed (bottom right) because
the AGN feedback heats a smaller mass of gas compared with the
lower-resolution run. Consequently, the steady state that is reached is at a
higher temperature and lower density. The analogous effect can be
observed for \gadgetthree, but in this case, there is still some
resolution dependence even when the feedback mass is kept constant.

Fig. \ref{fig:FB_test_merger} shows a comparison of the different
methods for distributing the feedback energy for the \mwe merger.
For this model, the SFH differs most significantly in the \arepo run
in which an equal cell number is used; in this case, star formation
is quenched more effectively by the AGN after the starburst
than in any of the other simulations, and the SFH in this regime
differs most significantly from that of the lower-resolution run.

For the \gadgetthree simulations, the final BH mass of the lower-resolution
run agrees slightly better with that of the higher-resolution run in
which the mass over which the feedback energy is distributed is
kept constant, as was observed for the \smc case. However, for the
\arepo simulations, the final BH mass in the higher-resolution run
in which the number of cells over which the feedback energy is
distributed is kept constant agrees better with the final BH mass
of the lower-resolution simulation. The reason for this behaviour
is that in the equal-mass run, the initial stage of rapid BH growth
($t \la1$ Gyr) is terminated slightly earlier and at a slightly lower
BH mass than in the equal-cell-number run, primarily because
the rapid decline in the gas density occurs earlier in the equal-mass
run. Consequently, during the final-coalescence phase, in which the
BH undergoes Eddington-limited accretion, the less-massive BH grows less.
This unexpected behaviour demonstrates the difficulty of predicting
the effects of the sub-resolution model in the significantly more complex
context of galaxy mergers and highlights the aforementioned conclusion
that the final BH masses should only be considered robust to within
a factor of a few. Furthermore, the difference in the final BH masses
of the two higher-resolution simulations is less significant than the
difference between either of these masses and that of the lower-resolution
run.

The purpose of the sub-resolution model is to encapsulate
in a simple manner physics that is not included in the simulations. The implicit assumption in our model is that the feedback
energy deposited by AGN is thermalised over some physical scale, and this physical scale does not and should not depend on the resolution
of our simulations. Thus, in our production runs, we kept the feedback mass constant.

\section{Discussion} \label{S:discussion}

In earlier work \citep{Springel:2010arepo, Bauer:2012,
  Vogelsberger:2012, Keres:2012, Sijacki:2012, Torrey:2012disks,
  Nelson:2013}, we have compared results from \gadgetthree and \arepo
and, in some cases, identified significant differences that we
attributed to numerical issues with the conventional formulation of
SPH.  In the present paper and in studies of the Ly-$\alpha$ forest
\citep[e.g.,][]{Regan:2007, Bird:2013}, it has been found that SPH can
produce results in certain regimes that agree well with grid-based
codes.  To understand this situation, we first briefly review the
primary limitations of the SPH approach that have become clear in
recent work.  This then allows us to discuss why SPH can be expected
to work reasonably well in some applications but not in
others. Finally, we comment on what this implies for the numerical
robustness of different types of previous work.

\subsection{Limitations of traditional SPH}
\label{S:limitations_of_SPH}

The traditional formulation of SPH used in this work has been -- and
still is -- being used for many astrophysical simulations.  It is thus
important to understand which of these results may be influenced by
numerical artefacts of the type discussed below. Whereas we
acknowledge that there have been impressive efforts to address
at least some of these issues \citep[e.g.,][]{Monaghan1997,
  Ritchie2001, Price:2008, Wadsley:2008, Read:2010,
  Abel:2011, Read:2012, Garcia-Senz:2012, Saitoh2013, Hopkins:2013SPH}, we feel
that there remain many lingering misconceptions about SPH that muddle
the interpretation of the reliability of simulations performed even
with updated versions of this algorithm.

\subsubsection{Noise in local estimates and  convergence to the continuum solution}
\label{S:local_noise}

The transition from the continuum equations of fluid dynamics to the
discrete form used by SPH involves a two-step procedure
\citep[e.g.,][]{Monaghan1992}.  First, the exact fluid quantities are
replaced by smoothed versions via a convolution with the smoothing
kernel.  Second, the integral forms of these convolutions are replaced
by discrete sums over the SPH particles such that they can be
evaluated numerically. The error made in the discretization step depends
not on the number of SPH particles, $N_{\rm SPH}$, but instead on
the number of neighbours in the discrete sums, $N_{\rm ngb}$.  If, as
is typically done, $N_{\rm ngb}$ is held fixed as $N_{\rm SPH}$ is
increased, there will be a {\it constant} source of error in the local
estimates even as the resolution of the simulation is nominally increased
\citep{Belytschko1998,Robinson2012}.

Consequently, local fluid estimates are often noisy and are not guaranteed to
approach their continuum values as $N_{\rm SPH}$ is increased.  This source of
noise, although small, may have a significant impact on flows in which
the energy content is dominated by internal energy rather than kinetic
or gravitational energy. The noise is particularly strong in gradients
of interpolated quantities, most notably in the pressure force
\citep{Read:2010}. Furthermore, if $N_{\rm SPH}$ is increased without
simultaneously increasing $N_{\rm ngb}$, the solution may asymptote
to a fixed result that is different from the true solution because of this constant
source of error \citep{Robinson2012}.

\subsubsection{Inaccurate treatment of fluid instabilities}
\label{S:inaccurate_instabilities}

Tests by \citet{Agertz:2007} demonstrated that conventional
implementations of SPH do not accurately describe jumps in physical
quantities across contact discontinuities because the pressure
effectively becomes multi-valued at the interface, thereby resulting
in artificial repulsive forces that act as a macroscopic surface
tension.  If two fluid phases in pressure equilibrium shear relative
to one another, the spurious surface tension inhibits the proper
growth of Kelvin-Helmholtz instabilities and the two phases will not
mix together correctly.  Instead, the colder, more dense phase can
fragment into clumps that retain their identity because of the
presence of the spurious surface tension
\citep[e.g.,][]{Kaufmann2007,Hobbs:2013}.

Various new formulations of SPH have shown promise in alleviating this
problem \citep[e.g.,][]{Price:2008, Read:2010, Read:2012, Saitoh2013,
  Hopkins:2013SPH}; thus, we will not dwell on it here.  However, we
mention it for completeness and because nearly all previous studies of
galaxy mergers using SPH have employed traditional formulations of SPH
that are susceptible to this surface tension issue.

\subsubsection{Treatment of small-scale mixing}
\label{S:inaccurate_continuity}

In most implementations of SPH, the mass continuity equation is not
integrated in detail; instead, estimates of the fluid density at any
given simulation time are made using kernel interpolation
\citep[e.g.,][]{Monaghan1992} applied to the current particle
positions.  Partly for this reason, SPH is often referred to as a
`Lagrangian' method.  However, SPH is only `pseudo-Lagrangian'
because the particle mass is fixed in time and particle shapes are
not allowed to become distorted arbitrarily by the flow (see
\citealt{Vogelsberger:2012} for a detailed discussion).
Consequently, SPH cannot properly describe fluid mixing on small scales,
whereas in \arepo, mixing is not suppressed because
the implied mass exchange between cells is computed
correctly according to the continuity equation.

\subsubsection{Shock capturing and spurious viscosity}
\label{S:shock_capturing}

In essentially all widely used SPH codes, shocks are captured using
some form of artificial viscosity.  Ideally, this viscosity should
operate only in and near shocks because it can have unintended
consequences for the properties of the flow in other regions if it is
not accurately controlled; for example, it can cause spurious cooling
in cosmological applications \citep[e.g.,][]{Hutchings2000}.
Moreover, the action of the viscosity within shocks is to locally broaden
them over several smoothing lengths, thereby degrading the effective spatial
resolution in shocked regions. 

Many grid-based codes, including \arepo, instead treat shocks by
solving the Riemann problem across all cell-cell interfaces.  This
treatment has several advantages because it implies that such codes
minimize the additional source of unphysical diffusion that would
arise from artificial viscosity and because shocks can be spatially
resolved more precisely than in SPH.

\subsection{Why SPH works reasonably well for some applications but not others} 
\label{S:reasons_for_agreement}

From the discussion in Section \ref{S:limitations_of_SPH}, we can now
provide arguments as to why SPH yields results that are reliable in
some situations and why it fails in others.  For definiteness, we consider
four applications: (1) idealized tests of driven supersonic and
subsonic turbulence, (2) the intergalactic medium (IGM),  (3) gas
accretion on to galaxies, and (4) starbursts and AGN activity in
galaxy mergers, as described in this paper.

\subsubsection{Driven turbulence}

\cite{Bauer:2012} compared \gadgetthree and \arepo for idealized
simulations of isothermal turbulence in periodic boxes subject to
large-scale forcing.  Their tests demonstrate that the traditional
formulation of SPH, as incorporated in \gadgetthree, does not yield a
proper cascade of energy to small scales when the turbulence is
subsonic, as illustrated in their fig. 4.  However, otherwise
identical simulations performed with \arepo and the Navier-Stokes
version of \arepo \citep{Munoz2013} produced a well-developed
turbulence spectrum, not only in the inertial range but also through
the dissipational range. In contrast, SPH was found to perform significantly more reliably for
supersonic turbulence \citep{Price2010, Bauer:2012}, yielding results
that are in reasonable agreement with \arepo independent of the motion
of the mesh.

These apparently contradictory conclusions can be
readily explained by the limitations discussed above in Section
\ref{S:local_noise} and Section \ref{S:shock_capturing}.  In the
supersonic limit, the energy density of the fluid is dominated by
kinetic energy.  Although noise is still present in the local fluid
quantities (see Section \ref{S:local_noise}), its influence is
subdominant in this regime.  Similarly, spurious entropy generation
from the artificial viscosity (see Section \ref{S:shock_capturing}) is
also a minor source of error. In contrast, for subsonic turbulence, the
internal energy is comparable in
magnitude to the kinetic energy; thus, the force errors from gradient
noise and excessive spurious dissipation corrupt the solution on small
scales such that the correct cascade of turbulent energy is not
reproduced.

\subsubsection{The intergalactic medium}

Early simulation results for the Ly-$\alpha$ forest obtained using
both SPH \citep{Hernquist:1996, Katz1996lya} and grid codes
\citep{Zhang1995, Miralda:1996}
agreed well.  More refined comparisons \citep[e.g.,][]{Regan:2007,
  Bird:2013} have demonstrated that when applied to the same initial
conditions, SPH and grid codes yield statistical measures for the
Ly-$\alpha$ forest, such as flux probability distribution functions and
power spectra, that agree at the $\sim 1\%$ level.

The reasons for this agreement can be understood based on
the discussion in Section \ref{S:limitations_of_SPH}.
The energy density of the IGM gas responsible for the Ly-$\alpha$ forest
is dominated by kinetic and gravitational energy; thus, errors in the
local fluid quantities due to noise are subdominant.
Furthermore, the physical state of the gas is simple, in the sense
that different phases of gas are not in close proximity.  Thus, errors
such as those discussed in Sections \ref{S:inaccurate_instabilities}
and \ref{S:inaccurate_continuity} will not greatly affect the IGM, and
hence the SPH results are fairly reliable. Whether this conclusion also extends
to metal lines, for which issues of mixing of galactic outflows with
pristine IGM gas become important, has not been investigated thus far.

\subsubsection{Gas accretion on to galaxies from hot haloes}

Within galaxy haloes, however, the physical state of the gas can differ
significantly between, e.g., \gadgetthree and \arepo
\citep{Vogelsberger:2012, Nelson:2013}.  Why should SPH predict the
physical state of the gas responsible for the Ly-$\alpha$ forest so reliably
yet fail so spectacularly within the haloes of galaxies? 

This can be understood by realizing that the internal energy in
approximately hydrostatic gas is no longer negligible compared with
the kinetic and gravitational energy. Noise in the local SPH estimates
can then significantly affect the physical state of the halo gas by, for
example, producing spurious viscous heating effects that reduce
cooling flows \citep[e.g.,][]{Nelson:2013}.  Moreover, the gas within
haloes can exhibit complex phase structure, with cold, dense gas in
close proximity to and shearing relative to shock-heated diffuse gas.
When simulated with traditional SPH, the different gas phases will not
mix correctly, as demonstrated by \citet{Torrey:2012disks}, because
fluid instabilities (Section \ref{S:inaccurate_instabilities}) and
mixing due to fluid motions (Section \ref{S:inaccurate_continuity})
are not treated properly.  Instead, the cold gas can fragment into
clumps that remain intact \citep[e.g.,][]{Agertz:2007, Kaufmann2007},
thereby delivering an artificial supply of cold, low-angular-momentum
gas to the central galaxy, which in turn can inhibit the formation of
a rotationally supported disc.

\subsubsection{Star formation and AGN activity in galaxy mergers}
\label{S:merger_discussion}

In this paper, we have demonstrated that the results of idealised
(i.e., non-cosmological) numerical experiments involving isolated disc
galaxies and galaxy mergers are relatively similar between SPH and the
moving-mesh approach, in contrast with cosmological simulations of
forming galaxies. This finding especially holds for simulations in
which AGN feedback is not included or, more generally, during early
stages of mergers when the gas structure is relatively simple.  Later,
once the gas becomes virialized and feedback from BH growth drives
large-scale outflows, some detailed differences do however appear. The
discussion in Section \ref{S:limitations_of_SPH} can again be used as
a guide to understand this behaviour.

In the simulations presented here, we construct models of disc
galaxies that evolve in isolation or merge.  In these models, the gas
is initially rotationally supported in the galaxy potential; thus, its
internal energy is small compared with its kinetic and gravitational
energy and the flows are effectively supersonic.  In this regime, similar
to the gas that produces the Ly-$\alpha$ forest, we expect that noise
in the SPH estimates (see Section \ref{S:local_noise}) should not be a
significant source of error.  Moreover, in the approach adopted here,
as formulated originally in \citet{Springel:2003}, there is no effort
made to resolve the multiphase structure of the star-forming gas.
Instead, the ISM is described using a sub-resolution model.  Because
of this, as for the gas in the IGM, the gas locally has a simple
structure and different phases of gas do not exist in close proximity.
As long as the gas can be characterised well in this manner, we do not
expect errors associated with an inaccurate treatment of fluid
instabilities (Section \ref{S:inaccurate_instabilities}) or mixing
(Section \ref{S:inaccurate_continuity}) to affect our model galaxies.
These conditions can start to be violated in a galaxy merger as gas is
shock heated and virialized and as different gas phases start to shear
relative to one another because of, e.g., the action of AGN
feedback. We note that these conditions would also be violated in
simulations that have enough resolution to truly resolve the
multi-phase structure of the ISM. In this case, extremely strong local
density constrasts would be prevalent.

Furthermore, in contrast with cosmological simulations, in which gas
cooling is a crucial determinant of how and when gas is supplied to
galaxies, the gas flows in galaxy merger simulations are driven
primarily by gravitational torques. In both the \gadgetthree~and
\arepo~simulations presented here, we use the same accurate tree-based
gravity solver. Consequently, the gravitational forces are treated
equally accurately in both codes. This is another reason that the
results agree relatively well.  For the same reason, the dark matter
halo mass functions of \arepo~and \gadgetthree~cosmological
simulations agree very well despite the significant differences in the
distributions of baryons \citep{Vogelsberger:2012}.

As stressed above, the agreement is less good during the starburst and
post-starburst phases of the merger simulations.  During the
starburst, gas is shock-heated and a hot halo forms. If AGN feedback
is included, the amount of gas in the hot halo is increased. Once this
hot halo forms, cooling becomes important for the post-starburst SFR
and gas morphology and thus the relevant differences between SPH and
grid-based approaches become manifest. As in cosmological simulations
\citep{Vogelsberger:2012,Keres:2012} and idealised simulations of
`inside-out' disc formation \citep{Sijacki:2012}, gas cools more
effectively from the hot haloes formed in the mergers in \arepo~than
in \gadgetthree.  Furthermore, some gas that is ejected during the
starburst and AGN activity falls back on to the remnant during the
post-starburst phase. In the \arepo~simulations, infalling clumps and
filaments are effectively disrupted, whereas in the
\gadgetthree~simulations they survive (see Section
\ref{S:inaccurate_instabilities}); the differences in this regard are
the same that explain why `cold flows' are less prominent in
cosmological simulations performed with \arepo~\citep{Nelson:2013}, as
discussed above.

\subsection{Implications for previous work}
\label{S:implications_for_previous_work}

It is worthwhile considering the implications of the differences we
have demonstrated, regarding both the robustness (or lack thereof) of
previous work and comparisons of simulations with observations. We
restrict ourselves to comments on the rich literature of simulations
of isolated galaxies and galaxy mergers, which forms the central topic
of this paper.

For the reasons discussed above, we in general expect many of these
past works to be at most weakly affected by the inaccuracies of traditional
SPH provided they use cold gas from the start and do not have
sufficient resolution to resolve a truly multi-phase ISM.  However,
the differences that we found are much more significant once hot haloes are
present in the simulations. Hence, we expect that the results of
simulations that initially include hot haloes in the progenitor
galaxies \citep[e.g.,][]{Moster:2011hot_halo,Moster:2012hot_halo}
could differ significantly if a moving-mesh technique rather than
traditional SPH were used. Furthermore, conclusions regarding hot
haloes produced by gas shocking in mergers \citep[e.g.,][]{Cox:2004},
the X-ray emission from hot haloes \citep[e.g.,][]{Cox:2006X-rays},
and the properties of starburst- and AGN-driven winds
\citep[e.g.,][]{Narayanan:2008turtlebeach,Hopkins:2013merger_winds}
may depend on the numerical method employed. Also, because the
re-formation of discs in mergers depends on gas cooling from the hot
halo of the remnant, studies of disc re-formation
\citep[e.g.,][]{Springel:2005disks,Robertson:2006disk_formation,Robertson:2008,
Hopkins:2009disk_survival,Puech2012}
might also be affected by the spurious suppression of cooling that is
inherent in SPH. Indeed, we have already noted that the discs that
re-form in the mergers presented here are very different in the
\gadgetthree~and \arepo~simulations (see the fourth columns of
Figs. \ref{fig:NB_gas_comparison} and \ref{fig:gas_comparison}).

The differences between traditional SPH and moving-mesh simulations
may also have important implications for studying AGN feedback with
merger simulations. In particular, the differences in the BH masses
yielded by \arepo~and \gadgetthree can affect the strength of the
feedback, the AGN luminosity and duty cycle
\citep[e.g.,][]{Hopkins:2005quasar_evolution,Hopkins:2005quasar_lifetimes,Hopkins:2006unified_model},
and the $M_{\mathrm{BH}}-\sigma$ relation of the merger remnants
\citep[e.g.,][]{Robertson:2006}.\footnote{The details of the BH
  accretion and AGN feedback models can also affect the BH masses
  \citep[e.g.,][]{Wurster:2013,Newton:2013}. The code-dependent
  differences are an additional source of uncertainty.} However, we
caution that for the thermal AGN feedback model presented here, the
final BH masses depend primarily on the feedback coupling efficiency
$\epsilon_{\rm f}$ \citep{Springel:2005feedback}. Because this number
is very uncertain, it may be possible and reasonable to simply use a
slightly reduced value of $\epsilon_{\rm f}$ in \arepo~to make the
final BH masses more consistent with those yielded by \gadgetthree.
Furthermore, the scatter in the BH masses for different resolutions
suggests that the values of the final BH masses are only robust to
within a factor of a few \citep[see also][]{Newton:2013}.

Recently, \citet{Hopkins:2011self-regulated_SF,Hopkins:2013mergers}
have presented idealised isolated disc and galaxy merger simulations
performed with \gadgetthree that include models for star formation and
stellar feedback that are significantly more sophisticated than the
EOS approach used here. This work attempts to directly resolve the ISM
multi-phase structure. Hence, one possible concern is the use of
traditional SPH.  However, \citet{Hopkins:2013mergers} also demonstrate
that the results of their simulations agree rather well with the
results of \gadgetthree simulations that employ the much simpler EOS
approach.  Also, \citet{Hopkins:2013merger_winds} compared simulations
performed using the traditional density-entropy formulation of SPH
used in \gadgetthree with simulations performed using an alternative pressure-entropy
formulation of SPH \citep{Hopkins:2013SPH} that is designed to overcome the inaccurate treatment
of fluid instabilities; they found that the spurious cold clumps
present in the outflows in the standard \gadgetthree simulations were
not formed when the new pressure-entropy flavour of SPH was used.
Thus, many of the conclusions of \citeauthor{Hopkins:2013mergers} are
likely robust to the hydrodynamical solver employed because the
rotational support is still dominant over pressure forces, although
some details, such as the post-starburst gas morphologies, are clearly
affected.

\section{Conclusions} \label{S:conclusions}

We have compared a suite of idealised isolated disc and galaxy merger
simulations performed using the SPH code \gadgetthree~and the moving-mesh
hydrodynamics code \arepo, both with and without BH accretion and
AGN feedback. To isolate the effects of the hydrodynamical solver
used, we have kept all other aspects of the simulations (i.e., the
gravity solver, the treatment of cooling, and the sub-resolution
models for star formation, supernova feedback, BH accretion, and AGN
feedback) as similar as possible. Our principal conclusions are the following:

\begin{enumerate}
\item Unlike for cosmological hydrodynamical simulations, the results
  of idealised (non-cosmological) isolated disc and merger simulations
  performed with \arepo~and \gadgetthree~are similar because (1) in
  these simulations, the gas is already initialised by hand in a
  rotationally supported disc, and thus the flow is supersonic in
  character and noise in the SPH estimates is not a significant source
  of error; (2) the gas phase structure is relatively simple; and (3)
  gravitational torques (rather than cooling) are the primary
  determinant of the gas inflows.
\item When BH accretion and AGN feedback are not included, the results
  are quite insensitive to the code used. The SFHs and cumulative
  stellar mass formed versus time are remarkably similar.
\item When BH accretion and AGN feedback are included, the results of
  the two codes are qualitatively similar, but the quantitative
  differences are more significant. In particular, the BH masses can
  differ by as much as an order of magnitude. The
  \arepo~simulations typically yield larger BH masses.
\item The gas morphologies and phase structures are also similar but
  differ in detail. Primarily, \arepo yields smoother, less clumpy
  morphologies and less prominent hot gaseous haloes in the
  post-starburst phase.
\item As for cosmological simulations, the differences between the SPH
  and moving-mesh results are primarily caused by more efficient
  cooling of hot halo gas and stripping of gas clumps and filaments in
  \arepo. 
\item Much of the previously published results of idealised isolated
  disc and galaxy merger simulations are likely robust to the
  inaccuracies that are inherent in traditional SPH.  However, for
  some studies, such as simulations in which the progenitors are
  initialised with hot haloes and studies of disc re-formation in
  mergers, the \arepo~results may differ qualitatively from those
  yielded by traditional SPH. Consequently, it would be interesting to
  revisit such studies using \arepo.
\end{enumerate}

It is certainly reassuring that the bulk of our results for merger
simulations exhibit good quantitative agreement between SPH and the
very different moving-mesh technique. Interpreting this as a
vindication of using SPH for merger simulations would nevertheless be
incorrect.  As we have discussed, unsolved conceptual problems with
the accuracy of SPH and its convergence rate remain even in the most
recent incarnations of the proposed improved versions of SPH. We
therefore think that more accurate numerical techniques, such as our
moving-mesh approach, should clearly be preferred over traditional SPH
to perform such simulations, especially in future calculations for
which a higher numerical precision and the resolution of multi-phase
media is desired.

\acknowledgments

We thank Shy Genel, Phil Hopkins, Federico Marinacci, R\"{u}diger Pakmor, Ewald Puchwein, and Debora Sijacki for useful discussion;
Jorge Moreno and Fran\c{c}ois Schweizer for detailed comments on the manuscript; and the anonymous referee for a critical report
that helped to improve the paper. CCH is grateful to the Klaus Tschira Foundation for financial support,
acknowledges the hospitality of the Aspen Center for Physics, which is supported by the National Science Foundation Grant No. PHY-1066293,
and heartily thanks Andreas Bauer for many {\sc python} tips, which were very helpful for (almost completely) curing a long-standing {\sc idl} dependency.
VS acknowledges support from the European Research Council under ERC-StG grant EXAGAL-308037.
This research has made use of NASA's Astrophysics Data System Bibliographic Services.
\\
\footnotesize{
\bibliography{arepo_comparison,std_citations}
}

\label{lastpage}

\end{document}